\documentclass[useAMS,usenatbib,usegraphicx]{mn2e}

\usepackage{times}

\def\rchisq{\ensuremath{\chi_{\nu}^{2}}}

\newcommand{\tim}[1]{\ensuremath{\times 10^{#1}}}
\newcommand{\plmin}[3]{\ensuremath{{#1}^{#2}_{#3}}}
\def\deg{\ensuremath{^{\circ}}}

\def\mekal{{\sc mekal}}
\def\gauss{{\sc gauss}}
\def\bbody{{\sc bbody}}
\def\brems{{\sc brems}}
\def\pcfabs{{\sc pcfabs}}
\def\phabs{{\sc phabs}}
\def\xspec{{\sc xspec}}

\def\cms{cm$^{-2}$}
\def\cps{counts s$^{-1}$}

\newcommand{\kms}{km\,s$^{-1}$}
\newcommand{\myemail}{vitaly@neustroev.net}
\newcommand{\IRAF}  {{\sc iraf}}
\newcommand{\MIDAS}  {{\sc midas}}
\newcommand{\ROSAT}  {\textit{ROSAT}}
\newcommand{\Halpha} {H$\alpha$}
\newcommand{\Hbeta}  {H$\beta$}
\newcommand{\Hgamma} {H$\gamma$}
\newcommand{\HeII} {He\,{\sc ii}}

\newcommand{\fsaur} {FS~Aur}

\title[Optical and X-ray observations of FS Aurigae]
{Steps towards a solution of the FS Aurigae puzzle --\\
II. Confirmation of the intermediate polar status}
\author[V.\,V.\, Neustroev et al.]{V.\,V.\, Neustroev$^{1}$\thanks{E-mail:
\myemail},
G.\,H.\, Tovmassian$^2$, S.\,V.\, Zharikov$^2$, and George Sjoberg$^{3,4}$\\
$^{1}$Astronomy Division, Department of Physics, PO Box 3000, FIN-90014 University
of Oulu, Finland\\
$^{2}$Instituto de Astronomia, Universidad Nacional Autonoma de Mexico, Apdo. Postal 877,
Ensenada, Baja California, 22800 Mexico\\
$^{3}$The George-Elma Observatory, Mayhill, New Mexico, USA\\
$^{4}$American Association of Variable Star Observers, 49 Bay State Road, Cambridge,
MA 02138, USA}
\begin{document}

\date{Accepted 2013 April 10.  Received 2013 March 15; in original form 2012 August 8}

\pagerange{\pageref{firstpage}--\pageref{lastpage}} \pubyear{2013}

\maketitle

\label{firstpage}

\begin{abstract}

\fsaur\ is famous for a variety of uncommon and puzzling periodic photometric and
spectroscopic variabilities. It was previously proposed that the precession of a
fast-rotating magnetically accreting white dwarf can successfully explain these
phenomena.
We present a study of \fsaur\ based on two extensive sets of optical photometric
observations and three X-ray data sets in which we intended to verify whether
the observational properties of the long period modulations observed in \fsaur\
and V455 And are similar in appearance to the spin modulation in ordinary IPs.
These new optical observations have revealed, for the first time in photometric data,
the variability with the presumed precession period of the white dwarf, previously
seen only spectroscopically. We also found that the modulations with the precession
and orbital periods are evident in X-ray data.
We show that the observed properties of FS Aur closely resemble those of
other intermediate polars, thus confirming this cataclysmic variable as a member
of the class.

Our analysis of multicolour observations of intermediate polars has shown that
time-series analysis of colour indices appears to be a powerful technique for revealing
hidden variabilities and shedding light on their nature. We have found that the ($B-I$)
power spectrum of V1223~Sgr indicates the presence in the data of the spin pulsation
which is not seen in the optical light curve at all. Also, the analysis of the
colour indices of V455~And revealed the presence of the photometric variations
which, similarly to \fsaur, was previously observed only spectroscopically.
\end{abstract}


\section{Introduction}

Cataclysmic Variables (CVs) are close interacting binary stars which consist of
a white dwarf (WD) accreting material from a Roche-lobe-filling companion, usually
a late main-sequence star. Intermediate polars (IP) are an important subset of CVs
in which the magnetic field of the WD is strong enough to disrupt the inner accretion
disc or even prevent disc formation completely and to force the accreting material to
flow along field lines onto one or both magnetic poles.

CVs are known to be very active photometrically, showing variability on time scales
from seconds to years (see review by \citealt{Warner}). However, the only strictly
periodic system clocks in CVs are associated with the binary orbital period and the
rotation of the WD. They may also produce modulation sidebands (beat periods) often
observed in light curves of IPs (see \citealt{Warner-Beats} for the origin of optical
pulsations in CVs). In addition to these strictly periodic oscillations, unstable
quasi-periodic oscillations and superhumps may appear in the light curves of CVs.
We note that all these variabilities have periods much shorter, equal, or very
close to the orbital period of the binary system.

FS Aurigae represents one of the most unusual CV to have ever
been observed. The system is famous for a variety of uncommon and puzzling periodic
photometric and spectroscopic variabilities which do not fit well into any of the
established sub-classes of CVs.
\fsaur\ was discovered by \citet{Hoffmeister}, who classified it as a dwarf nova.
H$\alpha$ velocity variations with a period of 85.7 min ascribed to the binary orbital
period (OP), have been reported by \citet{Thorstensen}. This period was confirmed by
follow-up spectroscopic observations \citep{Neustroev2002,Tovmassian2003,Tovmassian2007}.
\citet{Neustroev2002} also detected substantial orbital variability of emission line profiles
and their equivalent widths.

The light curve of \fsaur\ is highly variable in its appearance, only sometimes
showing the OP.
The outlandish peculiarity
of \fsaur\ is the existence of well-defined photometric optical modulations with the
amplitude of up to $\sim0.5$ mag and a very coherent long photometric period (LPP)
of 205.5 min that exceeds the OP by 2.4 times \citep{Neustroev2002,Tovmassian2003}.
Furthermore,
in late 2004 the optical brightness of the system dropped by 2 magnitudes for
a few months. The spectral observations made in December 2004, revealed a second
long spectroscopic period (LSP) of 147 minutes, appearing in the far wings of the
emission lines. Frequency of this new period is equal exactly to the beat between
the OP and LPP: $1/P_{beat}=1/P_{orb} - 1/P_{phot}$ \citep{Tovmassian2007}.
It is interesting to note that the LPP has never been detected spectroscopically
whereas the LSP has not been seen in the photometric data.

The photometric and spectroscopic behaviour of \fsaur\ somewhat resembles that of IPs
in which several periodic signals are often observed. Nevertheless, the principal
difference between \fsaur\ and IPs is in the time-scales of their variabilities:
(a) most IPs are relatively long orbital period systems: only 5 of the 36 confirmed/ironclad
IPs\footnote{The catalog of IPs and IP candidates (Version 2011a) by Koji
Mukai, http://asd.gsfc.nasa.gov/Koji.Mukai/iphome/catalog/alpha.html}
have an OP below the period gap, and probably one more system
lies in the period gap; \\
(b) a spin period of WD in all IPs is shorter than the orbital one, typically
  $P_{spin} \leq 0.1 P_{orb}$
(for a detailed review on IPs, we refer the reader to \citealt{Patterson1994} and
\citealt{Hellier1996}). In contrast, the period which is proposed to be the OP of \fsaur,
is the shortest of all the detected periods in the system, and is only a little longer than
in SDSS J233325.92+152222.1 -- the IP with the shortest-known  orbital period.

This fact may call into question whether the orbital period has been correctly identified
in \fsaur, and why, for instance, are 205.5m, 147m and 85.7m not the orbital, beat
and spin periods, respectively? None the less, we claim that the 85.7 m period is
indeed the OP and our arguments are the following:
\begin{enumerate}
 \item It is commonly accepted that radial velocities are a much more reliable indicator
       of the orbital period than photometric variability. The LPP is the only period of
       the three detected in \fsaur\ which is not seen spectroscopically, and as such
       can be definitely ruled out as an orbital period.
 \item Both the 85.7 m period and the LSP are observed spectroscopically. The emission
       lines in many IPs also vary at the orbital and spin periods. In few cases,
       emission-line fluxes, equivalent widths and radial velocity variations with
       the spin period dominate over variations with the orbital period. We note,
       however, that the character of these two types of variabilities is very different.
       The profiles of broad emission lines found in most of CVs are generally determined
       by the macroscopic motion of the emitting gas in the binary system. The periodic
       motion, due to Doppler shifts, generates spectral components with a sinusoidal
       modulation of wavelength which revealed themselves as coherent S-waves in a
       trailed spectrogram. Participating in the orbital motion, the main contributors
       to the emission lines -- the accretion disc, secondary star, gas stream and
       bright spot -- superpose many such S-waves with a wide range of velocity
       amplitudes and create a characteristic multicomponent trailed spectrogram.
       A good collection of such spectrograms is provided by \citet{TrailedSpectra}.
       The trailed spectra of \fsaur\ folded with the 85.7 m period closely resemble
       many of them.
       In contrast, the sources of the spin spectral variability are located in a more
       compact area around the accretion curtain and/or the innermost part of the
       accretion disc and as such dominate in the far wings of emission lines
       \citep{Hellier1999}. The LSP variability in \fsaur\ has much more similarities
       with such spin variability. Thus, we have no reason to believe that the LSP is
       the true orbital period, instead of the 85.7 m one.
 \item In V455 And, another cataclysmic variable which, similarly to \fsaur, exhibits
       two very different spectral periods -- a short one of 81 min and a longer one of
       210 min \citep{Tovmassian2007} -- eclipses in the system's light curve
       unambiguously confirms the OP to be 81 minutes.
\end{enumerate}

We therefore conclude that the weight of evidence favours 85.7 m as the OP of \fsaur.
The nature of the two other periods is thus still an open question. Neither the LPP
nor the LSP can be accepted as the period of rotation of the WD. On theoretical grounds,
the spin period of the WD in IPs is determined by the combined action of accretion and
magnetic field \citep{Norton2004}. Fast rotation can be expected given the large amount
of angular momentum transferred by accreting matter. The WD is expected to spin-up to
a value near to the rotation period at the inner edge of the accretion disc
\citep{KingLasota,WarnerWickramasinghe}. From the observational point of view,
no CVs have been found with the primary rotating so slowly.

Thus, the discovered multiple periodic phenomena in \fsaur\ represent a real
challenge to the theory of accretion processes in low mass close binaries.
Moreover, several other systems are now known to show
variabilities with periods much longer than the orbital one (GW Lib, V455 And),
and the origin of these modulations also remains unclear. It seems that \fsaur\ and
similar objects can represent a new type of CVs, which still needs an explanation
of their nature.

\begin{figure*}
\begin{center}
\hbox{
\includegraphics[width=17cm]{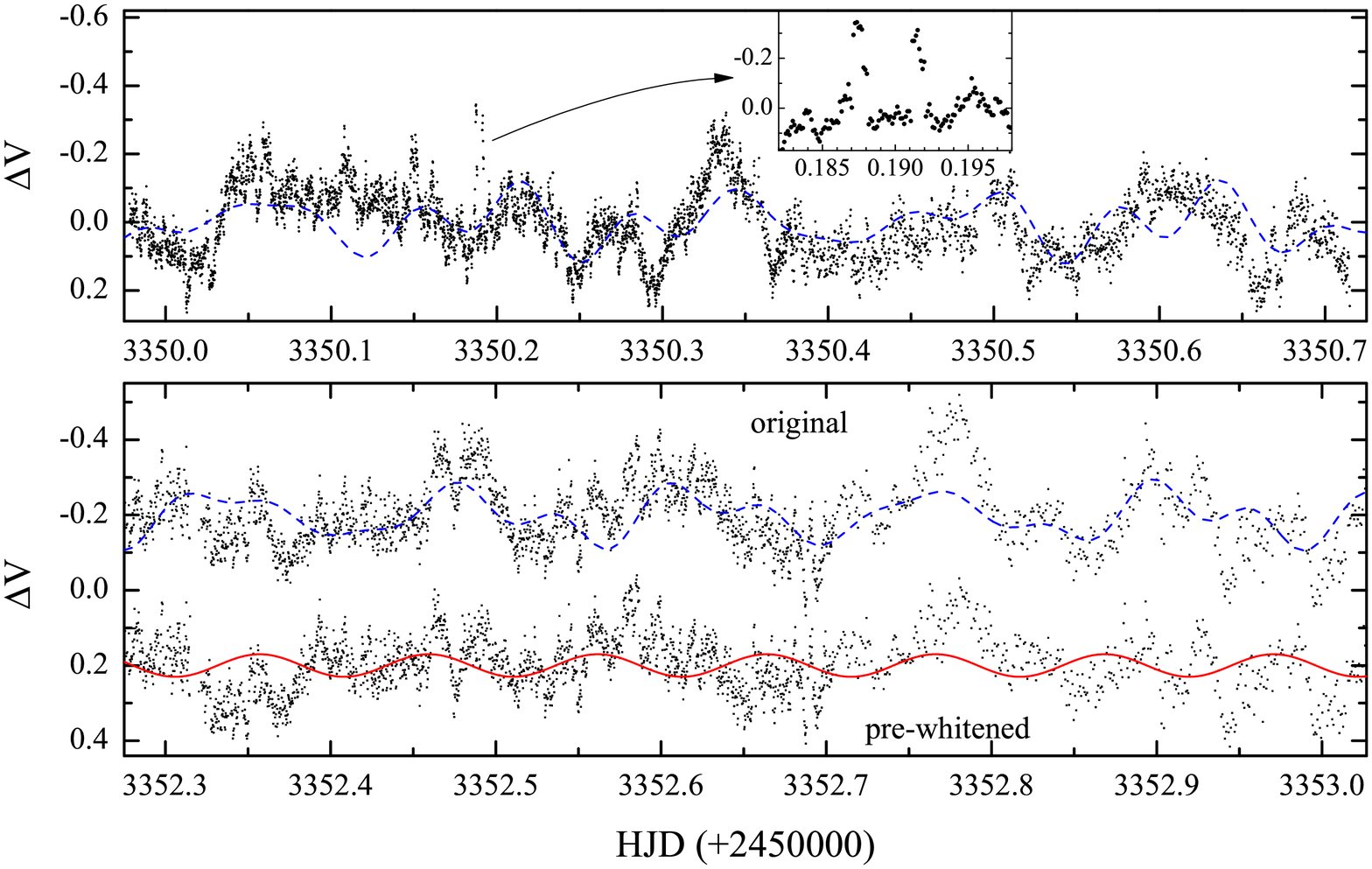}
}
\end{center}
\caption{Two 18 hour samples of the $V$ light curve of FS Aur from \textit{set-04}.
In the inset of the upper panel two half-magnitude flares with durations shorter
than 100 sec are shown.
In the lower panel the original light curve (upper) is shown along with the light curve
(bottom) pre-whitened by the LPP frequency and its first harmonic and by the OP
frequency. The dashed blue lines are the sum of sinusoidal fits of the LPP and its
first harmonic and the OP which yielded the ephemeris~(\ref{ephemerisLPP}) and
(\ref{ephemerisOP}). The solid red line is a fit of the LSP corresponding to the
ephemeris~(\ref{ephemerisLSP}).}
\label{fig:LightCurves}
\end{figure*}

\begin{table}
\caption{Journal of photometric observations of FS~Aur (\textit{set-2004}). The horizontal
line divides two subsets, \textit{set-04} and \textit{set-05}.}
\begin{tabular}{ccccc}
\hline
\hline
 HJD start  & Duration  &  Exp. Time  &   Band    & Telescope$^*$     \\
 2450000+   & (hours)   &     sec.    &           &                \\
\hline
 3347.010   &  8.69     &     10      &    V      &   BOAO         \\
 3347.691   & 16.54     &     10      &    V      &   SPM--BOAO    \\
 3348.676   & 15.89     &     10      &    V      &   SPM--BOAO    \\
 3349.978   & 17.69     &  5, 10      &    V      &   BOAO--Loiano \\
 3350.984   & 17.78     &  5, 10      &    V      &   BOAO--Loiano \\
 3351.719   &  7.37     &      60     &   white   &   CBA          \\
 3352.275   & 10.30     &     10      &    V      &   Loiano       \\
 3352.682   &  8.27     &      60     &   white   &   CBA          \\
 3353.659   &  8.82     &      60     &   white   &   CBA          \\
 3354.671   &  8.52     &      60     &   white   &   CBA          \\
 3355.652   &  8.97     &      60     &   white   &   CBA          \\
 3356.661   &  8.52     &      60     &   white   &   CBA          \\
 3358.681   &  8.06     &      60     &   white   &   CBA          \\
 \hline
 3383.615   &  7.85     &      60     &   white   &   CBA          \\
 3388.635   &  6.88     &      60     &   white   &   CBA          \\
 3390.633   &  6.92     &      60     &   white   &   CBA          \\
 3394.620   &  6.72     &      60     &   white   &   CBA          \\
 3403.589   &  6.96     &      60     &   white   &   CBA          \\
 3404.624   &  6.13     &      60     &   white   &   CBA          \\
 3405.631   &  5.96     &      60     &   white   &   CBA          \\
 3406.612   &  6.40     &      60     &   white   &   CBA          \\
\hline
\end{tabular}

\begin{tabular}{l}
$^*$ Telescopes:\\
BOAO -- Bohyunsan Optical Astronomy Observatory, South Korea \\
CBA -- Center of Backyard Astronomy  \\
Loiano -- Osservatorio Astronomico di Bologna, Stazione di Loiano, Italy \\
SPM -- Observatorio Astronomico Nacional San Piedro Martir, Mexico \\
\end{tabular}

\label{ObsPhotTab}
\end{table}


In order to explain the puzzling behaviour of \fsaur, \citet{Tovmassian2003} proposed,
and \citet{Tovmassian2007} enhanced the IP scenario with a fast-rotating, magnetically
accreting WD which precesses with the LSP (see Section 3.2 in \citealt{Tovmassian2007}).
Due to the magnetic nature of the WD, the accreting material  in \fsaur\ should
be controlled and channelled by the strong magnetic field of the WD within its
magnetospheric radius, as it is in an ordinary IP. However, as the WD is involved
in two periodic motions -- rotational and precessional -- one could expect to observe
stable modulations in the optical and X-ray light curves with both the periods.
Observational properties of these modulations should depend, among other things, on
geometrical factors such as the angles between the magnetic, rotational and
precessional axes, but they should both be similar in appearance to the spin
modulation in IPs.

The search for a WD spin modulation in \fsaur\ is still to be concluded.
The first attempt to detect this modulation was made by \citet{Neustroev2005}.
Even though they found a signature of $\sim$101 and/or 105 sec oscillation in
the optical power spectra, there is still no conclusive evidence.

It appears that \fsaur\ may have even longer photometric period, which was described
and discussed by \citet{FS2012}. A dynamical solution for that very long photometric
period (VLPP) may have an influence on the stochastic variability, but most likely
is not directly related to the shorter, strictly periodic variabilities discussed in
this paper.

In this paper, we present the analysis of two extensive sets of optical photometric
observations and three X-ray data sets of \fsaur. We also present the analysis
of optical multicolour photometric observations of V455 And.
We intended to verify whether the observational properties of the long period
modulations observed in \fsaur\ and V455 And are similar in appearance to the spin
modulation in ordinary IPs. We compared our results with the multicolour observations
of eight ironclad and confirmed IPs, generously provided to us by our colleagues
worldwide. For one more well-known IP V1223~Sgr, we used our own multicolour observations.

\begin{table*}
\caption{Journal of photometric observations of FS~Aur (set-2011).}
\begin{tabular}{||cccc||cccc||cccc||}
\hline
\hline
 HJD start  & Duration  &   Band   & Comments &  HJD start  & Duration  &   Band   & Comments & HJD start  & Duration  &   Band   & Comments  \\
 2450000+   & (hours)   &          &          &  2450000+   & (hours)   &          &          & 2450000+   & (hours)   &          &           \\
\hline
 5536.611   &  10.00    &    V     & Outburst &  5575.564   &  9.09     &    V     &          & 5621.586   &  5.80     &    BVRI  &   \\
 5537.608   &  10.02    &    V     & Outburst &  5576.651   &  6.96     &    V     &          & 5622.667   &  3.61     &    BVRI  &   \\
 5538.606   &  10.00    &    V     &          &  5577.565   &  2.58     &    V     &          & 5623.590   &  5.64     &    BVRI  & Outburst \\
 5539.603   &  9.99     &    V     &          &  5581.823   &  1.96     &    BVRI  &          & 5625.591   &  4.44     &    BVRI  & Outburst \\
 5541.610   &  10.16    &    V     &          &  5582.568   &  6.07     &    BVRI  &          & 5626.588   &  5.15     &    BVRI  & Outburst \\
 5542.596   &  10.66    &    V     &          &  5583.571   &  8.64     &    BVRI  &          & 5629.589   &  5.07     &    BVRI  &   \\
 5543.593   &  10.61    &    V     &          &  5584.570   &  5.05     &    BVRI  &          & 5648.660   &  2.10     &    BVRI  & Outburst \\
 5544.591   &  10.67    &    V     &          &  5587.583   &  8.11     &    BVRI  & Outburst & 5649.598   &  3.20     &    BVRI  &   \\
 5545.588   &  10.64    &    V     &          &  5588.596   &  7.65     &    BVRI  &          & 5650.598   &  3.62     &    BVRI  &   \\
 5548.715   &  7.61     &    V     &          &  5589.572   &  8.45     &    BVRI  &          & 5651.601   &  3.25     &    BVRI  &   \\
 5551.703   &  5.12     &    V     & Outburst &  5590.572   &  8.46     &    BVRI  &          & 5652.601   &  3.86     &    BVRI  &   \\
 5552.595   &  9.94     &    V     & Outburst &  5591.572   &  8.32     &    BVRI  &          & 5654.631   &  2.68     &    BVRI  &   \\
 5554.591   &  9.28     &    V     &          &  5596.592   &  7.37     &    BVRI  &          & 5865.708   &  7.33     &    BVRI  &   \\
 5555.566   &  9.81     &    V     &          &  5597.576   &  6.48     &    BVRI  &          & 5866.705   &  7.58     &    BVRI  &   \\
 5556.557   &  10.78    &    V     &          &  5598.576   &  7.39     &    BVRI  & Outburst & 5867.701   &  4.68     &    BVRI  &   \\
 5557.578   &  10.61    &    V     &          &  5600.575   &  3.10     &    BVRI  & Outburst & 5868.698   &  7.41     &    BVRI  &   \\
 5559.597   &  9.81     &    V     &          &  5603.579   &  7.08     &    BVRI  &          & 5870.695   &  7.78     &    BVRI  & Outburst \\
 5562.727   &  6.79     &    V     &          &  5604.608   &  6.44     &    BVRI  &          & 5871.712   &  7.26     &    BVRI  & Outburst \\
 5563.656   &  7.97     &    V     &          &  5605.579   &  5.85     &    BVRI  &          & 5872.688   &  7.91     &    BVRI  & Outburst \\
 5564.557   &  10.82    &    V     &          &  5607.581   &  6.93     &    BVRI  & Outburst & 5873.687   &  8.04     &    BVRI  &   \\
 5565.557   &  10.28    &    V     & Outburst &  5608.606   &  4.92     &    BVRI  &          & 5874.682   &  5.02     &    BVRI  &   \\
 5566.589   &  9.22     &    V     & Outburst &  5609.720   &  3.17     &    BVRI  &          & 5875.680   &  7.96     &    BVRI  &   \\
 5567.558   &  9.58     &    V     & Outburst &  5610.638   &  4.43     &    BVRI  &          & 5892.679   &  8.39     &    BVRI  &   \\
 5569.558   &  8.81     &    V     &          &  5612.606   &  4.22     &    BVRI  &          & 5894.637   &  9.50     &    BVRI  &   \\
 5570.561   &  10.20    &    V     &          &  5613.603   &  1.74     &    BVRI  &          & 5895.633   &  9.55     &    BVRI  & Outburst \\
 5571.560   &  3.98     &    V     &          &  5614.584   &  6.57     &    BVRI  &          & 5896.629   &  9.50     &    BVRI  & Outburst \\
 5573.561   &  7.71     &    V     &          &  5615.584   &  6.31     &    BVRI  &          & 5899.632   &  9.64     &    BVRI  &   \\
 5574.609   &  8.09     &    V     &          &  5616.585   &  6.31     &    BVRI  &          & 5913.581   &  10.09    &    BVRI  & Outburst \\
\hline
\end{tabular}
\label{ObsPhotTab2}
\end{table*}

\section{Observations}

\subsection{Optical observations}

\subsubsection{The 2004 multi-site observational campaign}

In late 2004, in order to obtain a long uninterrupted light curve of FS~Aur, we
acquired optical time-series photometry with the broadband Johnson--Cousins filter
V using three telescopes located in South Korea, Italy and Mexico. In South Korea,
the observations were made using the SITe 2048 $\times$ 2048 CCD camera attached
to the f/8 Cassegrain focus of the Bohyunsan Optical Astronomy Observatory (BOAO)
1.8 m telescope. In Italy, the observations were obtained using the imaging
spectrograph BFOSC with a 1300 $\times$ 1340 pixels EEV CCD
attached to the Cassini 1.5 m telescope at Loiano. In Mexico, the observations
were performed at the Observatorio Astronomico Nacional (OAN SPM) on the 1.5-m
telescope using the SITe SI003 CCD camera. A part of the latter observations were
obtained simultaneously with the spectroscopic observations presented in
\citet{Tovmassian2007} and which we analyze in this paper again.
Unfortunately, due to not-optimal weather conditions we were unable to obtain
any whole-day observations. We did acquire four long observations with
a duration of 16--18 hours each. Selected comparison and check stars from the
field were used for differential photometry performed using the standard
reduction systems \MIDAS\ and \IRAF.
Two 18 hour samples of this light curve are shown in Figure~\ref{fig:LightCurves}.

Additional photometry was acquired by members of the Center for Backyard
Astrophysics (CBA)\footnote{See http://cba.phys.columbia.edu}. For details of
their activity and procedures used to obtain and reduce the data, we refer to
\citet{CBA}.

Hereinafter we call this set of observations \textit{set-2004}.
Table~\ref{ObsPhotTab} provides a journal of these observations.

During this observational campaign, FS Aur was found to exhibit a decrease
in luminosity of 2 mags that lasted a few months until the end of 2005 January
\citep{Tovmassian2007}. As the character of the variability of the star has
changed somewhat during the campaign (see below), we divide these data
into two subsets. Observations taken between HJD 2453347--2453358 we call
\textit{set-04}, the rest of the data taken between HJD 2453383--2453406 we
call \textit{set-05}.

\subsubsection{The 2010-2011 observational campaign}
\label{Sec:campaign2010}

During the winter of 2010-2011 we initiated and conducted an observing campaign,
lasting more than 150 consecutive nights. During this campaign we observed 11
consecutive low-amplitude outbursts ($\leq$ 2 mag) \citep{NeustroevIAU}. The bulk of
these data will be discussed elsewhere. Here we mostly concentrate on observations
obtained in the quiescent state.

The data presented in this paper were taken using the 0.35-m Celestron C14
robotic telescope, located at New Mexico Skies in Mayhill, New Mexico.
Before the night of February 24, we used an SBIG ST-8XME CCD camera with
1530 $\times$ 1020 pixels, and since then it was replaced with an SBIG ST-10XME,
2184 $\times$ 1472 pixels. Both cameras were used with Johnson-Cousins $BV(RI)_C$
Astrodon Photometrics filters.

The observations were conducted every clear night from November 26, 2010 until May
3, 2011. Thus, more than 80 nights of time-resolved photometry were taken (38 of
them were in quiescence), and almost 14\,000 $V$-band data points were obtained.
Additionally, between January 20 and March 9, 2011, 31 nights of time-resolved multicolor
$BV(RI)_C$ photometry were taken (22 of them were in quiescence). Moreover, after
acceptance of our Swift ToO request (see below), we obtained 5 more nights of
multicolour observations (3 nights were in quiescence), between March 28 and
April 3, 2011. Depending on the weather conditions, we monitored the star for 6--8
hours per night in the beginning of the campaign and for 3--4 hours in the end.

The following autumn we continued monitoring FS Aur. Between the nights of October 31
and December 4, 2011, 15 nights of time-resolved multicolour photometry were
taken (8 nights were in quiescence).

During all quiescent observations,
exposure times were 240 seconds for the $VRI$ filters and 400 seconds for $B$.
The observations were taken quasi-simultaneously in the $V\,B\,V\,R\,I$ sequence.
The reduction procedure was performed using the \IRAF\ environment and the software
AIP4Win v. 2.4.0 \citep{AIP4Win}. All of the \fsaur\ photometry used the secondary
standards found in \citet{AAVSO} to establish the zero points. \citet{AAVSO} do not
provide R magnitudes, so instead we used R magnitudes from \citet{Misselt}. The
typical accuracy of our measurements varied between 0.01 and 0.10 mag depending
on the brightness of the object and the weather conditions. The median value of
the photometric errors in the quiescent state was 0.03 mag in the $VRI$ filters
and 0.04 mag in $B$.

Hereinafter we call this set of observations \textit{set-2011}.
Table~\ref{ObsPhotTab2} provides a journal of this set of observations.

\begin{table}
 \centering
  \caption{Log of the \emph{Chandra} and \emph{Swift} observations of \fsaur.}
   \begin{tabular}{@{}lllll@{}}
\hline
\hline
Observatory &    ObsId      & Obs. date  & HJD start & Exp.     \\
Dataset     &               &            & 2450000+  & (sec)    \\
\hline
Chandra     &        5189	& 2005-01-05 & 3375.612  & 25180    \\
\hline
Swift-2007  & 00030873001	& 2007-01-17 & 4117.921  & 6690     \\
            & 00030873002	& 2007-01-18 & 4118.525  & 11797    \\
            & 00030873003	& 2007-01-19 & 4119.528  & 10693    \\
\hline
Swift-2011  & 00030873004	& 2011-03-29 & 5650.304  & 4055     \\
            & 00030873005	& 2011-03-30 & 5651.034  & 3934     \\
            & 00030873006	& 2011-03-31 & 5652.237  & 4323     \\
            & 00030873007	& 2011-03-31 & 5652.509  & 8148     \\
\hline
\end{tabular}
\label{ObslogXrayTab}
\end{table}

\subsection{X-ray data}
\label{xrays}

From previous X-ray observations we know that \fsaur\ is a rather hard and relatively
bright X-ray source with a count-rate of 0.13 counts/s as observed with \ROSAT\ PSPC
\citep{Tovmassian2003},
suggesting that the system might be an IP. Most IPs show modulations at the spin and
orbital periods. Motivated by our findings, described in the following Section, we have
found that it is instructive to compare the optical and X-ray light curves of \fsaur.
A log of the X-ray observations is presented in Table~\ref{ObslogXrayTab}.

\begin{figure}
\centerline{\includegraphics[width=8.5cm]{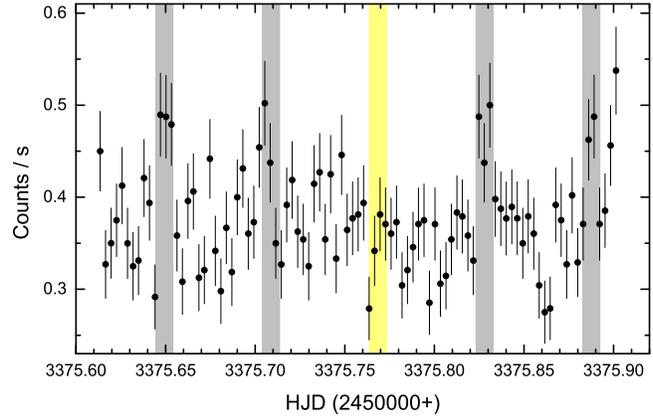}}
\caption{The X-ray light curve of \fsaur\ obtained with Chandra and binned at 264 s
intervals. The shaded areas are centered on the orbital phase interval 0.27--0.43
during which strong and narrow periodic flares were observed. One of these peaks
(marked by yellow area instead of grey) was seem to be missing.}
\label{fig:Chandra_LC}
\end{figure}

\begin{figure}
\centerline{\includegraphics[width=8.5cm]{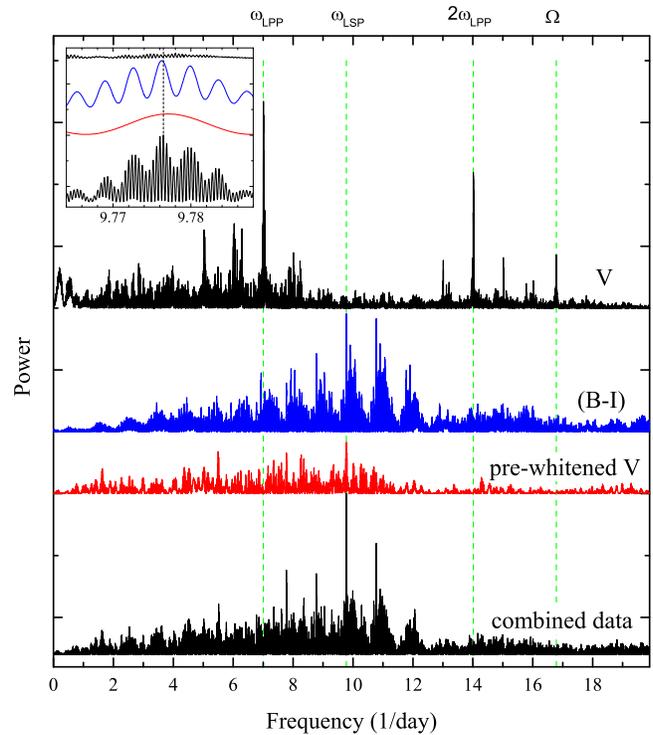}}
\caption{The Lomb-Scargle power spectra for different sets of optical data of \fsaur.
The upper black line represent the spectrum for the entire V photometry. The strongest
peaks at $f$ = 7.0077 day$^{-1}$ and 14.0161 day$^{-1}$
correspond to the LPP and its first harmonic. Another relatively strong peak at
16.7848 day$^{-1}$ corresponds to the orbital frequency. The three bottom
spectra were calculated from the ($B-I$) colour-index data from the \textit{set-2011}
(blue), from the pre-whitened $V$ light curve from the \textit{set-2004} (red), and
from the combined data set consisting of the former two and the ULTRACAM colour data
(black, see Section~\ref{sec:LSP} for details).
The strongest peaks in these spectra at $f$ = 9.77644 day$^{-1}$ exactly
coincide with the LSP which is the presumed precession period of the WD.
The inset shows the enlarged region around the LSP frequency.}
\label{fig:FS_Aur_PS}
\end{figure}


\begin{figure*}
\includegraphics[width=8.5cm]{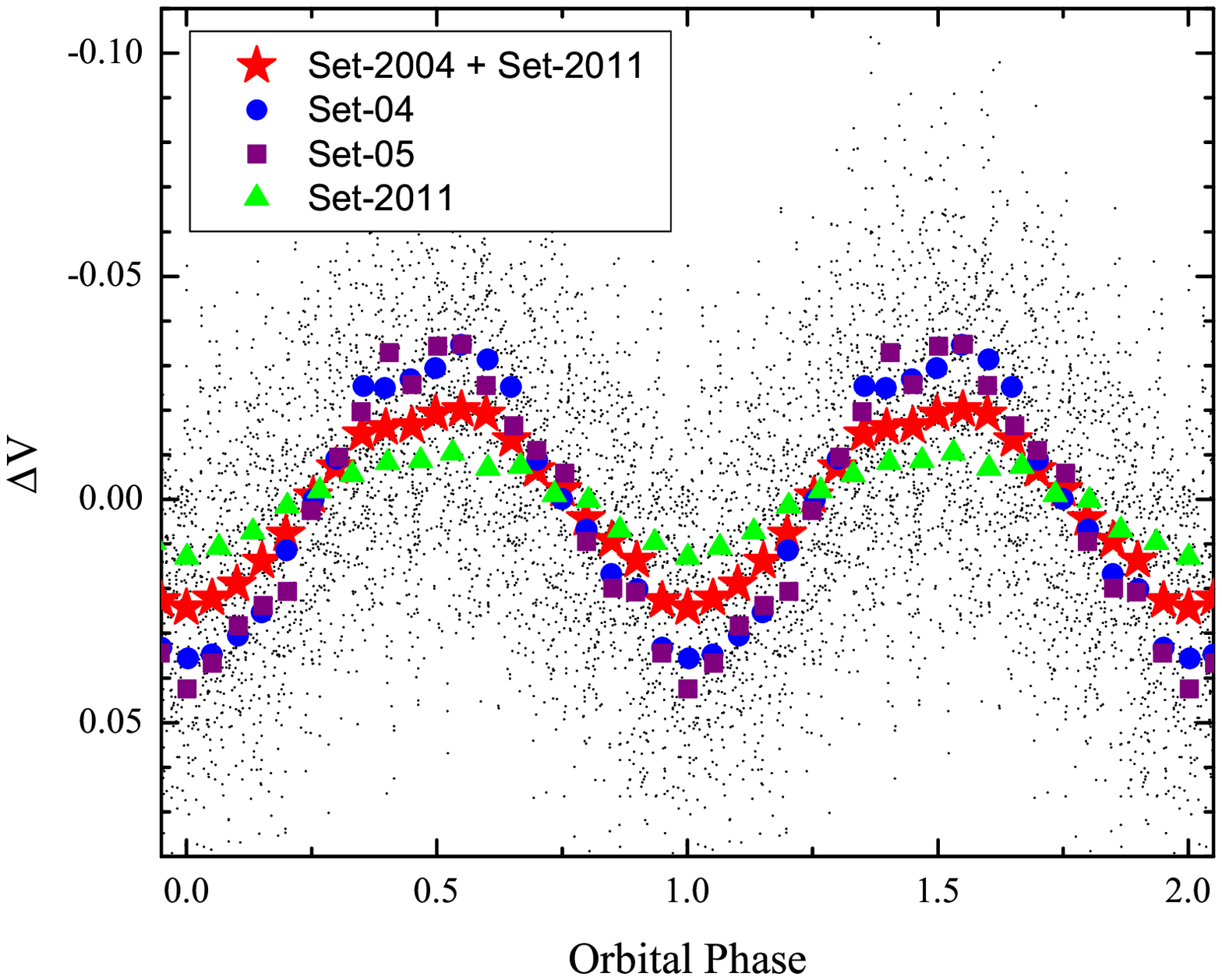}
\hspace{5 mm}
\includegraphics[width=8.5cm]{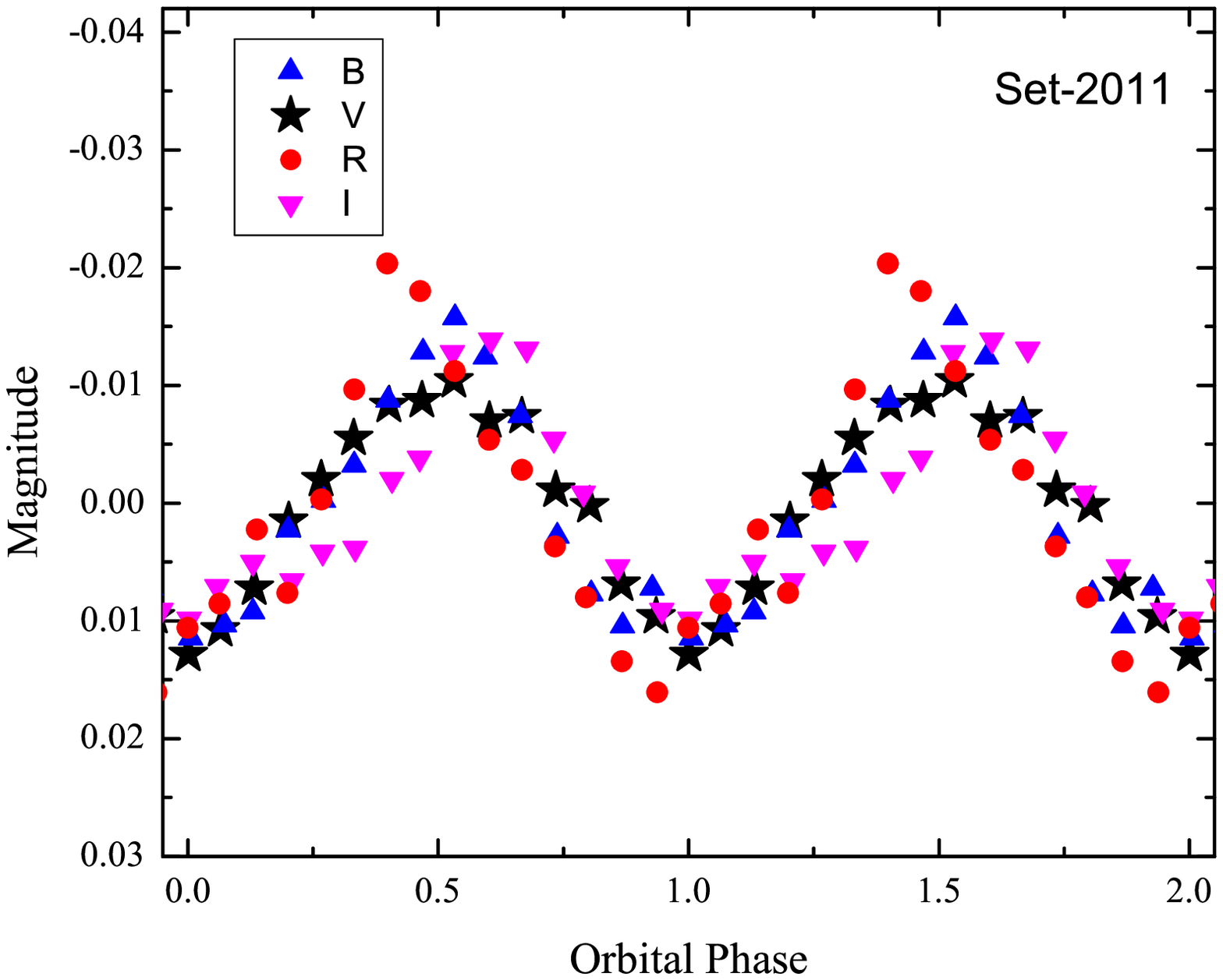}
\caption{\textit{Left:} the entire pre-whitened $V$ phase curve (with the means,
and linear trends subtracted for each night of observations) folded with the OP
according to the ephemeris~(\ref{ephemerisOP}). The large symbols represent data from
different subsets averaged in 20 phase bins. \textit{Right:} the folded light
curves in different colours of the \textit{set-2011}. All data are plotted twice
for continuity.}
\label{fig:FoldedOP}
\begin{center}
\hbox{
\includegraphics[height=7.3cm]{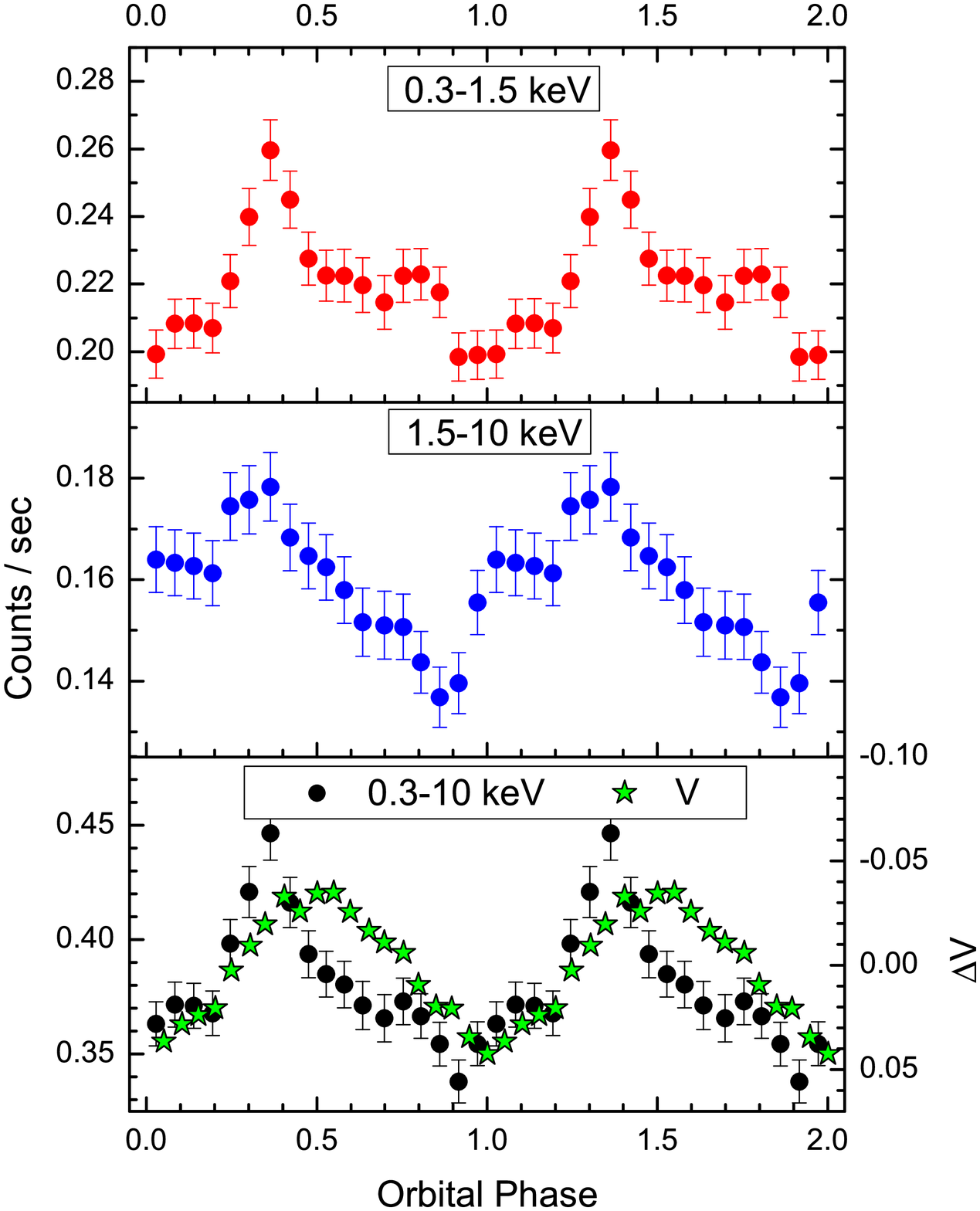}
\includegraphics[height=7.3cm]{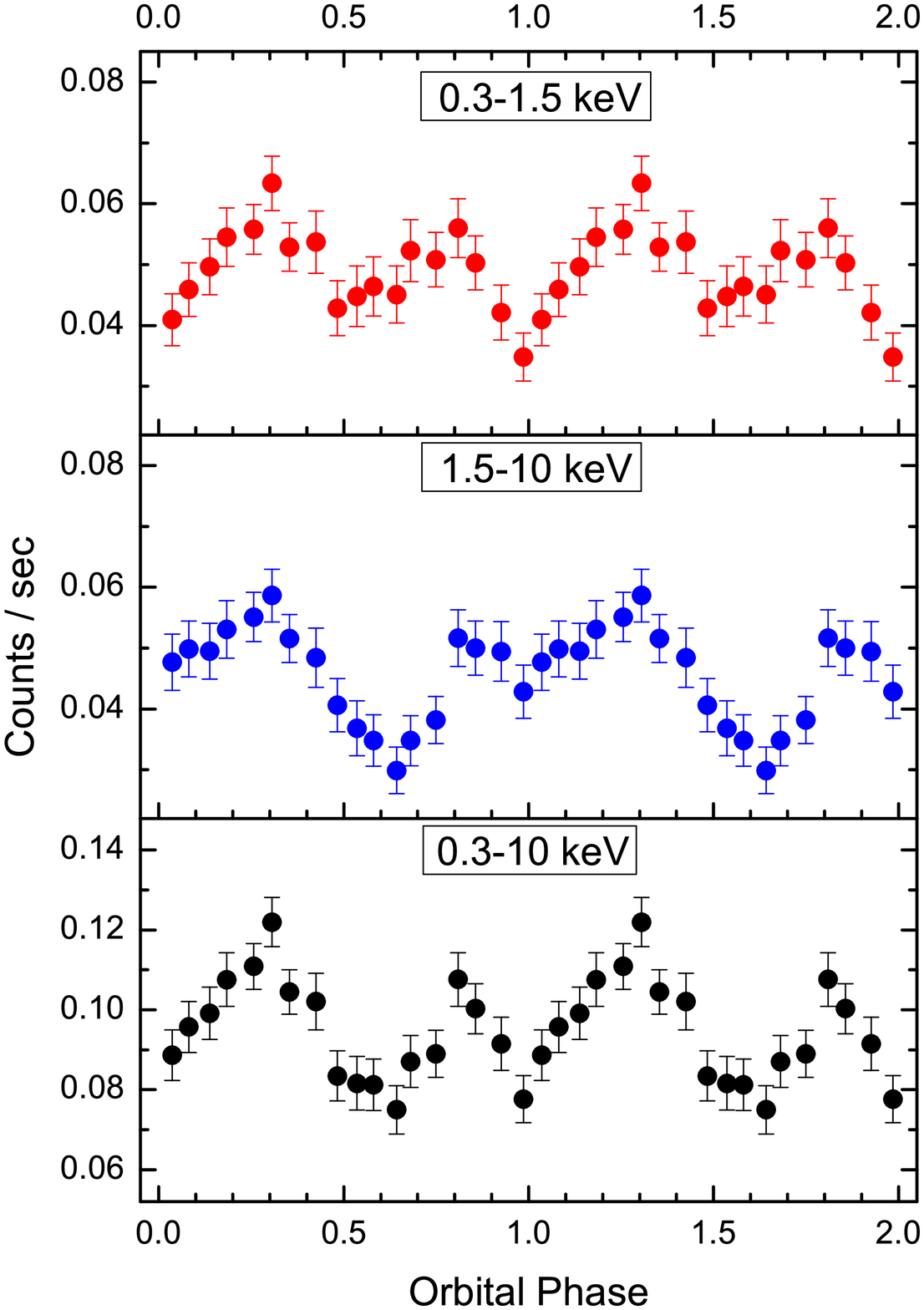}
\hspace{5mm}
\includegraphics[height=7.3cm]{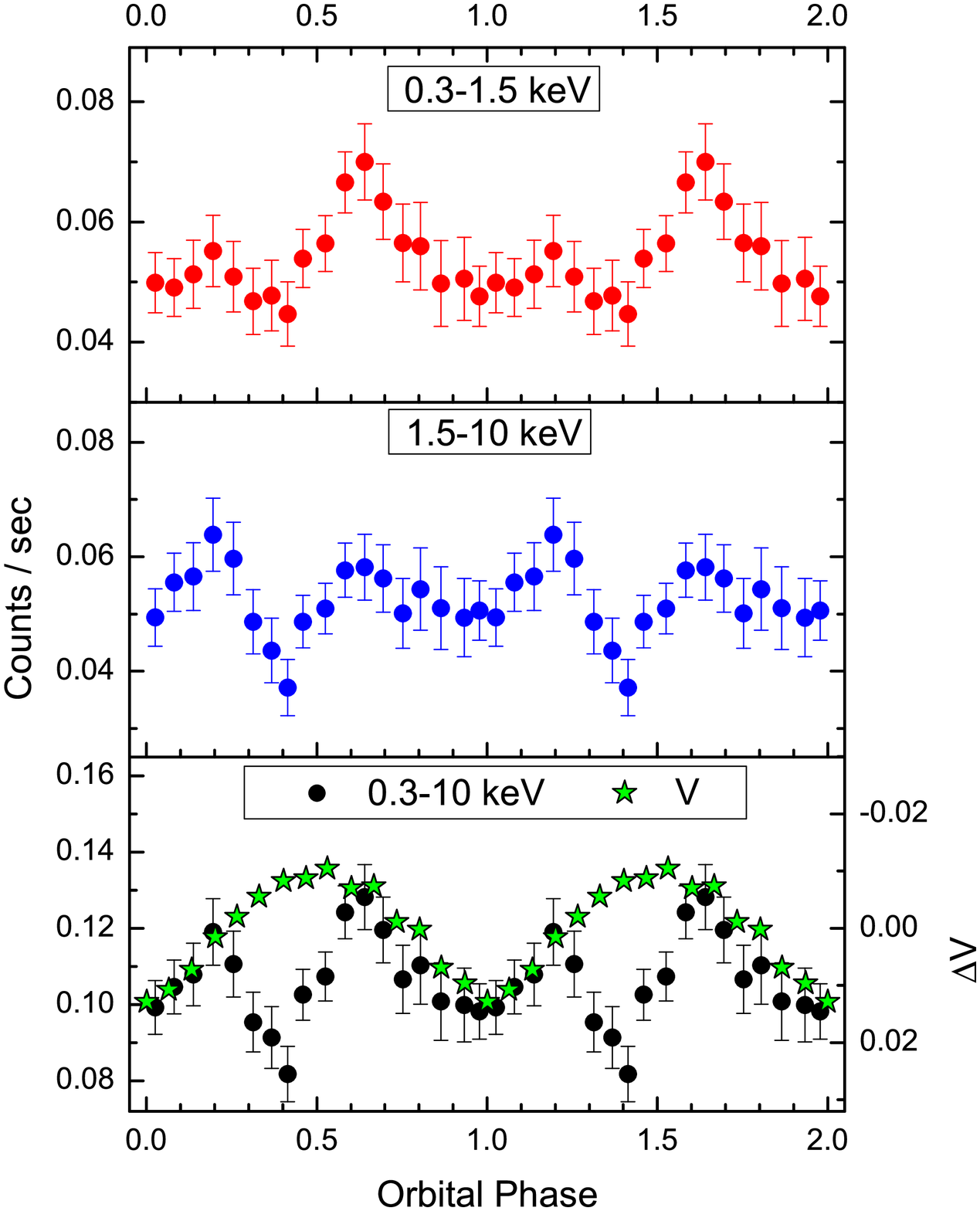}
}
\end{center}
\caption{The Chandra (left panel), Swift-2007 (middle) and Swift-2011 (right)
data folded with the OP  according to ephemeris~(\ref{ephemerisOP}). The panels
are from top to bottom: the soft range (0.3-1.5 keV), hard range (1.5-10 keV),
and the total X-ray flux (0.3-10 keV). The latter are shown together with the
corresponding $V$ light curves (the Chandra data are compared with the optical
\textit{set-05} whereas the Swift-2011 is compared with the \textit{set-2011}).
Two cycles are shown for continuity.}
\label{fig:xraysOP}
\end{figure*}

\begin{figure*}
\begin{center}
\hbox{
\includegraphics[width=5.6cm]{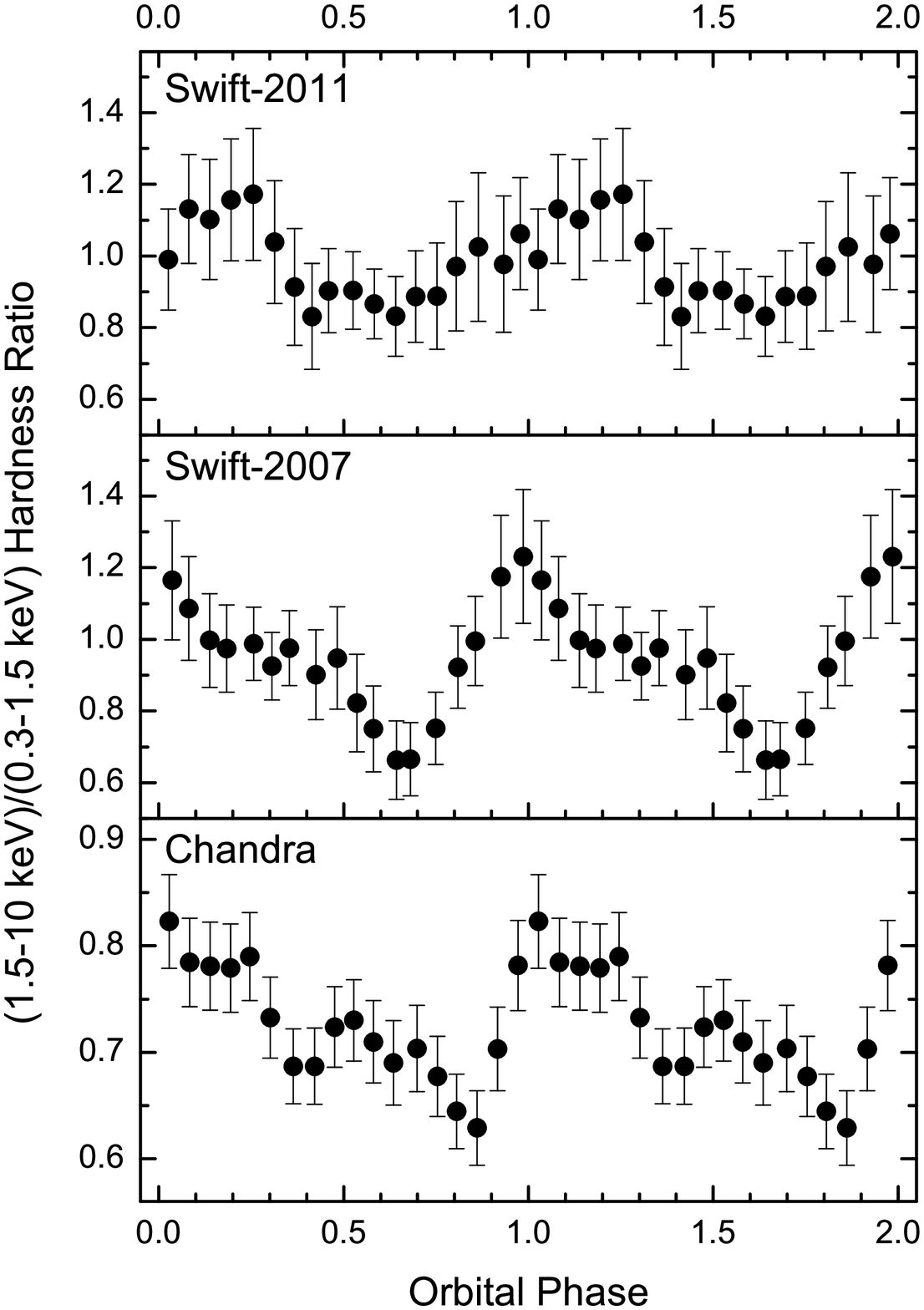}
\hspace{2 mm}
\includegraphics[width=5.6cm]{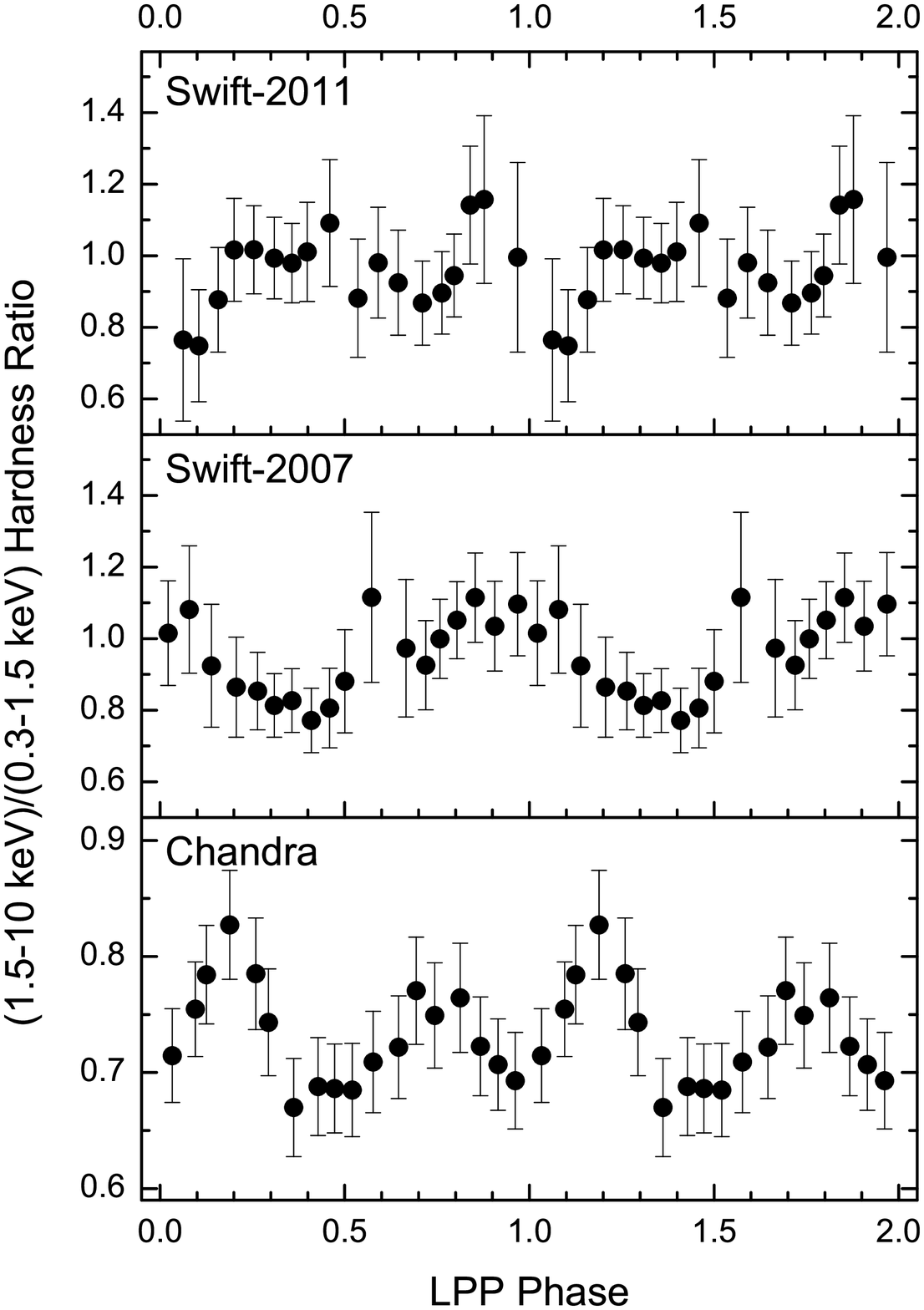}
\hspace{2 mm}
\includegraphics[width=5.6cm]{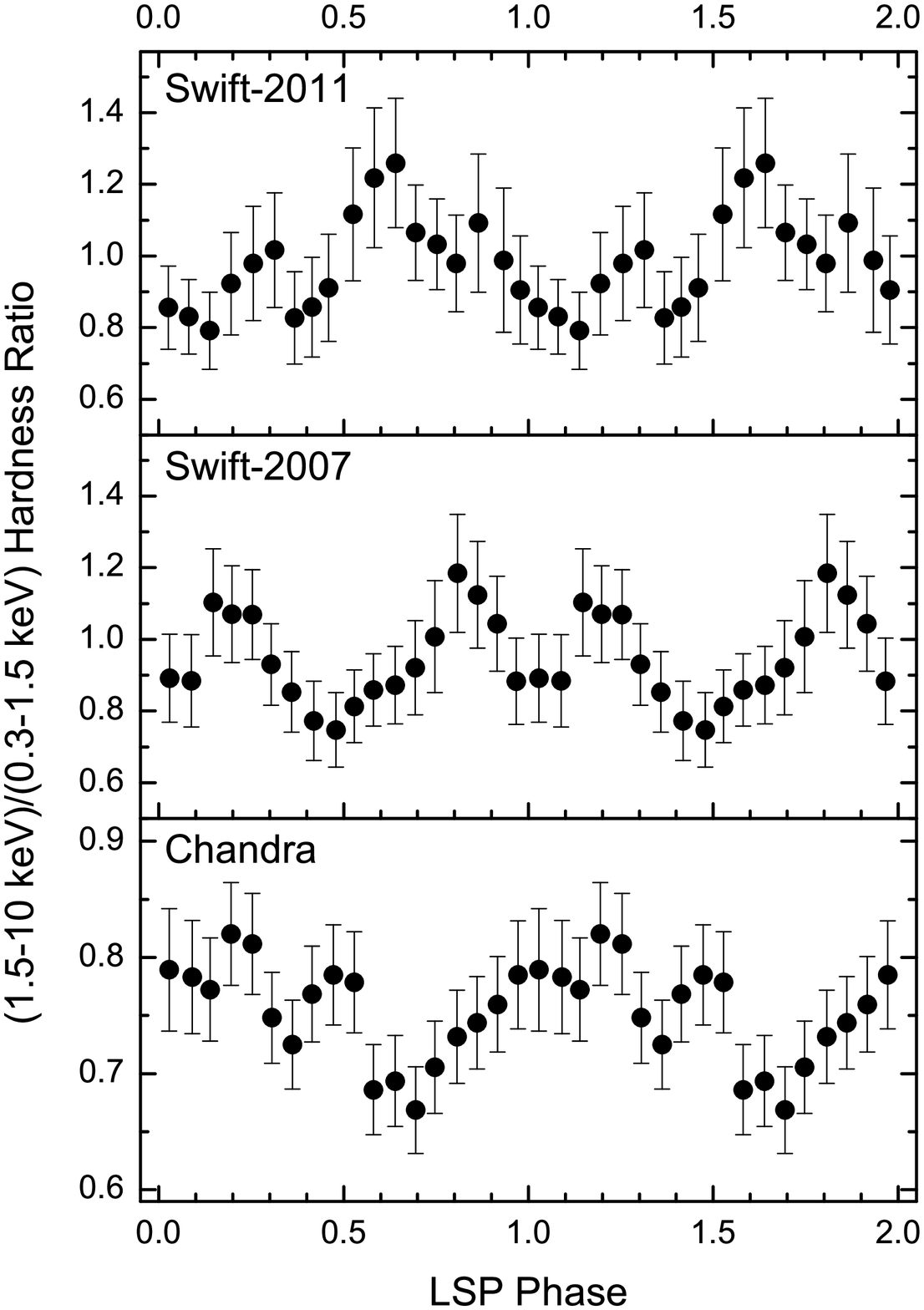}
}
\end{center}
\caption{
Hardness ratio curves of FS Aur folded with the OP (left panel), the LPP (middle) and
the LSP (right) according to the ephemerides~(\ref{ephemerisOP}), (\ref{ephemerisLPP})
and (\ref{ephemerisLSP}), respectively.}
\label{fig:xraysHR}
\end{figure*}

The Chandra ACIS-S observations of FS Aur which were performed on January 5, 2005 with
a total on-source exposure of 25 ksec, allow a direct comparison with our optical
\textit{set-2004}. These data were analyzed following the standard procedures using
CIAO\footnote{http://cxc.harvard.edu/ciao/} (Version 4.3) provided by the Chandra
X-ray Center (CXC).

During the 2010-2011 observational campaign, we also performed a ToO observation
of \fsaur\ with the X-ray telescope (XRT) and the UV-Optical Telescope (UVOT) onboard
the Swift X-ray satellite \citep{Swift}. These observations were taken on March
29 -- April 1, 2011 for a total of about 20.4 ksec. The object was also observed
by Swift on another occasion during January 17 -- 19, 2007 with a total on-source
exposure of 30 ksec. Hereinafter we call these sets of observations ``Swift-2011''
and ``Swift-2007''. We used the Swift Release 3.7
software\footnote{http://swift.gsfc.nasa.gov/docs/software/lheasoft/} together
with the most recent version of the Calibration Database
to analyze the Swift data. The XRT data were reduced in the standard fashion
with xrtpipeline (v. 0.12.6). In 2007, the UVOT observations in the uvw2
and uvm2 filters were taken in event mode. These data were used for evaluation of
the short-term variability. They were reduced following the procedure described in
\citet{UVOT}.

In order to analyze the X-ray variability with the fundamental periods, we extracted
soft (0.3-1.5 keV) and hard (1.5-10 keV) light curves. For the spectral analysis,
the background-subtracted spectra in the energy range 0.3-10 keV were extracted
and averaged for each dataset.

\begin{figure*}
\includegraphics[width=8.5cm]{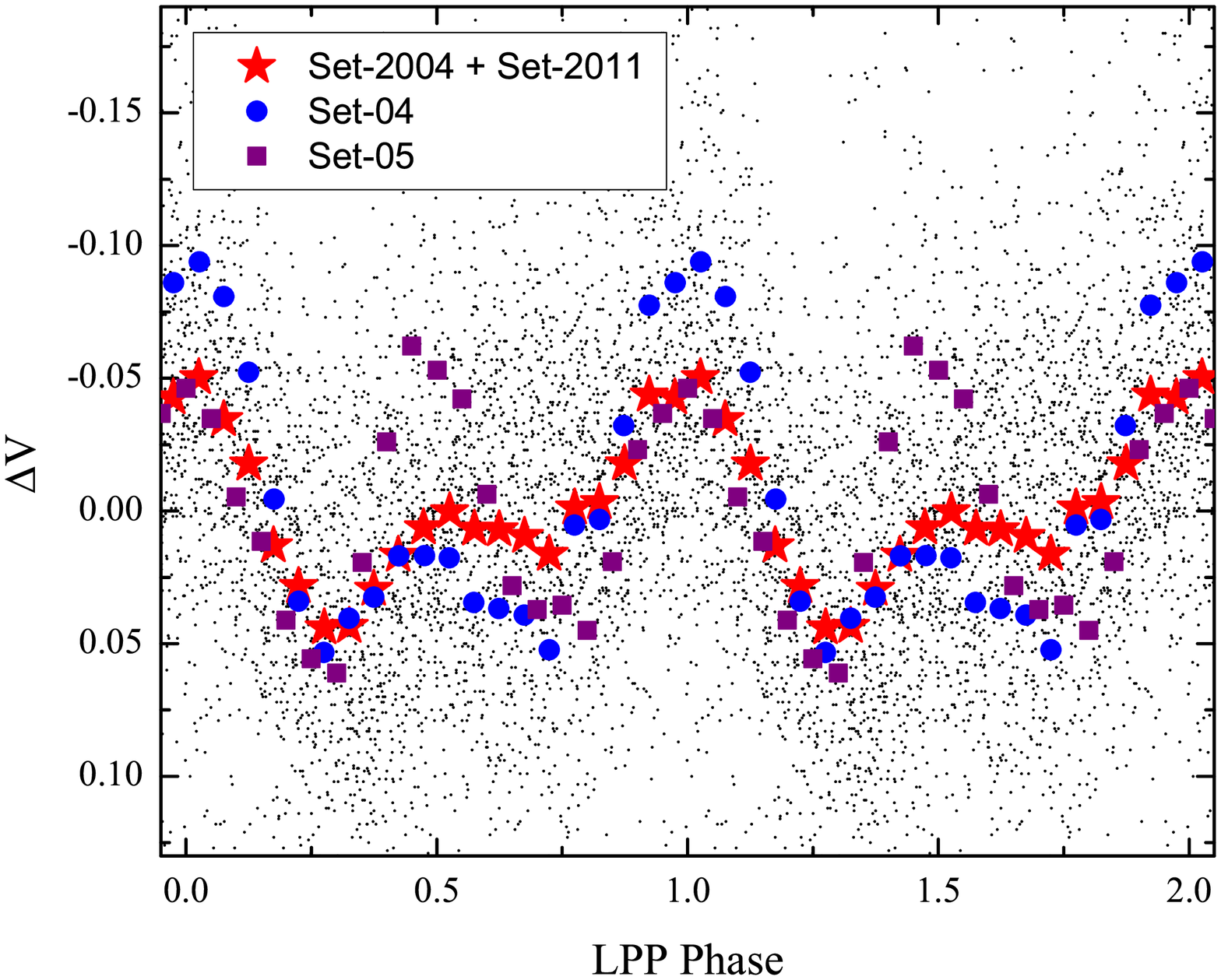}
\hspace{5 mm}
\includegraphics[width=8.5cm]{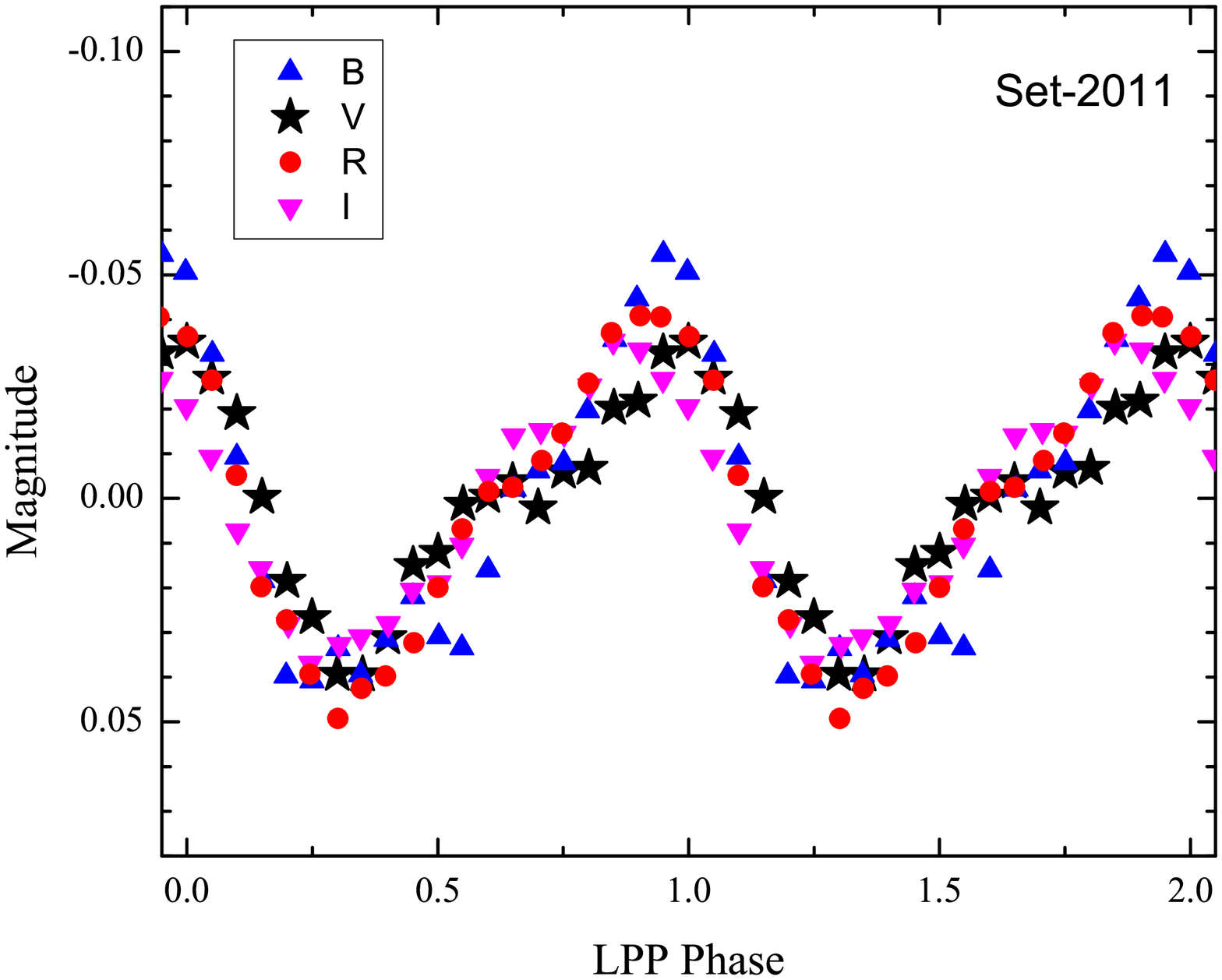}
\caption{\textit{Left:} the entire $V$ phase curve (with the means, and linear
trends subtracted for each night of observations) folded with the LPP according
to the ephemeris~(\ref{ephemerisLPP}). The large symbols represent data from different
subsets averaged in 20 phase bins. \textit{Right:} the folded light curves in
different colours of the \textit{set-2011}.
All data are plotted twice for continuity.}
\label{fig:FoldedLPP}
\begin{center}
\hbox{
\includegraphics[height=7.3cm]{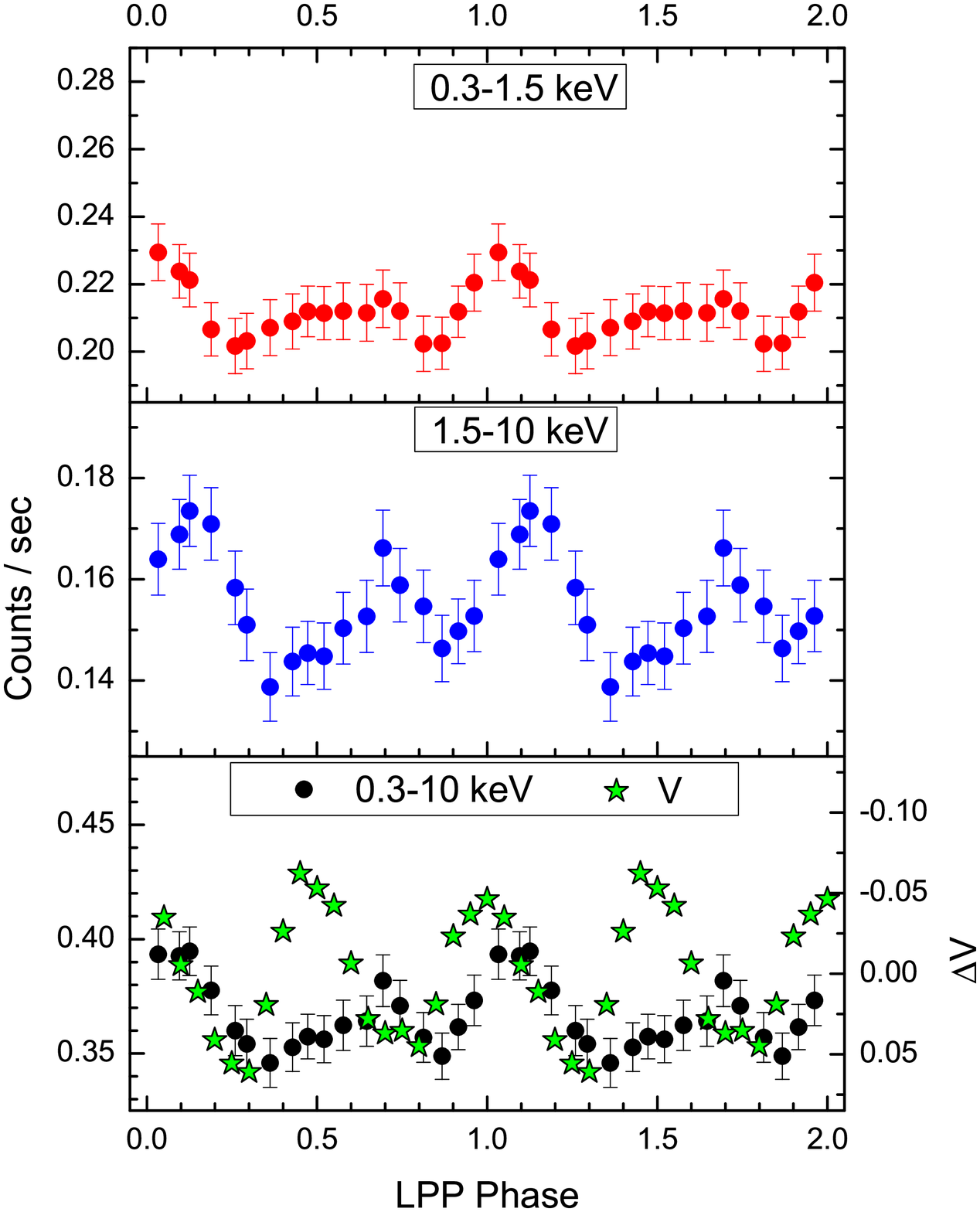}
\includegraphics[height=7.3cm]{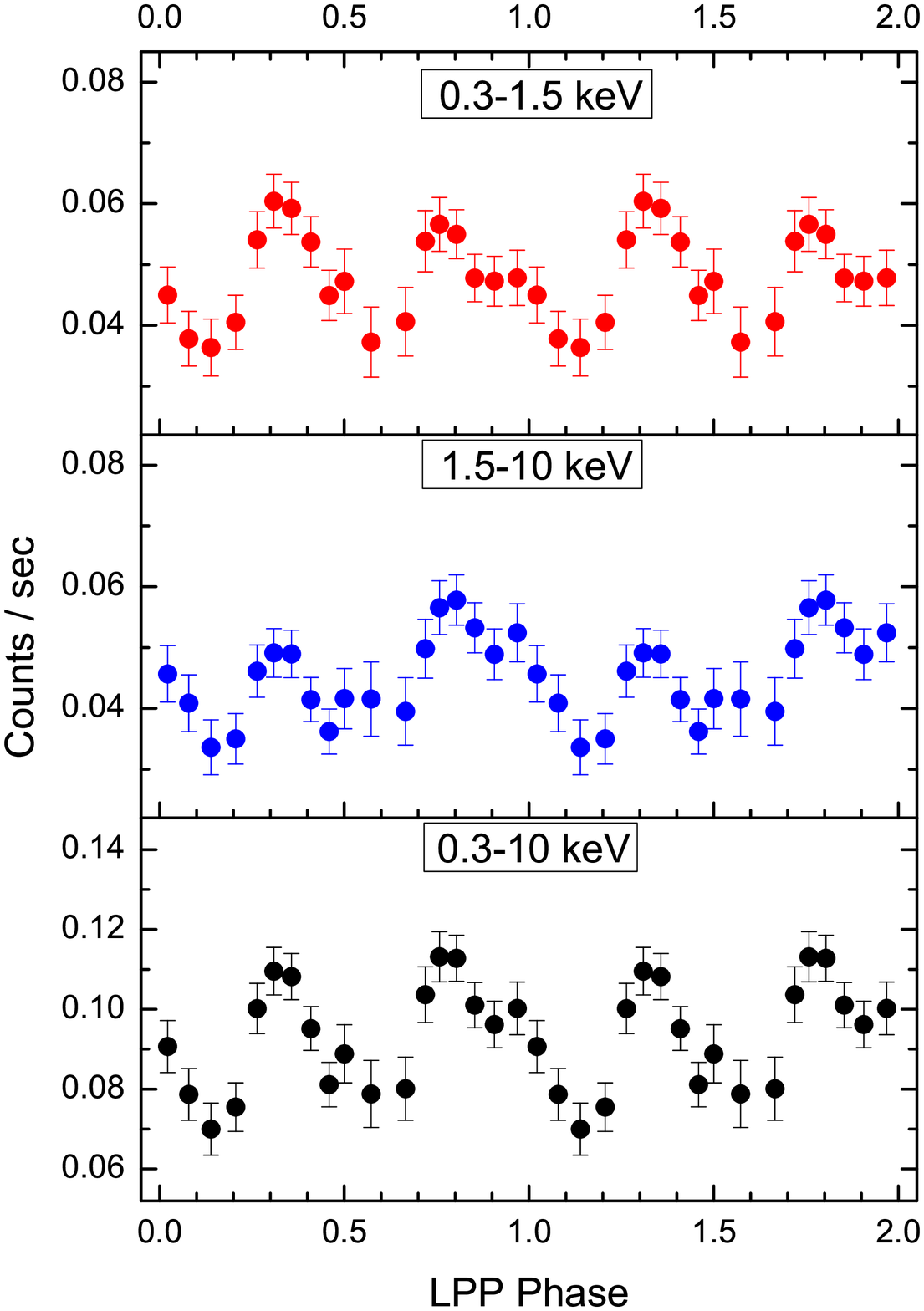}
\hspace{5mm}
\includegraphics[height=7.3cm]{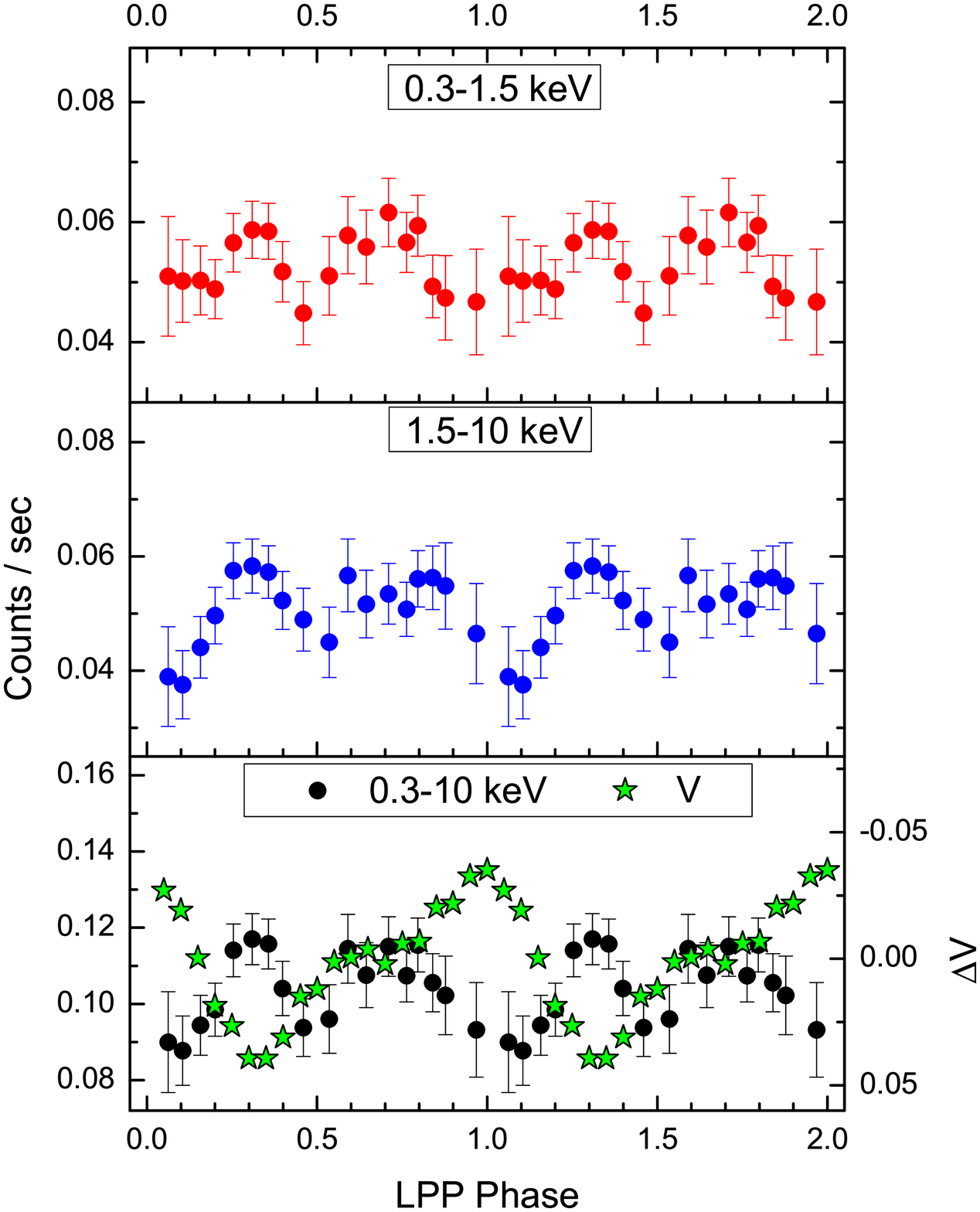}
}
\end{center}
\caption{The Chandra (left panel), Swift-2007 (middle) and Swift-2011 (right)
data folded with the LPP according to ephemeris~(\ref{ephemerisLPP}). The panels
are from top to bottom: the soft range (0.3-1.5 keV), hard range (1.5-10 keV),
and the total X-ray flux (0.3-10 keV). The latter are shown together with the
corresponding $V$ light curves (the Chandra data are compared with the optical
\textit{set-05} whereas the Swift-2011 is compared with the \textit{set-2011}).
Two cycles are shown for continuity.
}
\label{fig:xraysLPP}
\end{figure*}

\section{Data Analysis}

It has been previously shown that usually the most prominent features of the
optical light curve of FS Aur is the well-defined LPP modulation of 205.5 min
contaminated by strong stochastic variations \citep{Neustroev2002, Tovmassian2003,
Neustroev2005}. The orbital variability was previously seen only
occasionally while the LSP variability was never observed photometrically.
However, during the presented observations the photometric behavior of FS Aur
was significantly different yet consistent across both sets of the data.
In particular, the character of the LPP variability was changed, the OP
modulation was apparent most of the time, the LSP variability was observed
in colour curves, and stochastic variations were exceptionally strong.

The Lomb-Scargle periodogram for
the entire V photometry (with the means, and linear trends subtracted for each
night) is shown in Figure~\ref{fig:FS_Aur_PS}. Comparing this power spectrum with
the one presented in Figure~6 of \citet{Tovmassian2003}, one can also find apparent
differences: the strongest,
very sharp peak at $f = 7.0077 \pm$ 0.0001 day$^{-1}$ related to the LPP is now
accompanied by its first harmonic at $f = 14.0161 \pm$ 0.0001 day$^{-1}$
which was not observed previously, and the appearance of the relatively strong
peak at the orbital frequency $f = 16.7848 \pm$ 0.0001 day$^{-1}$.

The relative shortness of the X-ray observations obviously does not allow us to
perform a time series analysis. We note, however, that the most prominent feature
of the Chandra light curve is the modulation with the OP that can be seen even
with the naked eye (Fig.~\ref{fig:Chandra_LC}). In order to perform a more detailed
analysis, we folded the X-ray light curves with the OP, LPP and LSP.

\begin{figure*}
\begin{center}
\hbox{
\includegraphics[width=5.7cm]{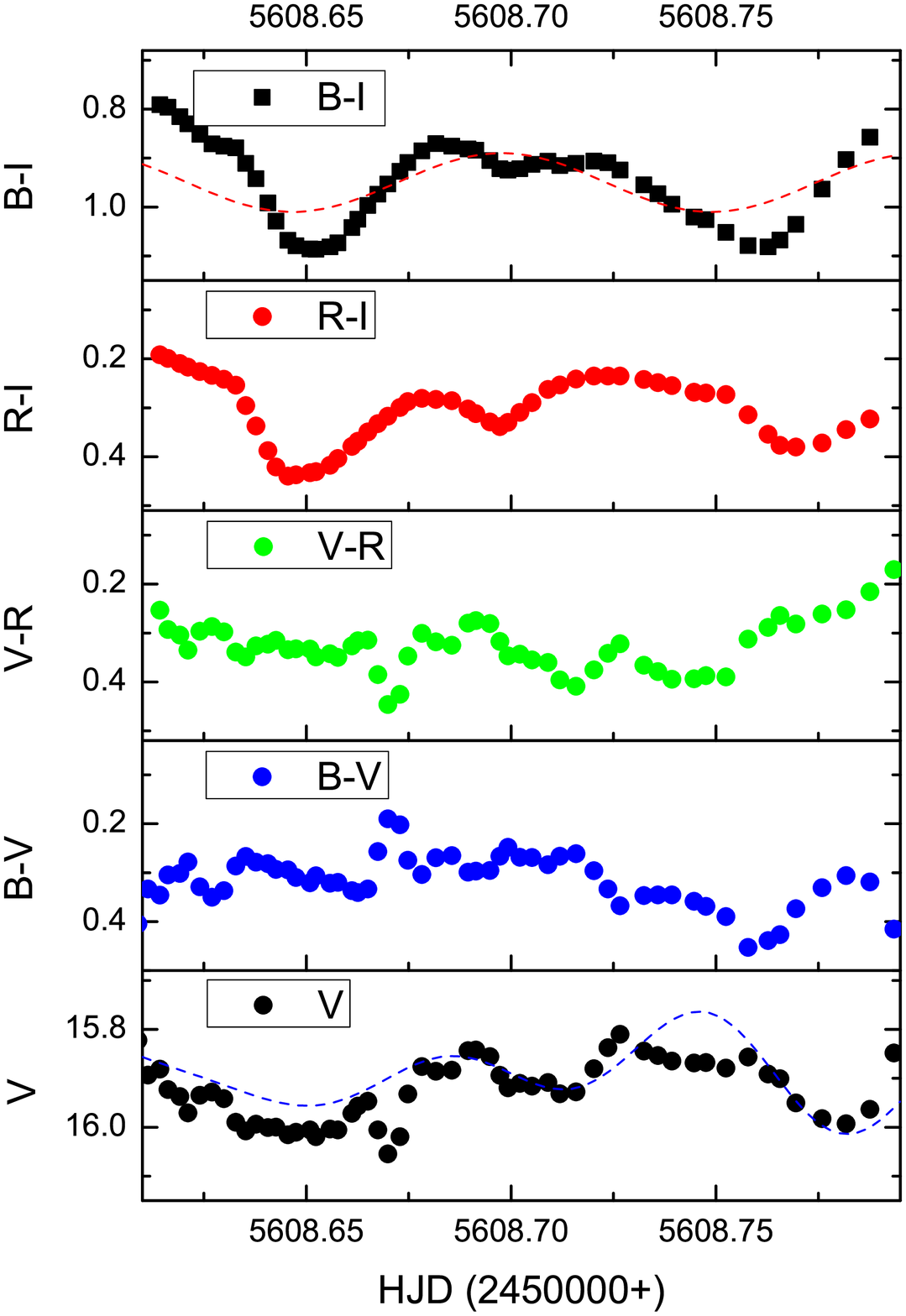}
\includegraphics[width=5.7cm]{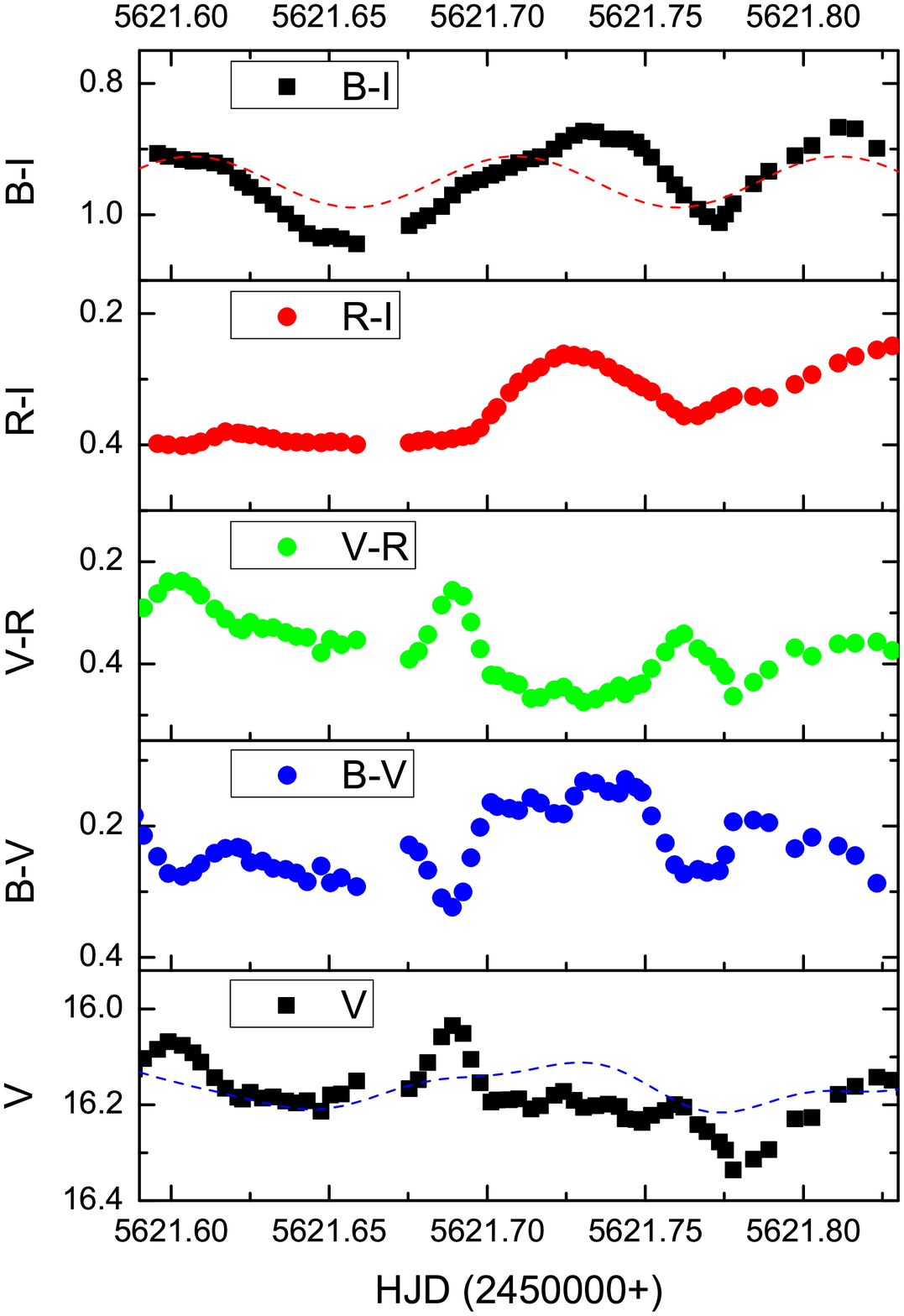}
\includegraphics[width=5.7cm]{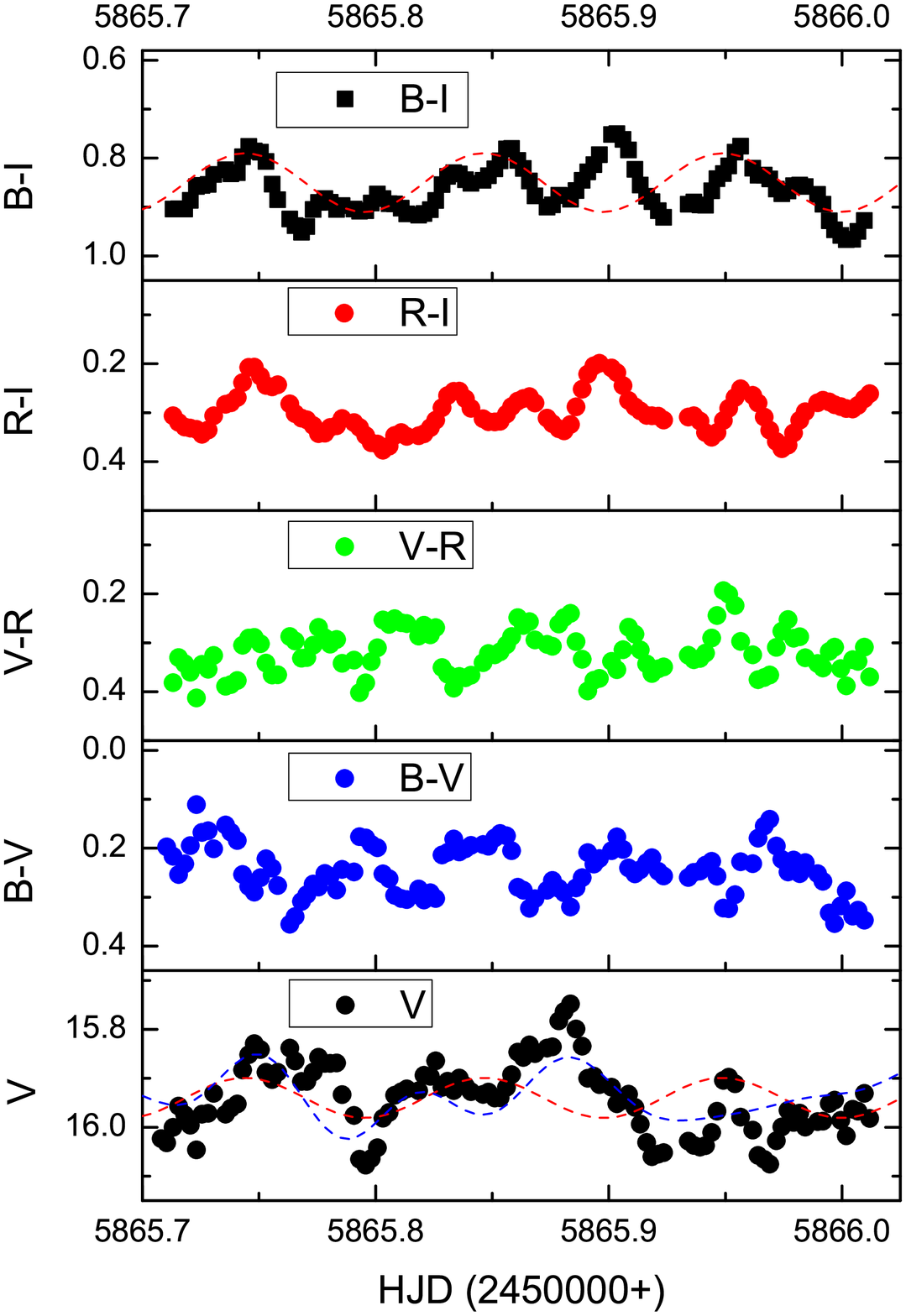}
}
\end{center}
\caption{
Sample $V$ light curves and ($B-V$), ($V-R$), ($R-I$) and ($B-I$) colour curves
of FS Aur from JD~2\,455\,608, JD~2\,455\,621 and JD~2\,455\,865. The dashed blue
lines are the sum of sinusoidal fits of the LPP and its first harmonic and the OP
which yielded the ephemeris~(\ref{ephemerisLPP}) and (\ref{ephemerisOP}). The
dashed red line is a fit of the LSP corresponding to the ephemeris~(\ref{ephemerisLSP}).
Note that on the night of JD 2\,455\,865 the LSP modulation is also visible in
the $V$ light curve.
}
\label{fig:colours}
\end{figure*}

\begin{figure*}
\includegraphics[width=8.5cm]{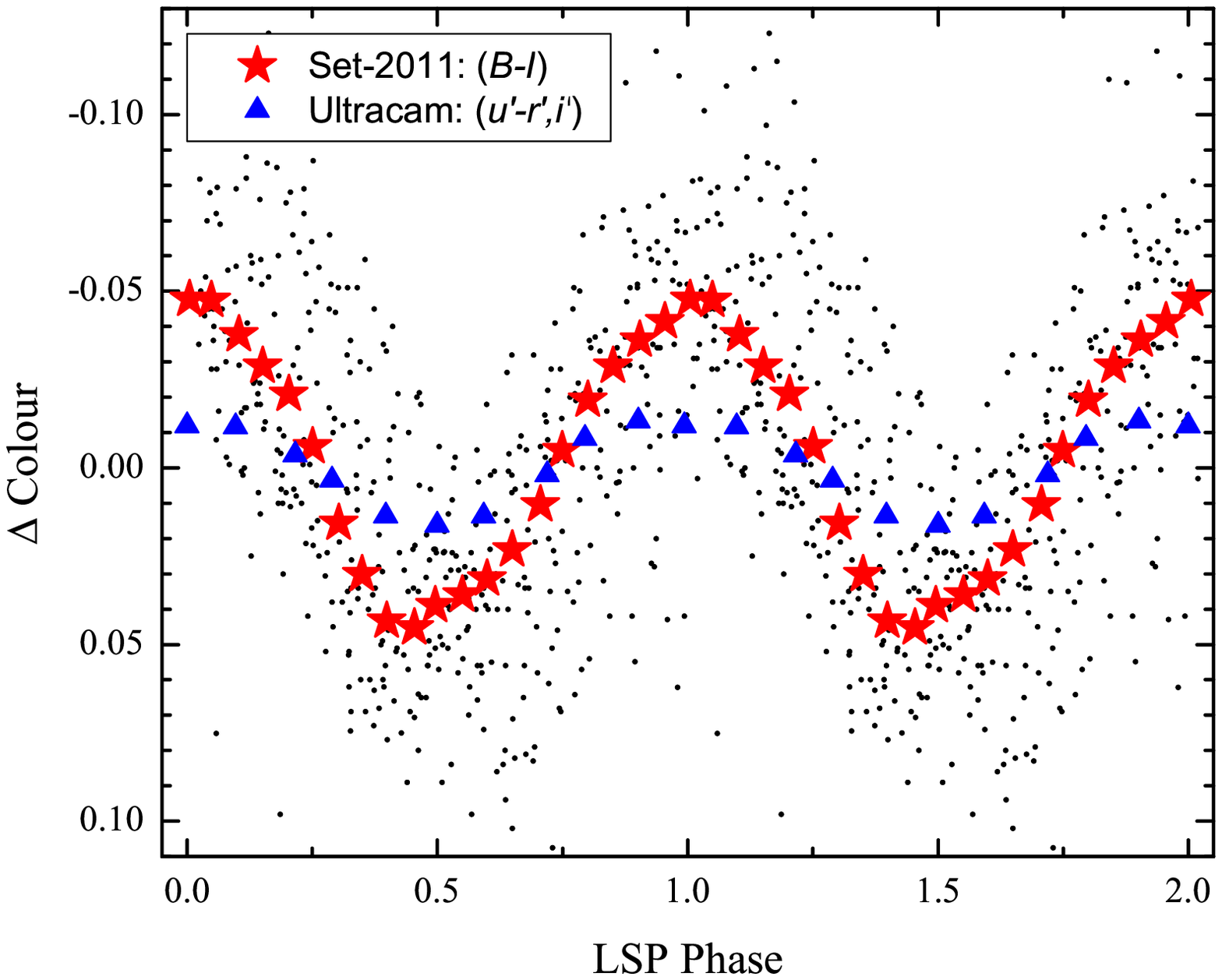}
\hspace{5 mm}
\includegraphics[width=8.5cm]{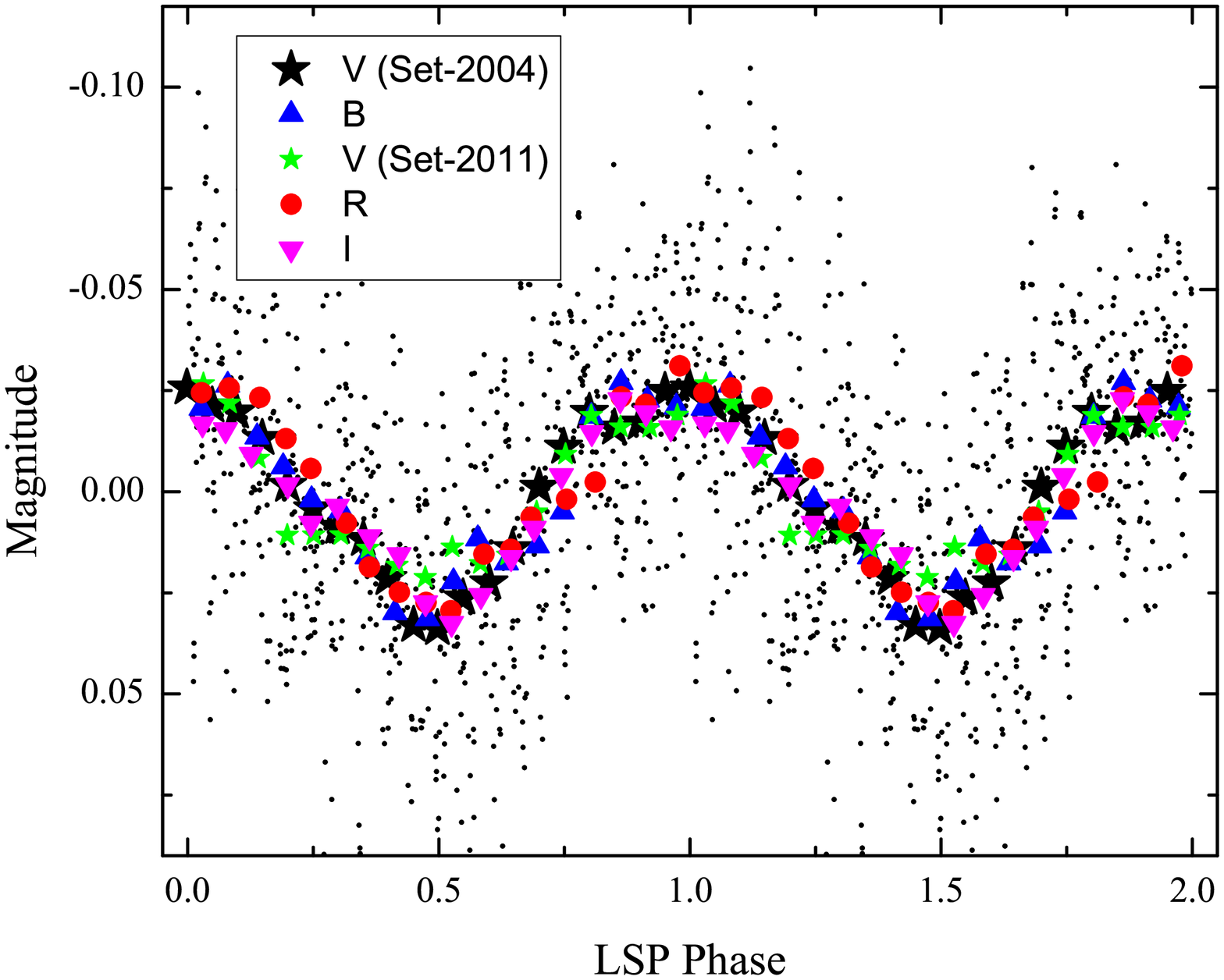}
\caption{\textit{Left:} The ($B-I$) colour data of FS Aur (with the means,
and linear trends subtracted for each night of observations) folded according to the
ephemeris (\ref{ephemerisLSP}). The large red stars represent the ($B-I$) data averaged
in 20 phase bins. The blue triangles represent the Ultracam data folded according to the
same ephemeris and averaged in 10 phase bins. \textit{Right:} the folded
light curves in different colours. The small dots represent the pre-whitened $V$ data
from the \textit{set-2004} whereas the large black stars show the same data averaged
in 20 phase bins. The other colour symbols represent the selected colour light curves
of the \textit{set-2011}. All data are plotted twice for continuity.}
\label{fig:FoldedLSP}
\begin{center}
\hbox{
\includegraphics[height=7.3cm]{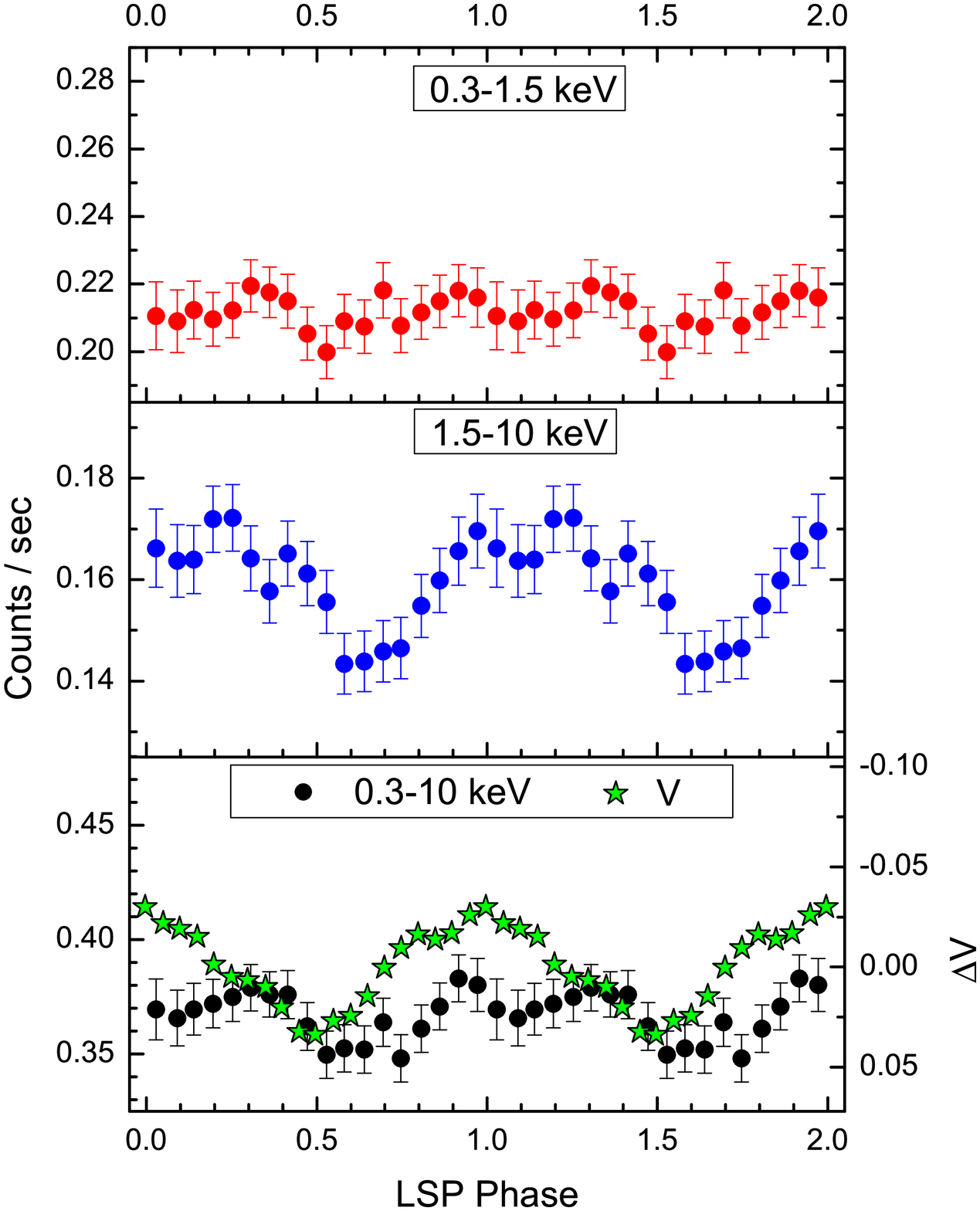}
\includegraphics[height=7.3cm]{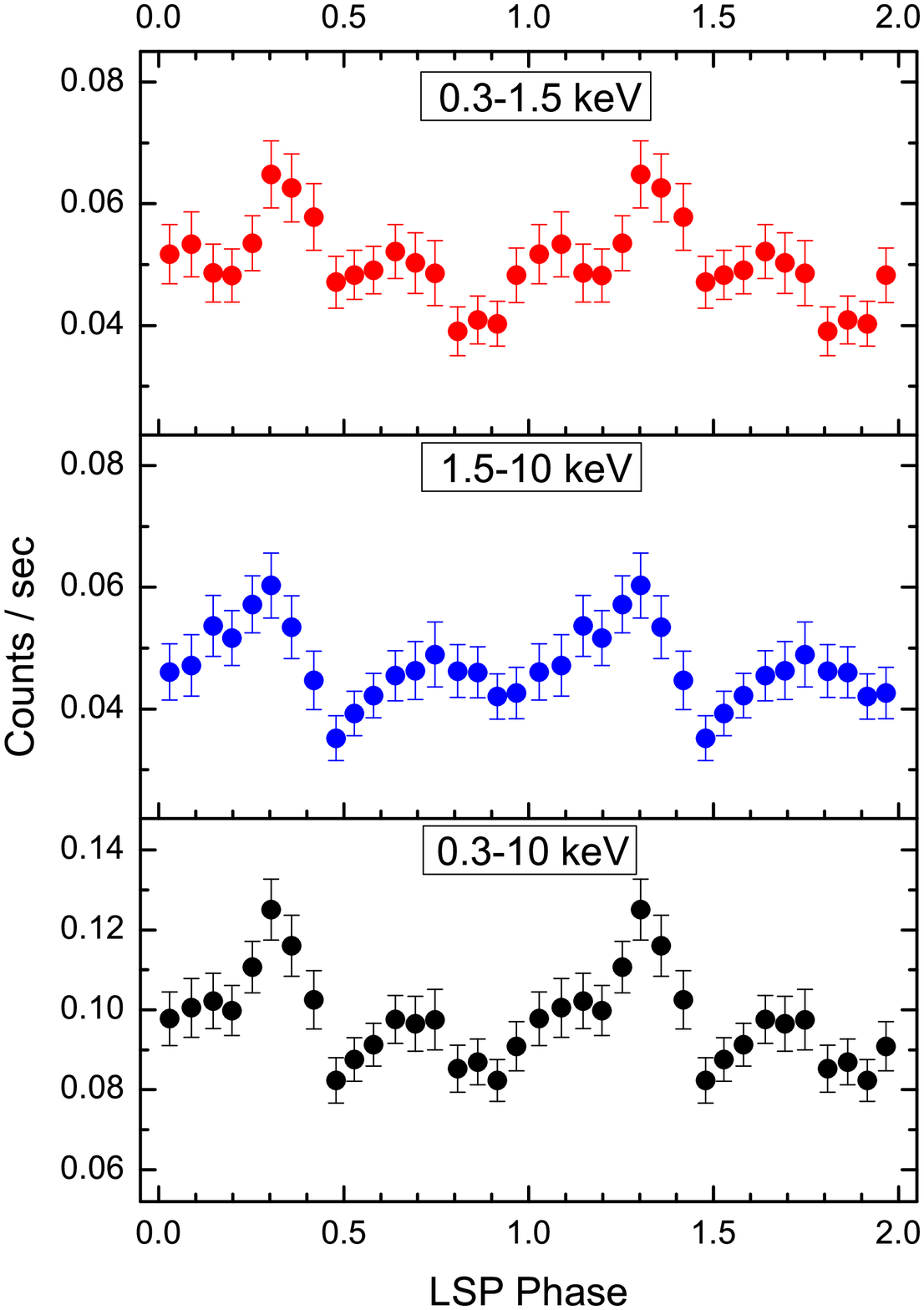}
\hspace{5mm}
\includegraphics[height=7.3cm]{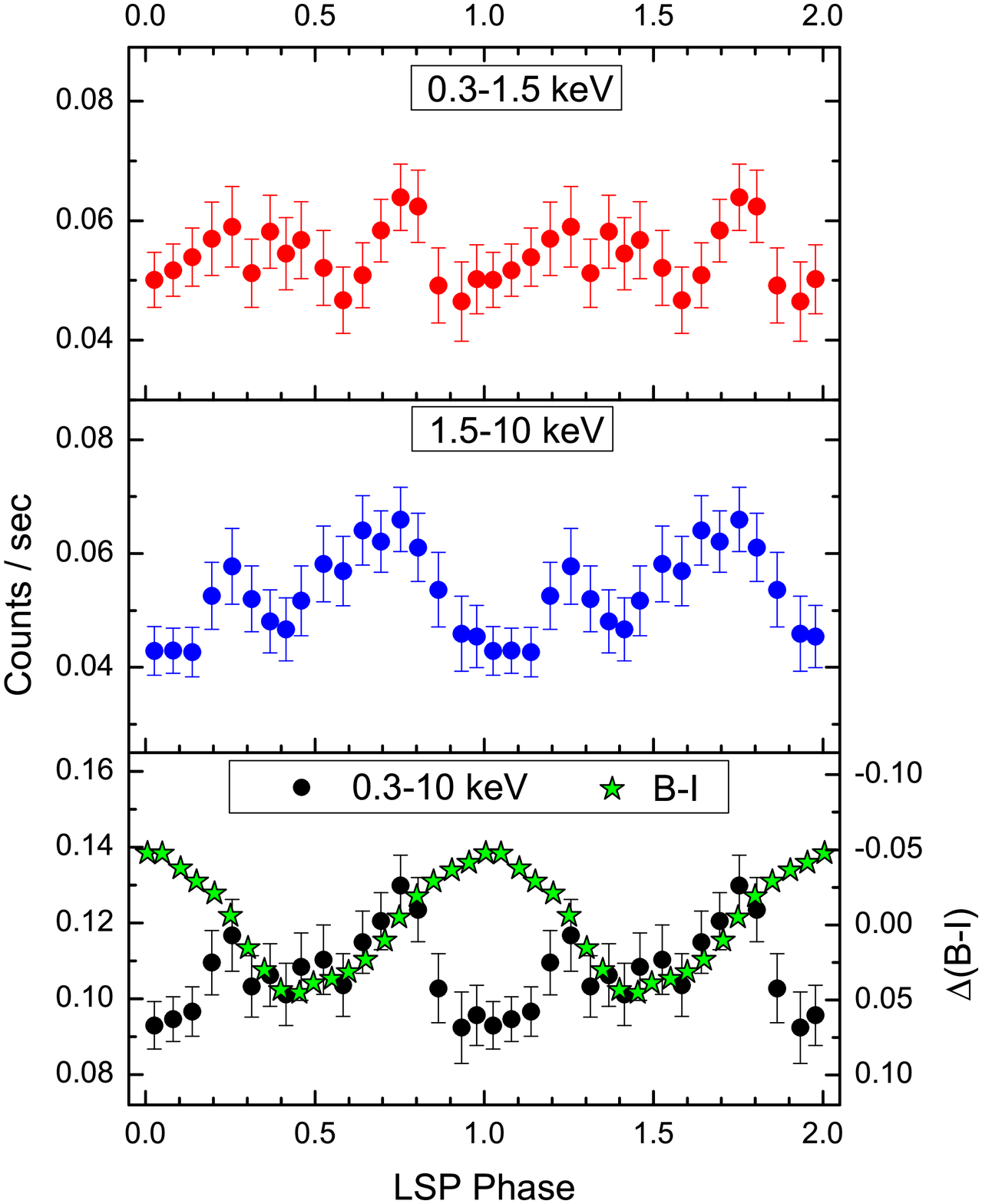}
}
\end{center}
\caption{
The Chandra (left panel), Swift-2007 (middle) and Swift-2011 (right) data
folded with the LSP period  according to the ephemeris~(\ref{ephemerisLSP}).
The panels are from top to bottom: the soft (0.3-1.5 keV), hard range (1.5-10 keV),
and the total X-ray flux (0.3-10 keV). The latter are shown together with the
corresponding colour or light curves (the Chandra data are compared with the optical
pre-whitened $V$ light curve from the \textit{set-2004} whereas the Swift-2011
is compared with ($B-I$) from the \textit{set-2011}). All data are plotted twice
for continuity.
}
\label{fig:xraysmod}
\end{figure*}

\subsection{The orbital variability}
\label{sec:OP}

In order to analyze the optical OP variability, we first  pre-whitened the light
curve by the LPP frequency and its first harmonic (see Section~\ref{sec:LPP}).
The entire 7 year set of photometry is of sufficient quality to compute an
accurate ephemeris:

\begin{equation}
T_{min} = 2453346.970(2)+0.05958096(5) \cdot E
\label{ephemerisOP}
\end{equation}

\noindent where the initial epoch is defined as the time of the minimum in the
$V$ light curve, as it is more likely to represent inferior conjunction of the
donor star in \fsaur. Figure~\ref{fig:FoldedOP} (left panel) shows the entire
pre-whitened $V$ light curve folded according to the ephemeris~(\ref{ephemerisOP})
and averaged in 20 phase bins.

Folding different subsets of the data over the OP results in the same smooth and
almost perfectly sinusoidal shape of the modulation, yet with slightly
different amplitudes. In the right panel of Figure~\ref{fig:FoldedOP} we show
the folded light curves in different colours of the \textit{set-2011}. All these
light curves have the same shape and amplitude, but the $R$ light curve
exhibits some phase shift in comparison with the $V$ one, whereas the $B$
and $I$ light curves are in phase with it. It is not easy to explain such
anomalous behaviour of the $R$ light curve. A possible reason might be
lower amplitude of the OP modulation during the \textit{set-2011} and relatively
high noise level in the corresponding data.

From the radial velocity study of the \Halpha, \Hbeta\ and \Hgamma\ emission
lines conducted in \citet{Tovmassian2007}, we can now compare the relative phasing
of the photometric and radial velocity modulations with the OP. Assuming that the
emission lines come from disc material orbiting the WD, the red-to-blue crossing of
the radial velocities provides an estimate of the moment of
inferior conjunction of the secondary star. Using radial velocity measurements made
with a double Gaussian separation of 600 \kms (for details see \citealt{Tovmassian2007}),
we estimated the moment of inferior conjunction of the secondary star in \fsaur\ to be
HJD $2453346.971\pm0.001$.

It is well known that measurements of emission-line radial velocities in CVs
often give quite uncertain or incorrect results, as the emission lines arising
from the disc may suffer several asymmetric distortions. However, almost perfect
coincidence between the initial epochs estimated by different methods allows us
to consider this moment as inferior conjunction of the secondary star.

In X-rays, the modulation with the OP is also strong in both energy ranges
(Fig.~\ref{fig:xraysOP}). However, there are significant differences in the X-ray
light curves between different X-ray sets and at different energies as seen from
the hardness ratio curves (Fig.~\ref{fig:xraysHR}, left panel).
The folded Chandra light curve is quasi-sinusoidal and displays a broad maximum
with an additional strong and narrow peak which is visible almost exclusively
in the soft band. This peak is so strong that it is evident even in the raw
data. Note, that one of these peaks was seem to be missing (Fig.~\ref{fig:Chandra_LC}).

The Swift light curves display a similar quasi-sinusoidal modulation which is
distorted by a broad depression which is deeper in the hard band. The peak
maximum in the Chandra data and the depression minimum in the Swift-2011 are
observed at nearly same phase 0.4, whereas in the Swift-2007 the minimum is
shifted toward later phases. The sinusoidal component in all the X-ray data sets
closely correlates with the optical light curves.

\subsection{The LPP variability}
\label{sec:LPP}

From the previous observations of FS Aur it was known that the LPP modulations
can be varied in strength and appearance, but on average the light curve was
smooth and almost perfectly sinusoidal (see Figures~4 and 7 in \citealt{Tovmassian2003}).
Nevertheless, the new photometric data reveal that the LPP modulation has changed
dramatically, being transformed from a sinusoidal shape to a double-hump one.
We defined the following accurate ephemeris:

\begin{equation}
T_{max} = 2453346.915(2)+0.14270191(9) \cdot E
\label{ephemerisLPP}
\end{equation}

\noindent where the initial epoch is defined as the time of primary maximum in
the binned V light curve. Figure~\ref{fig:FoldedLPP} (left panel) shows the
entire $V$ light curve folded according to this ephemeris and averaged in 20
phase bins.

Folding different subsets of the data over the LPP results in the slightly
different morphology of the modulation. Not only the amplitude of the modulation
but also the relative strength of the two humps varies substantially between
the different subsets. However, despite such dramatic changes from previous
observations, the LPP modulation kept an important feature -- its amplitude is
the same in different colour bands (Figure~\ref{fig:FoldedLPP}, right panel).

Similar double-hump modulation was also observed in each X-ray set
(Fig.~\ref{fig:xraysLPP}). The amplitude of variations is about the same in the
soft and hard X-ray ranges but there appears to be some energy dependence of this
modulation, most prominent in the Chandra data, as seen in the hardness ratio
curves (Fig.~\ref{fig:xraysHR}, middle panel).

Despite the similarity in morphology of the modulation profiles, we note
a considerable phase shift between optical and X-ray pulses, different in
the Chandra and Swift data sets.

\subsection{The LSP variability}
\label{sec:LSP}

\fsaur\ and V455~And compose a group of CVs in which the emission lines vary
with two very different periods, namely the OP and LSP. Such a behaviour is
very unusual for ordinary dwarf novae but it is common for Intermediate
Polars which often show more than one periodicity in the emission lines
\citep{Hellier1991,Hellier1999}. We note, however,
that those additional periods are usually quite clearly seen in optical light
curves of IPs, whereas neither in \fsaur\ nor in V455 And was the LSP ever
detected in optical photometry.

At first sight, the presented observations are not an exception. No significant
power is detected at the LSP frequency ($\sim$9.78 day$^{-1}$) neither in the total
power spectrum nor in the periodograms of the subsets (see the inset in
Figure~\ref{fig:FS_Aur_PS}).

As it was shown before,
the optical light curve of \fsaur\ is dominated by
the LPP and OP modulations, which amplitudes are approximately the same in all
filters. However, the colour curves ($B-V$), ($V-R$) and ($R-I$) also show
a noteworthy variation, more or less following the brightness variation: the
colour indices ($B-V$) and ($R-I$) are generally in anti-phase and
(${V-R}$) is in phase with the light curve (Fig.~\ref{fig:colours}).
The amplitudes of colour variations are $\sim0.1-0.2$ mag in all these colours.

Surprisingly, we found that the behavior of the colour index ($B-I$) is
notably different. The most noticeable feature in ($B-I$) is a modulation
with a period near 0.1 day, even though it is sometimes contaminated or even
completely replaced by the OP/LPP variations and flickering. After visual inspection
of the nightly ($B-I$) curves, we selected a total of 19 nights (half of the
observations taken in quiescence) when this variability is most evident.
The Lomb-Scargle periodogram of the combined ($B-I$) data (with the means, and
linear trends subtracted for each night) is shown in Figure~\ref{fig:FS_Aur_PS}.

The strongest peak at $f = 9.7762 \pm$ 0.0001 day$^{-1}$ exactly coincides
with the beat period between the OP and LPP: $1/P_{beat}=1/P_{orb} -
1/P_{phot}$, which was previously observed as the LSP and is the presumed
precession period of the WD in FS~Aur \citep{Tovmassian2007}.
This ($B-I$) data set is of sufficient length and photometric quality to compute
an accurate, current ephemeris:

\begin{equation}
T_{B-I} (min) = 2455582.124(1)+0.1022865(9) \cdot E
\label{ephemerisLSPbi}
\end{equation}

\noindent where $T_{B-I} (min)$ is the moment of the ($B-I$) colour index
minimum that corresponds to the hardest spectrum. Figure~\ref{fig:FoldedLSP}
(left panel) shows the ($B-I$) phase curve folded according to the
ephemeris~(\ref{ephemerisLSPbi}) and averaged in 20 phase bins.

For the following analysis it is useful to determine whether and how this colour
variability might be related to brightness and spectral variability, and we also
need to adjust the ephemeris (\ref{ephemerisLSPbi}) to be adequate for the
\textit{set-2004}. Unfortunately, the LSP modulation is scarcely visible in the
raw light curve, yet it was visually detectable in a few nights of both sets of
observations. On at least two nights of the \textit{set-2011} this modulation is
clearly seen in all the filters. It is also occasionally seen in lesser number of
filters on other nights\footnote{In this sense, the $V$ light curve seems worse than
others that may be because of its most complex shape.} (see the right panel of
Fig.~\ref{fig:colours}).

It seems that the LSP pulse profiles are nearly sinusoidal
in all colour bands, and they are in phase with each other and with the ($B-I$)
colour curve (Fig.~\ref{fig:FoldedLSP}). The power spectra of the combined ``good''
data also show the significant peak at the LSP frequency in all four colour bands.
We admit, however, that the timing analysis of these few manually selected nights
is somewhat speculative.

We now switch to the timing analysis of the \textit{set-2004} where it has more
credibility. Flickering and the LPP and OP pulsations were very strong during that
time, hiding other, possibly weak, modulations. Nevertheless, after pre-whitening
with the LPP and OP signals the LSP modulation becomes apparent, particularly in
the \textit{set-05}. The Lomb-Scargle periodogram of the pre-whitened light curve
shows the strongest peak at $f = 9.7770 \pm$ 0.0002 day$^{-1}$
(Figure~\ref{fig:FS_Aur_PS}).
We define the following current ephemeris for the moments of the pulse maximum:
 $T (max) = 2453347.064(2)+0.102281(7) \cdot E$.

Unfortunately, a direct comparison of these two data sets is ambiguous. Even
the more precise ephemeris (\ref{ephemerisLSPbi}) may not be accurate enough to
be valid over the time period of 7 years. Clearly, there is no cycle count ambiguity,
but the ephemeris prediction may be inaccurate as much as few tenths of the LSP.
To be more confident, we compared the \textit{set-2004} with the more recent
simultaneous multicolour observations taken with the high-speed camera ULTRACAM in
late 2003 \citep{Neustroev2005}. The most prominent feature of this light curve is
the LPP. This variability is evident in all filters. However, in both the ($u'-r'$)
and ($u'-i'$) colour indices a modulation with a shorter period of nearly 0.1 day has
appeared. Assuming that this variability has the same nature as observed in the
\textit{set-2011}, we fitted the combined ULTRACAM colour curve with a sine wave
of the LSP period and obtained the corrected value for the reference epoch
time of the colour index minimum $T_{u'-r',i'} (min)=2452942.518\pm0.001$. This
again suggests that the colour and light modulations are in phase with each other.

Thus, we assume that there is no time lag between the colour and light modulations.
We combined the pre-whitened \textit{set-2004} and both the colour curves together
into one large dataset. Its Lomb-Scargle periodogram is shown in
Figure~\ref{fig:FS_Aur_PS} (the bottom black line).
The sharpness and power of the LSP signal at
$f = 9.77644 \pm$ 0.00004 day$^{-1}$ indicates a very high degree of coherence.
Finally, we refined the ephemeris in order to be adequate for all our data:

\begin{equation}
T (max) = 2452942.417(2)+0.102286657(82) \cdot E
\label{ephemerisLSP}
\end{equation}

\noindent where $T (max)$ is the time of the optical pulse maximum, or the colour
index minimum. Figure~\ref{fig:FoldedLSP} shows different sets of our data folded
according to ephemeris~(\ref{ephemerisLSP}).

In X-rays, the modulation with the LSP is also evident in each X-ray set
(Fig.~\ref{fig:xraysmod}). There appears to be a clear signature of the energy
dependence of this modulation as to be directly visible in the light curves, and
as seen in the hardness ratio curves (Fig.~\ref{fig:xraysHR}). This dependence
is most prominent in the Chandra data where the soft photon flux is nearly
constant while the hard flux exhibits strong quasi-sinusoidal variations
(Fig.~\ref{fig:xraysmod}, left panel).

There are significant differences in the X-ray light curves between different X-ray
sets. Unlike the Chandra's nearly sinusoidal modulation, the Swift light curves
display complex double-peaked profiles. We suspect, however, that the latter also
are sine-like in shape but distorted by a depression around LSP zeroth phase.
This depression is deep and wide in \textit{Swift-2011}, narrower and shallower
in \textit{Swift-2007}, and is scarcely seen in the Chandra set. If this supposition
is correct then the sinusoidal component in all the X-ray data sets is nearly in
phase with the optical modulation.

We can now compare the relative phasing of the photometric and radial velocity
modulations with the LSP. We have used our previous radial velocity measurements
made with a double Gaussian separation of 1600 \kms (for details see
\citet{Tovmassian2007}). The LSP component, appearing in the far wings of
emission lines, has maximum blueshift at phase $0.74\pm0.02$, according to
ephemeris~(\ref{ephemerisLSP}). Thus, it crosses the line of sight in front of
the WD at the optical pulse maximum.

\begin{figure}
\includegraphics[width=8.0cm]{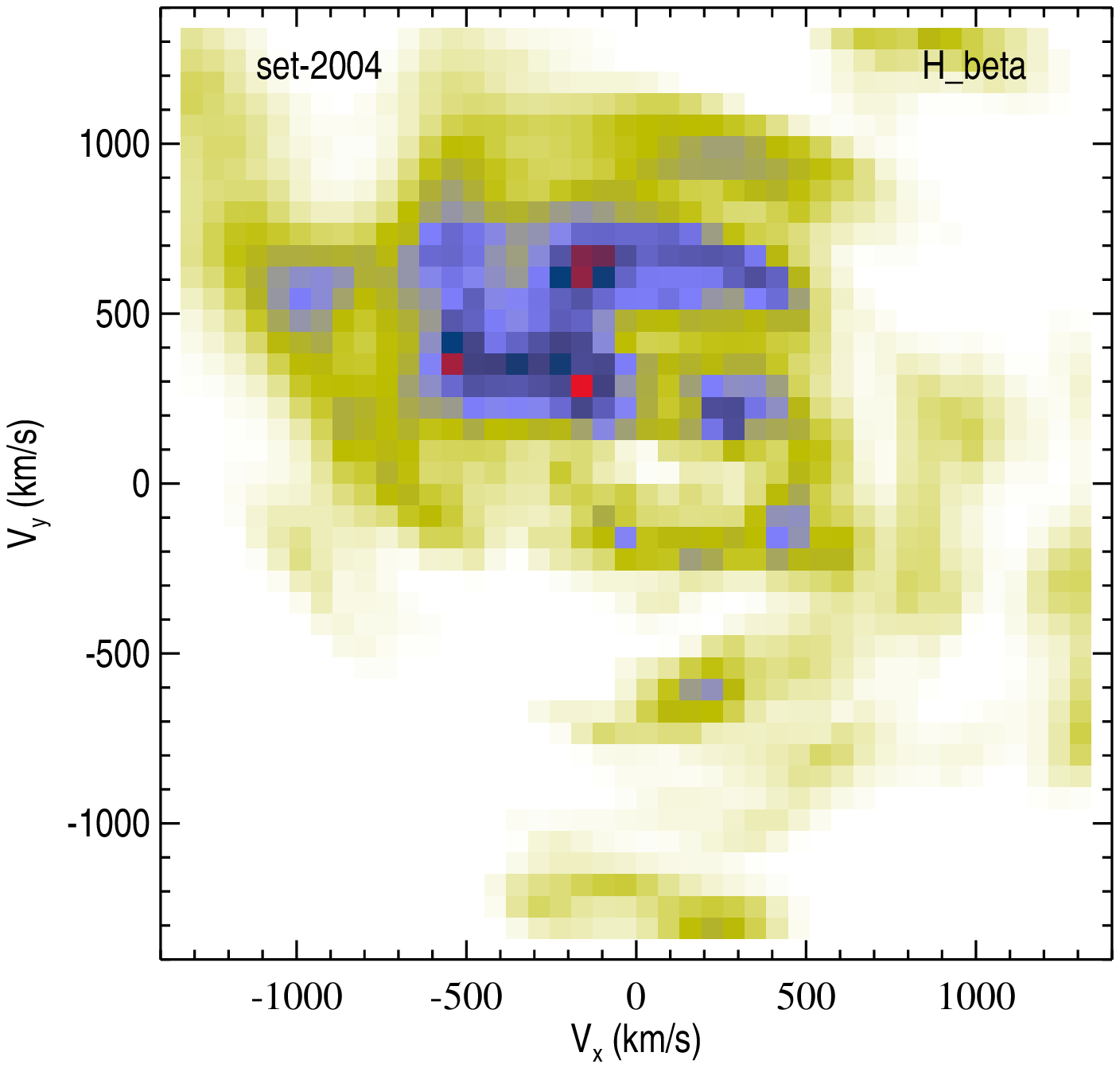}\\
\includegraphics[width=4.0cm]{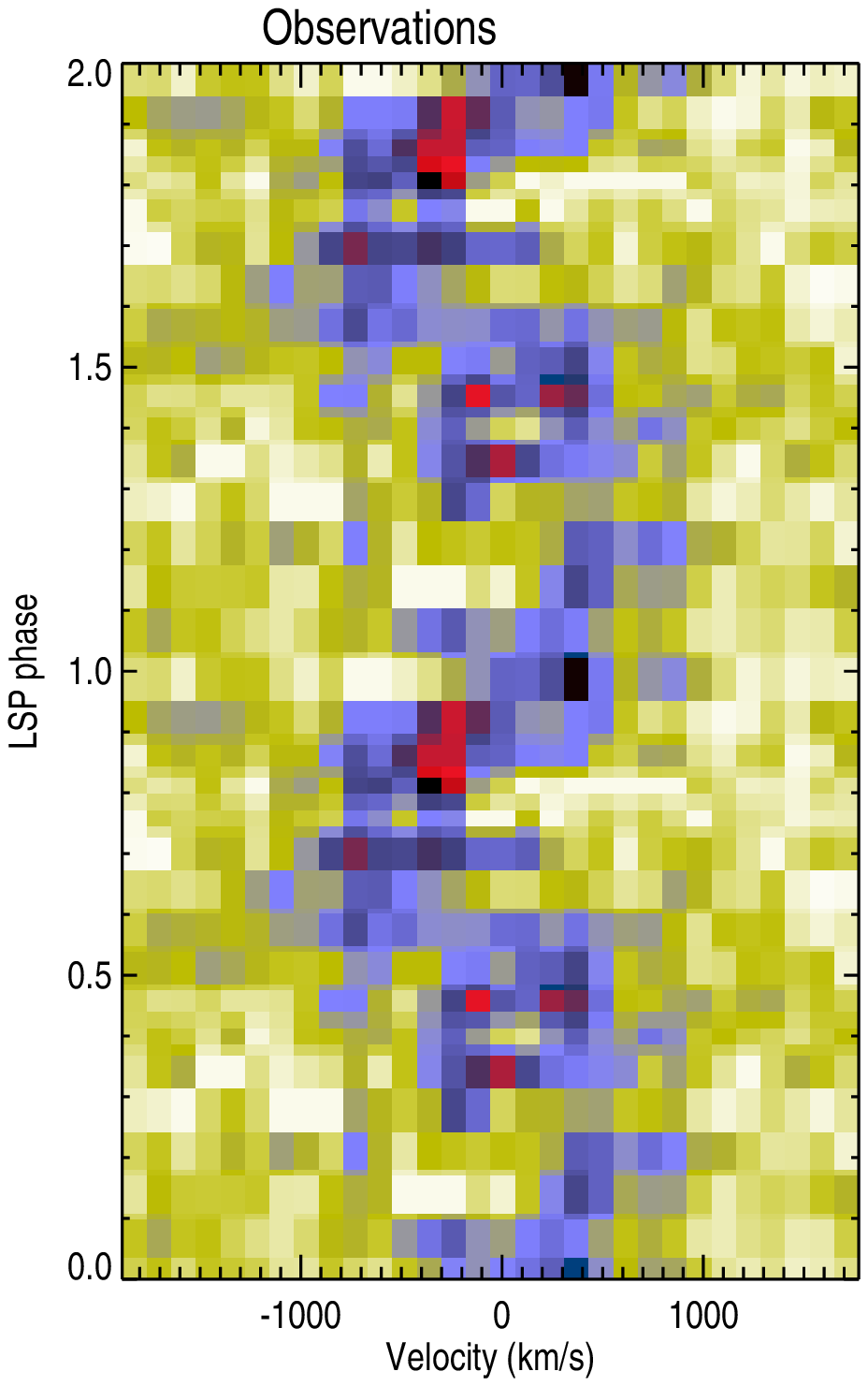}
\includegraphics[width=4.0cm]{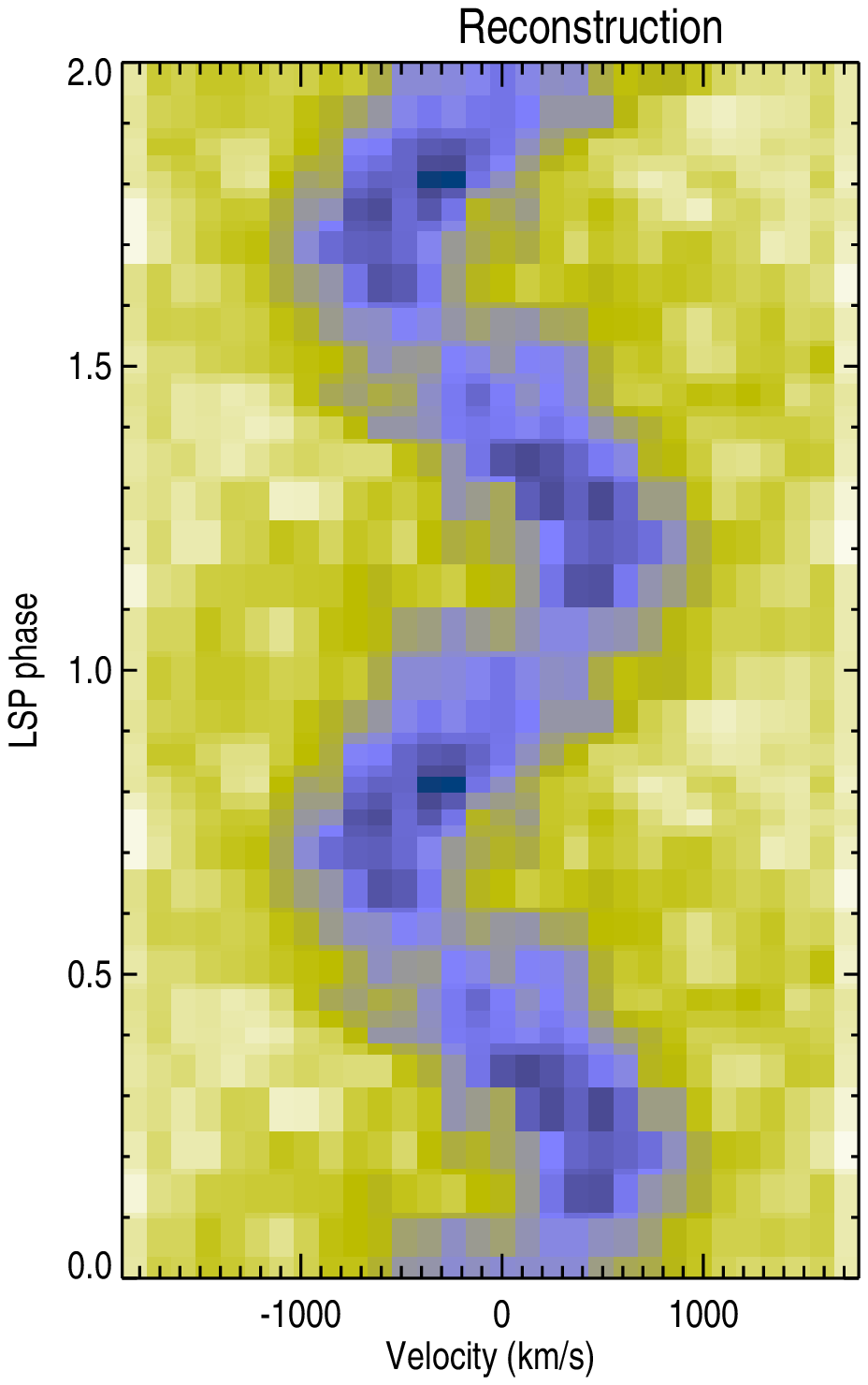}
\caption{Doppler tomography for the \Hbeta\ emission line from the \textit{set-2004}.
In the bottom panels are the line profile folded on the LSP cycle and then subtracted
of the phase-invariant profile (left), and reconstructed profiles (right). In the top
half the corresponding Doppler map is shown. For technical details see
\citet{Hellier1999}.}
\label{fig:dopmap}
\end{figure}

\subsubsection{Doppler tomography of the LSP cycle of \fsaur.}
In conclusion, we applied a Doppler imaging technique to calculate a tomogram on
the LSP cycle of \fsaur. The original implementation of Doppler tomography has
been used to map the emission regions in CVs using information stored in emission
line profiles as a function of orbital phase \citep{Marsh-Horne}. Such an approach
allows to reconstruct the distribution of emission locked in the binary rest-frame,
but other not orbital phase-locked emission components should they exist in the
system, will be smeared out during the reconstruction. The latter case is common
among the IPs which spectral lines often vary with additional to orbital periods,
such as the spin period and its orbital sideband. In order to map the structures
like the ``accretion curtains'' in IPs, one should calculate tomograms on the spin
cycle. This technique has been used, for example, by \citet{Hellier1999} who showed
that spin-cycle tomography is a useful diagnostic tool.

We found that it is instructive to apply a similar technique to image the LSP component
of emission lines in \fsaur. We exactly followed the approach of \citet{Hellier1999}
and the interested reader should refer to this work for the technical details.
The Doppler tomogram of the \Hbeta\ emission line was
calculated of 64 individual spectra obtained during two observing nights of Dec
8-9, 2004. The line profiles were folded on the LSP cycle, and then subtracted from
the phase-invariant profile to leave only the LSP-varying component (Fig.~\ref{fig:dopmap},
the bottom left panel). The spectra were phased according to ephemeris~(\ref{ephemerisLSP}) thus phase 1 coincides
with the optical pulse maximum.
The resulting Doppler map is presented in the top panel of Fig.~\ref{fig:dopmap},
whereas the reconstructed emission line profiles are shown in the bottom right panel of Fig.~\ref{fig:dopmap}.

The LSP component produced the diffuse area of emission at 12 o'clock in the tomogram,
the emission subtends $\sim 120 \deg$ at the origin. Its brightest part reaches
$\sim$800 \kms, with a possible extension up to $\sim$1100 \kms indicating that
the source of the LSP emission is located close to the WD.

\begin{figure*}
\includegraphics[height=6.5cm]{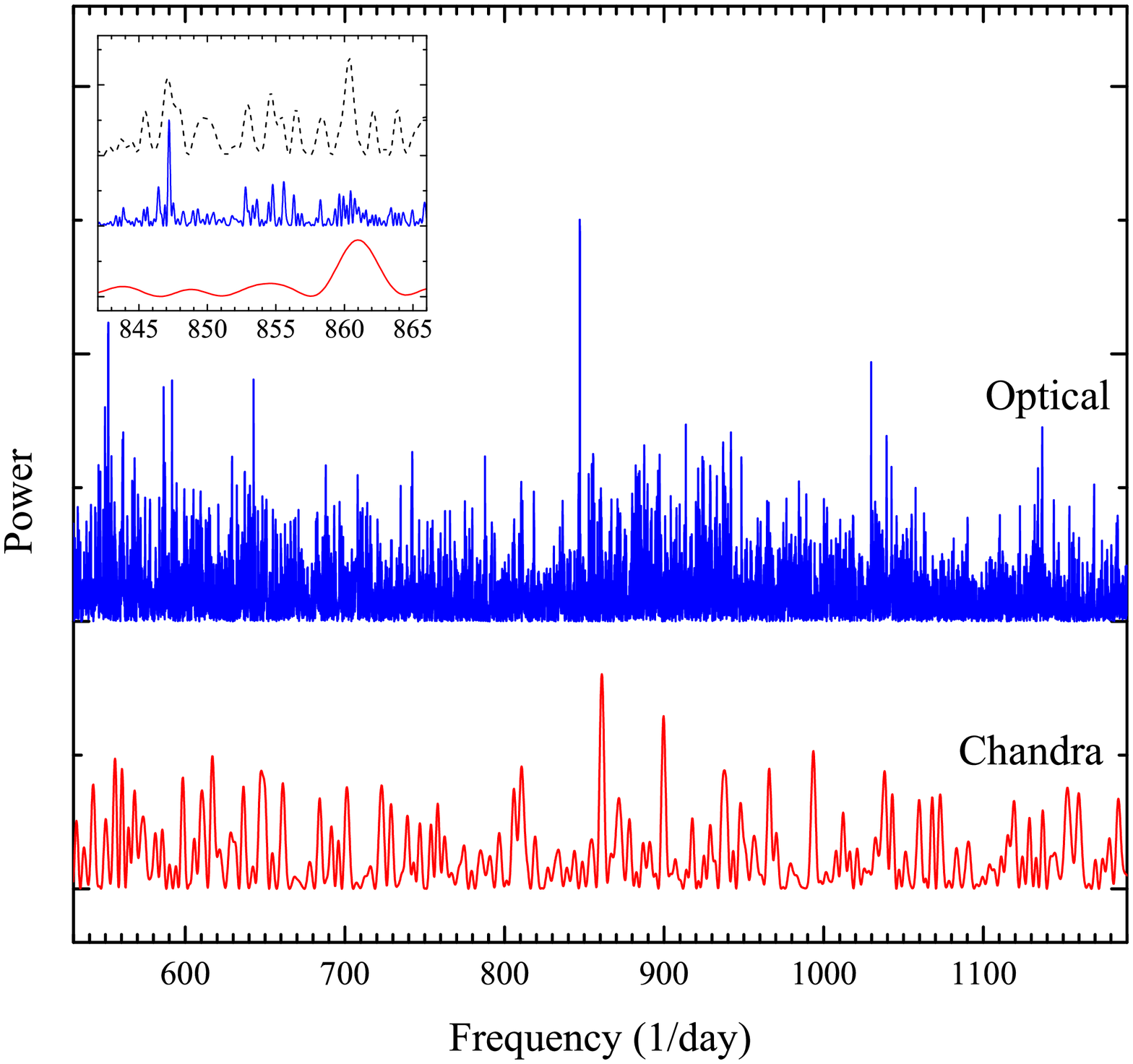}
\hspace{5 mm}
\includegraphics[height=6.5cm]{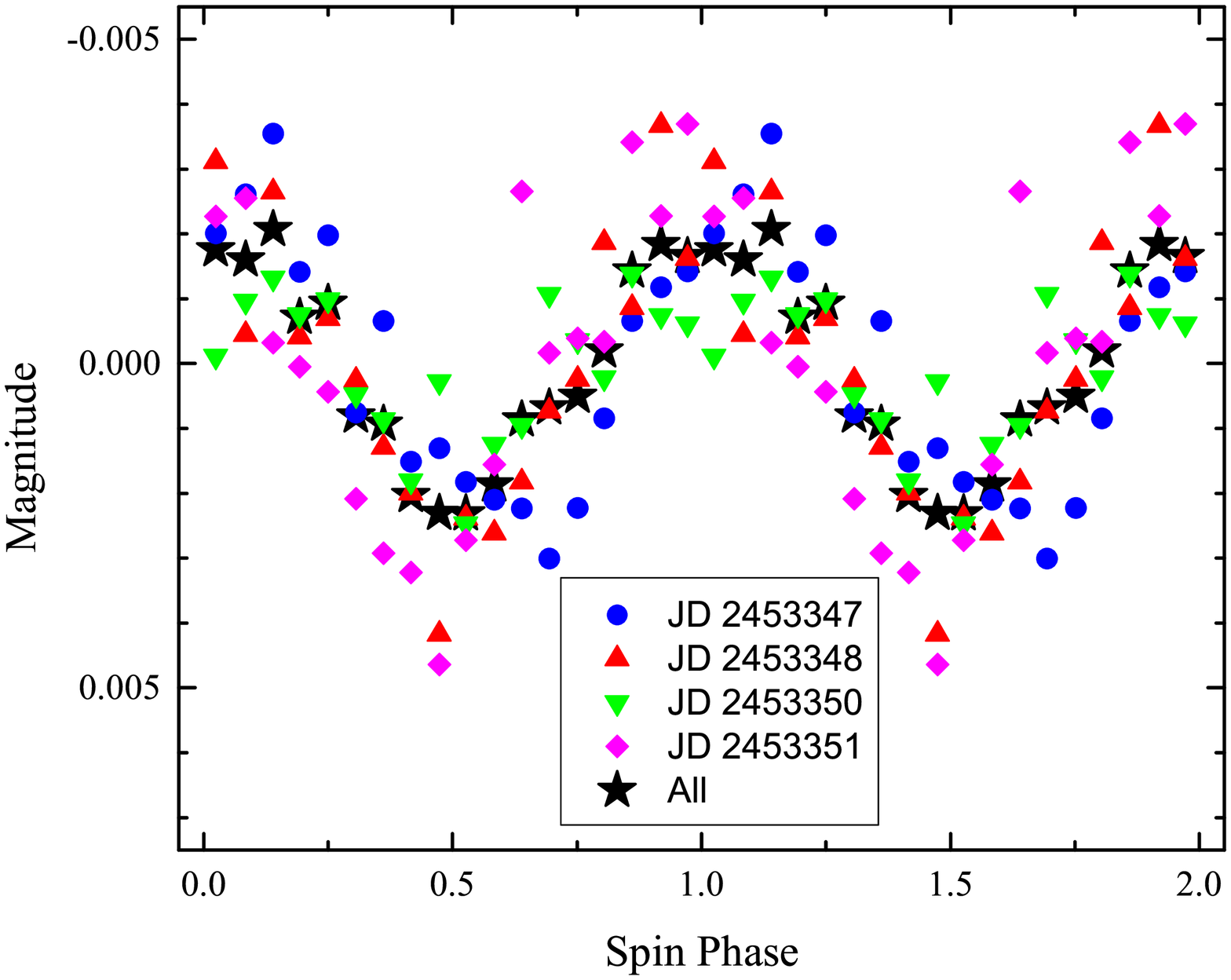}
\caption{\emph{Left panel:} The Lomb-Scargle power spectra for the Chandra data (red)
and the high time resolution observations from the \textit{set-2004} (blue).
The frequency difference between the strongest peaks in these spectra at 860.98 d$^{-1}$
(Chandra) and 847.218 d$^{-1}$ (optical) is close to the double LPP frequency,
$2\omega_{LPP}$. The inset shows the enlarged region around these frequencies. The dashed
line represents the periodogram of the night of JD 2453347 in which both peaks are
clearly seen. \emph{Right panel:} The pre-whitened $V$ light curves from the high
time resolution observations obtained between JD~2453347--2453352, folded with the
101.98-s period. The large stars represent the entire data set, whereas the smaller
symbols represent the individual nights.
All data are plotted twice for continuity.
}
\label{fig:Spin}
\end{figure*}

\subsection{Searching for the spin period of the WD}

A previous attempt to determine the spin period of the WD in \fsaur\ have proved
inconclusive. Basing on high-speed photometric observations taken with ULTRACAM,
\citet{Neustroev2005} reported the detection of oscillations with a period of
$\sim$101 and/or $\sim$105 s associated with the spin period of the WD. Nevertheless,
taking into account the short duration of those observations and the weakness of the
detected signal, the authors suggested that new high signal-to-noise ratio observations
of longer duration, and especially X-ray observations, would be very useful for
confirming this unconvincing result.

From the presented data, the Chandra observations and the high time resolution data
from the \textit{set-2004} are the most suitable for the search for the spin period
of the WD in \fsaur.

The Chandra light curve was extracted with a 22 s time resolution and has been Fourier
analyzed to detect periodic signals. Inspection of the power spectrum at the high-frequency
end (Fig.~\ref{fig:Spin}, left panel) reveals the strongest peak at 860.98 d$^{-1}$
($100.35 \pm 0.01$ s). Even though this peak is still statistically insignificant, its
frequency is very close to the one previously detected from the ULTRACAM photometry.

In order to probe if this signal is also present in the current optical data, we combined
the observations obtained between JD~2453347--2453352 with the use of the telescopes
located in South Korea, Italy and Mexico (Table~\ref{ObsPhotTab}). The Lomb--Scargle
periodogram of this long data set with total exposure time of almost 87 hours is
dominated (Fig.~\ref{fig:Spin}, left panel), at the higher frequencies, by a sharp and
relatively strong peak which frequency is however different than that detected in
the Chandra data: 847.218 d$^{-1}$ ($101.9808 \pm 0.0001$ s). To examine if this
signal is real, we calculated the power spectra on individual nights and their parts.
We found that the corresponding peak is present in all the power spectra except for
the half-night of JD 2453352. Moreover, folding these subsets at the 101.9808-s period
results in sinusoidal modulations of nearly identical amplitude of $\sim$0.002-0.003 mag
which are all in phase with each other (Fig.~\ref{fig:Spin}, right panel). All these
indicate that the periodic 101.98-s modulation really exists. Nevertheless, its cause
is still unclear.

If this is the spin period of the WD then we might expect to see the same modulation
in X-rays, yet the Chandra power spectrum shows no power at its frequency. On the
other hand, the $\sim$100.35-s X-ray signal was occasionally clearly seen in the
optical data. For example, in the periodogram of the night of JD 2453347 the 100.35-s
peak is even stronger than the 101.98-s one (the dashed spectrum in the inset in
Fig.~\ref{fig:Spin}, left panel). Furthermore, we call the reader's attention to the
fact that the frequency difference of these two modulations is very close to the
double LPP frequency, $2\omega_{LPP}$. This allows us to speculate that the real
spin period is 100.35-s whereas the 101.98-s modulation is the optical $2\omega_{LPP}$
sideband.

In conclusion, we also evaluated for the short-term variability the uvw2 and uvm2
data from the \textit{Swift-2007} observations which were taken in event mode.
We found that the power spectrum of the uvm2 data is again dominated by a number
of peaks around 0.01 Hz. However, due to the extremely complex spectral window of
the UVOT time-series, it is difficult to make a better conclusion on the exact
frequency(ies) of dominated oscillation(s). We applied the CLEAN procedure
\citep{CLEAN} to sort out the alias periods resulting from the uneven data sampling,
but still obtained several strong peaks in this frequency region. Two of them,
however, are located very close to the detected oscillations.

\subsection{X-ray Spectral Analysis}

\citet{FS2012} have recently shown that the optical brightness of \fsaur\ in quiescence
varies within a wide range (17.4--15.2 mag). The Chandra observations reported in this
paper were performed in the beginning of a significant decrease in luminosity. The Chandra
data were taken when the optical brightness in $V$ reached $\sim$16.5 mag, $\sim$0.8 mag
weaker than observed few months earlier whereas during the Swift observations \fsaur\ was
in the usual quiescent state and had
the average brightness of $V$=16.17 in 2007 and $V$=16.04 in 2011. Interesting to note that
the amplitude of variation of the average brightness in optical, UV and X-ray wavelengths
is nearly the same. The UVOT photometry shows that during the Swift observations \fsaur\
had the average brightness of $UVW1$=15.00 and $UVW2$=15.14 in 2007 and $UVW1$=14.84 and
$UW2$=15.03 in 2011, with the corresponding X-ray count rate of 0.087 cts/s in 2007 and
0.098 cts/s in 2011.
One can reasonably suspect that not only the X-ray flux, but also the X-ray spectrum
might be variable, and thus simultaneous spectral fitting of all available data by a
single model would not be feasible. Thus, we analyzed the Chandra and Swift spectra
independently (because of the low count rate for each Swift observation, they were merged
to improve the signal-to-noise ratio).

The X-ray spectrum of \fsaur\ can be characterized by a relatively smooth continuum with
strong interstellar absorption that practically leaves no counts below 0.3 keV, and with
several superimposed spectral lines, the most intense being the fluorescence Fe K$\alpha$
emission line at $\sim$6.4 keV. The extracted spectra were fitted with \xspec\ v12.7.1
\citep{XSpec}. The spectra were rebinned to have a minimum of 20 counts per bin to allow
the use of the $\chi^2$-statistics.

The results of simple model fits, including single-component blackbody, thermal bremsstrahlung
or collisionally ionized thermal equilibrium plasma models (\bbody, \brems\ and \mekal\ in
\xspec, respectively), along with a Gaussian to fit the Fe lines near 6.4 keV, have
been shown to be inadequate to describe the spectrum, but multiple-temperature model fits
are good. We found the best fits with composite models that consist of a combination of a
number of \mekal\ components  with different temperatures but an identical abundance,
absorbed by a partial covering absorption (\pcfabs).

It is well known that in most IPs the spectra undergo strong photoelectric absorption
which usually cannot be characterized by a single column density. Multiple absorption
components are often required to adequately fit the spectra. We also tried using both
simple photoelectric absorption (\phabs) and a partial covering absorber (\pcfabs). We
found, however, that there was no noticeable improvement with a multiple-absorber model,
whereas a single partial covering absorber gave a slightly better result  in most cases
than a simple photoelectric absorber. We thus removed a simple absorber with no effect
on the fit quality.

\begin{table*}
\label{XspecTab}
\begin{center}
\caption{Model components and parameters fitted to the phase-averaged spectrum
of \fsaur. The errors are given to the same power of ten as the values.}
\begin{tabular}{llllll}
\hline
Component     &  Parameter                & Chandra                     & Chandra                     & Chandra                      & Swift                        \\
              &  (Units)                  & (average)                   & (unpulsed)                  & (pulsed)                                                    \\
\hline
Part. Absn.   & $N_{\rm H}$ (\tim{21}\cms)& \plmin{1.70}{-0.63}{+1.05}  & \plmin{1.55}{-0.26}{+0.36}  & \plmin{10.19}{-2.74}{+2.50}  & \plmin{1.11}{-0.12}{+0.12}   \\
              & CvrFract                  & \plmin{0.70}{-0.08}{+0.15}  & \plmin{0.70}{-0.05}{+0.0.09}& \plmin{0.75}{-0.07}{+0.05}   & 1.0 (frozen)                 \\
Mekal         &  $kT$ (keV)               & \plmin{0.64}{-0.05}{+0.03}  & \plmin{0.64}{-0.06}{+0.03}  & \plmin{0.63}{-0.05}{+0.05}   & \plmin{0.52}{-0.07}{+0.10}   \\
              &  Abundance                & \plmin{0.69}{-0.20}{+0.23}  & 0.69 (frozen)               & 0.69 (frozen)                & \plmin{0.34}{-0.22}{+0.22}   \\
              &  Norm (\tim{-3})          & \plmin{0.15}{-0.04}{+0.07}  & \plmin{0.14}{-0.02}{+0.02}  & \plmin{0.59}{-0.20}{+0.24}   & \plmin{0.12}{-0.02}{+0.02}   \\
Mekal         &  $kT$ (keV)               & \plmin{1.68}{-0.40}{+0.55}  & \plmin{1.45}{-0.27}{+0.62}  & \plmin{2.40}{-0.38}{+1.16}   & \plmin{1.63}{-0.29}{+0.33}   \\
              &  Norm (\tim{-3})          & \plmin{0.20}{-0.09}{+0.14}  & \plmin{0.16}{-0.05}{+0.16}  & \plmin{2.12}{-0.57}{+0.66}   & \plmin{0.30}{-0.11}{+0.13}   \\
Mekal         &  $kT$ (keV)               & \plmin{11.8}{-2.1}{+3.5}    & \plmin{13.7}{-2.6}{+7.7}    & 13.7 (frozen)                & \plmin{12.3}{-2.6}{+5.3}     \\
              &  Norm (\tim{-3})          & \plmin{1.54}{-0.12}{+0.10}  & \plmin{1.54}{-0.10}{+0.05}  & \plmin{0.76}{-0.71}{+0.44}   & \plmin{2.23}{-0.14}{+0.15}   \\
\hline
\rchisq (dof) &                           & 1.04 (190)                  & 0.98 (178)                  & 0.92 (60)                    & 0.92 (165)                   \\
Flux (obs), 0.3-10 keV    & (erg cm$^{-2}$ s$^{-1}$)  & 3.22\tim{-12}               & 3.2\tim{-12}               & 3.2\tim{-12}                & 4.34\tim{-12}                \\
\hline
\end{tabular}
\end{center}
\end{table*}

First the models were fitted to the strongest spectrum i.e. the averaged Chandra spectrum.
The two-temperature \mekal\ model with solar abundance (plus a Gaussian and absorption) gave
a fit with $\chi_{\nu}^2=1.13$ with temperatures 0.67 and 8.35 keV. The fit further improves
($\chi_{\nu}^2=1.11$) when the metal abundance is left to free to vary. A third \mekal\
component further improved the fit to $\chi_{\nu}^2=1.04$. This fit gave
temperatures of 0.64, 1.68 and 11.8 keV. The inclusion of additional \mekal\ components
gave a negligible improvement in $\chi^2$, so we show the parameters of the best-fit
three-temperature \mekal\ model in Table~4.

Even though a good fit to this averaged Chandra spectrum was obtained, the detected X-ray
energy-dependent orbital variability of \fsaur\ (Section~\ref{sec:OP}) suggests the distinct
spectral variations over the orbital period. In order to investigate it further we extracted
the pulsed and unpulsed spectra. The former was extracted for the orbital phase interval
0.27--0.43,
approximately centered on the orbital spikes, the data from the time interval of the missing
pulse were also included in this spectrum. The unpulsed spectrum consists the rest of the
data. The count-rates of the pulsed and unpulsed spectra differ notably: 0.42 and 0.37 \cps,
respectively.

\begin{figure*}
\includegraphics[height=6.5cm]{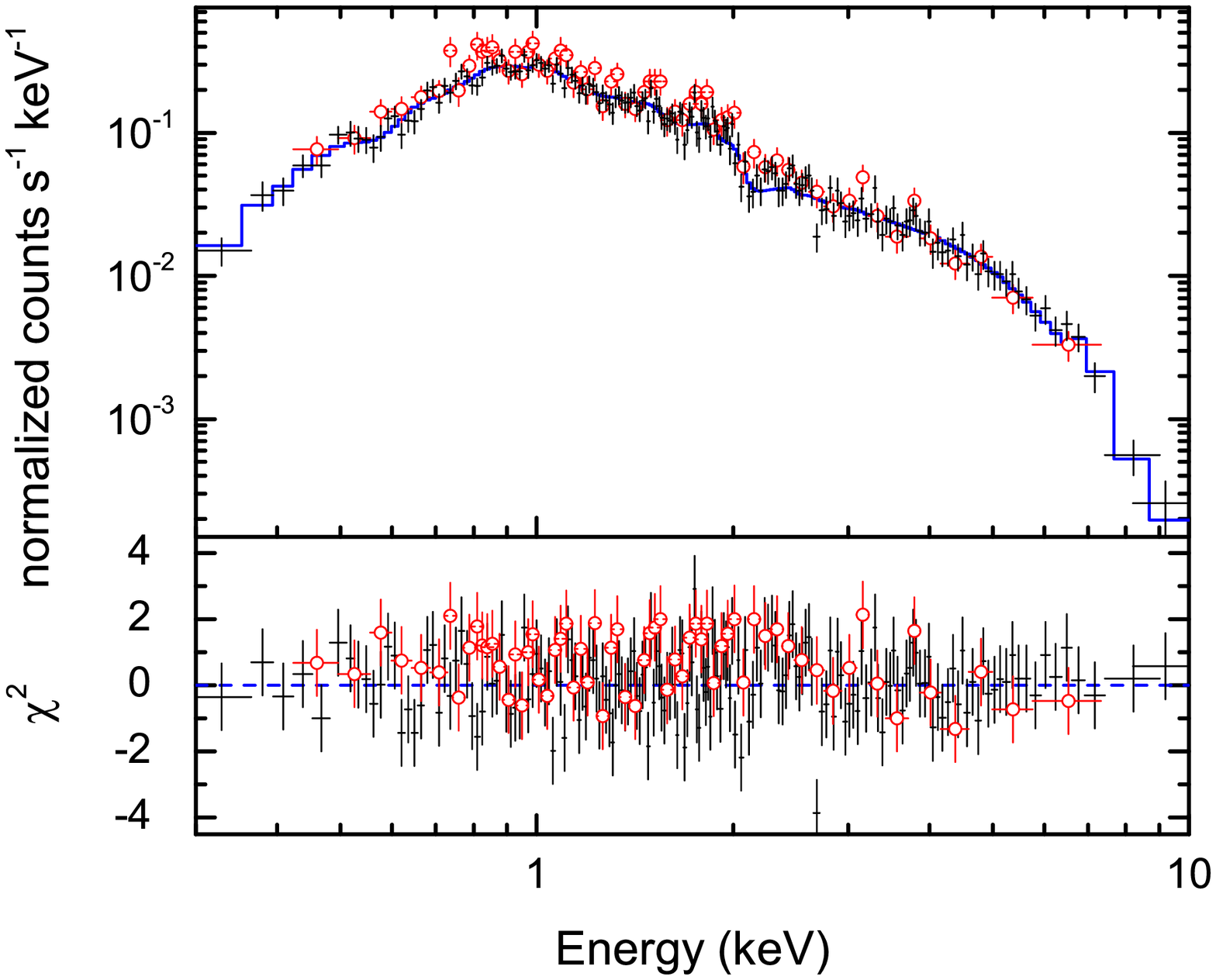}
\hspace{5 mm}
\includegraphics[height=6.5cm]{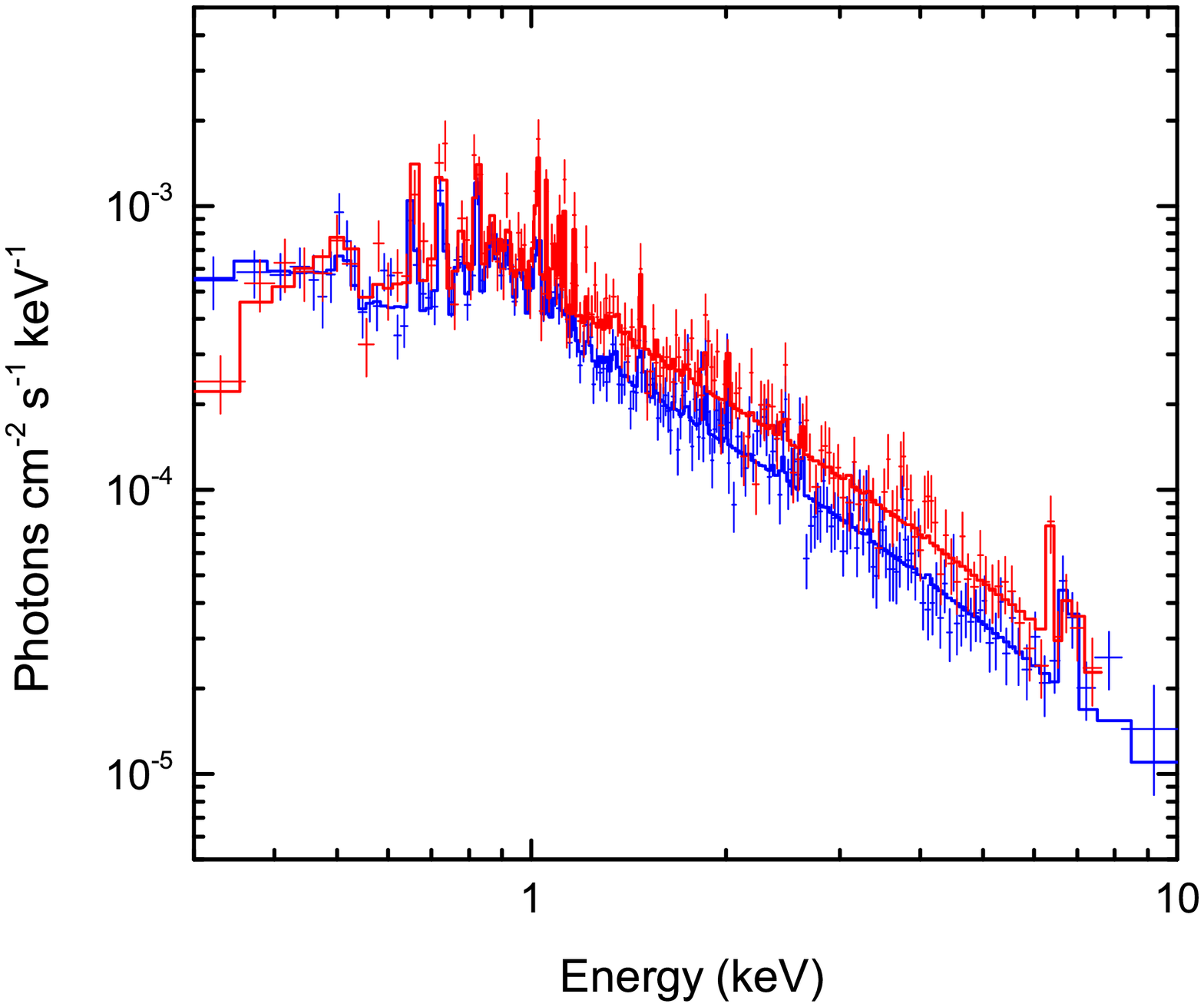}
\caption{\emph{Left panel:} The Chandra unpulsed (blue) and pulsed (red) spectra of
\fsaur\ together with the best fit model for the unpulsed spectrum. The bottom panel
shows the residuals. The pulsed spectrum shows an excess of counts, most notable in
the 0.5--2.0 keV band. \emph{Right panel:} The averaged Chandra (blue) and Swift (red)
spectra with the best fit models. The flux determined from the Swift spectrum is
$\sim$40 per cent larger than determined from the Chandra spectrum.}
\label{fig:Xspec}
\end{figure*}

The unpulsed spectrum can be well fitted with the three-temperature \mekal\ model, giving
$\chi_{\nu}^2=0.98$ with temperatures 0.64, 1.45 and 13.7 keV (we fixed the metal abundance
at a value of 0.69). Even though the pulsed spectrum consists of only 1581 counts and
suffers from their deficiency in the hardest ($\ga$8 keV) and softest ($\la$0.5 keV)
energies, it looks very similar to the unpulsed spectrum, yet it shows a notable flux
excess in the 0.8--2.5 keV energy range. This spectrum was fitted with the same model in
which we fixed the hardest \mekal\ component temperature at the value determined for the
unpulsed spectrum (13.7 keV). The fit gave $\chi_{\nu}^2=0.92$ with the same temperature
for the softest \mekal\ component (0.63 keV) whereas the second \mekal\ component has
a higher temperature of 2.40 keV and a much larger normalization than for the unpulsed
spectrum. The pulsed and unpulsed spectra together with the best fit model for the unpulsed
spectrum and the residuals are shown in Fig.~\ref{fig:Xspec} (left panel).

The Swift averaged spectrum, when fitted with the same three-temperature \mekal\ model,
gives fit parameters very similar to those obtained for the Chandra data. We notice,
however, that this model was not able to reproduce the iron K$\alpha$ emission line at
6.4 keV in the Swift spectrum, a separate Gaussian line (\gauss) was required. The fit
gave $\chi_{\nu}^2=0.92$ with temperatures 0.52, 1.63 and 12.3 keV. The averaged Swift
and Chandra spectra with the best fit models are shown in Fig.~\ref{fig:Xspec} (right
panel).

The unabsorbed flux determined from the spectra in the 0.3-10 keV range was 4.34\tim{-12}
erg cm$^{-2}$ s$^{-1}$ during the Swift observations, $\sim$40 per cent larger than
during the Chandra observation (3.22\tim{-12} erg cm$^{-2}$ s$^{-1}$). This ratio is also
kept in shorter energy subranges along the X-ray spectrum as well as in the optical band.
Thus, despite the different states in which we found \fsaur\ during the Chandra and Swift
observations, the shape of its spectrum from optical to X-rays shows little variations.

\section{Discussion}

\subsection{FS Aur as an Intermediate Polar}
Since the first detailed investigations, FS Aur immediately attracted attention
for having the very coherent, long periods (LPP and LSP) which exceed the orbital
period. In order to explain such a discrepancy between the periods in \fsaur,
\citet{Tovmassian2007} proposed the IP scenario with a rapidly rotating magnetic
WD precessing with the LSP. The period of precession depends on the star parameters
such as mass, size, shape, and the dynamics of its interior.
According to existing models \citep{Leins}, the precession period of a WD is
proportional to the third power of the spin period. In a slowly rotating compact
star, the precession period is extremely long. However, in a rapidly rotating WD
the precession period can explain observed long periodicities.
In order to have the proposed precession period, the WD in FS Aur should be
a fast rotator with the spin period of the order of 50-100 sec, according
to calculations of \citet{Leins}.

Because of the magnetic nature of the WD, one could expect to observe a strong and
stable modulation in the X-ray and optical light curves with such a period. This
hypothesis has received an observation confirmation after our detection of the $\sim$100~s
modulations in both the optical and X-ray data of \fsaur, even though the detected
signal is rather weak, especially in X-rays. In this context, it is interesting to
note that V455 And, another cataclysmic variable with many similar to \fsaur\ properties
\citep{Tovmassian2007}, has also been proven to possess a rapidly rotating WD with the
spin period of 67.6~s \citep{Gaensicke2007, BloemenV455And}. Similarly to \fsaur,
the spin period of the WD in V455 And was found through extensive  optical observations
whereas the power at the spin frequency in X-rays is very low \citep{Tovmassian2012a}.

This is slightly confusing but not really surprising. It was shown that if the magnetic
and rotational axes of a WD are closely aligned, then spin modulations might be
undetectable \citep{Ramsay2008}. Nevertheless, if such a WD in addition experiences
a precession then another modulation with the precession period may appear. Other
specific observational manifestations of precession are still to be understood. 
We guess, however, that as the primary reason for the temporal and spectral variability
of the IPs with the spin period is the variable geometrical factor due to the sweeping around
of the X-ray beam, a binary system with a precessing WD should share most of the
observational properties with ordinary IPs and even might be virtually indistinguishable
from them.
On the whole, the closer the magnetic and rotational axes of a precessing WD are aligned,
the more similarities such a binary will share with a classical IP.
The difference will only be apparent in the time scales of the variability:
instead of modulations with the \textit{spin and orbital-spin-beat} periods we expect
to observe modulations with the \textit{precession and orbital-precession-beat} periods.
We believe that FS Aur is such a case.

Before discussing the nature of FS Aur in more detail, we briefly summarize
the key observational properties of FS Aur which are relevant in this context:
\begin{enumerate}
 \item It has been proven that FS Aur exhibits a stable modulation in the optical
       and X-ray light curves with a period that is different from an orbital
       one (LSP).
 \item The emission lines also vary with this period.
 \item The optical light curve is normally dominated by variations at the beat
       period between the OP and the LSP: $1/P_{LPP}=1/P_{OP}-1/P_{LSP}$.
 \item \fsaur\ is a rather hard X-ray source with low-energy absorption and the presence
       of Fe K$\alpha$ emission.
 \item \citet{Polarization} reported the detection of a non-zero circular
       polarization in \fsaur.
\end{enumerate}

As seen, these properties of \fsaur\ satisfies most of Patterson's conditions for
being an IP \citep{Patterson1994}. They conflict only in the sense that in the canonical
model of the IPs a period of optical and X-ray modulations due to the WD rotation is
expected to be shorter than the OP. Nevertheless, as it was shown before, this restriction
can be overtaken if a role of the WD rotation is played by the precession. One additional
X-ray property of the IPs which was proposed by \citet{Norton_IPs} and which is also
fulfilled by \fsaur\ -- the presence of strong Fe K$\alpha$ emission.
Furthermore, the small amplitude outbursts observed in \fsaur, are another good
argument in favor of a truncated accretion disk in the system, i.e. the IP nature
of \fsaur.

We now proceed to compare other properties of FS Aur such as the phasing of the
optical and X-ray LSP pulses and their energy dependence with other well-studied IPs.

\subsection{Comparison with other IPs}

The spin period modulation is unambiguously detected in all the confirmed IPs as
this is the defining characteristic of the class. This modulation is a direct
consequence of the magnetically confined accretion flow onto the magnetic poles
of the WD whose magnetic field is of sufficient strength to disrupt the accretion
disc and to control the flow before it reaches the surface of the star.
At some distance from the WD surface, the
infalling material undergoes a strong shock, releasing X-rays as it cools by
thermal bremsstrahlung and Compton cooling processes. Thus, it is widely supposed
that the X-ray modulation reflects this physical process most clearly whereas the
optical range is heavily affected by X-ray reprocessing. This, however, is not
always the case. For example, DQ\,Her -- the prototype of the subclass of
IPs -- is characterized by a very low X-ray flux and the strong spin modulation
in the optical band.

IPs show a large variety of photometric and spectroscopic behaviour, both in
X-rays and the optical range. In X-rays, for example, some IPs show a
single-peaked pulsation whereas others show a double-peaked pulsation.
There is also observed a strong dependence on the photon energy (usually,
increasing modulation depth with decreasing energy).

The optical spin modulation, if observed, is usually correlated with the X-ray
one. However, the optical variability is caused by the reprocessing of variable
X-ray irradiation that often resulted in significant sideband modulations. These
sidebands and spin modulations can also have a strong wavelength-dependence.
Furthermore, these modulations are observed not only in
continuum but also in the hydrogen and \HeII\ emission lines. In many IPs the
spin component, appearing in the far wings of the emission lines, has maximum
blueshift near optical and/or X-ray pulse maximum \citep{Hellier1999}.

Despite such complex behaviour, our understanding of IPs is now quite good.
Most observable properties can be understood within the context of the widely
accepted accretion curtain model \citep{Rosen1988,Hellier1991}. In this model,
the material flows towards the magnetic poles of the WD in an arc-shaped curtain,
and the largest X-ray and optical flux is seen when the upper pole is on the far
side of the WD. It is not our intention to assess the strong and weak aspects of
the model. Instead, we aim to compare the key, most reliable observable properties
of \fsaur\ with those of the ``ironclad'' IPs. First of all, we pay attention to
the optical brightness and colour variability with a spin period, and a mutual
phasing between them and the emission line spin variability.

\begin{figure*}
\includegraphics[width=7.9cm]{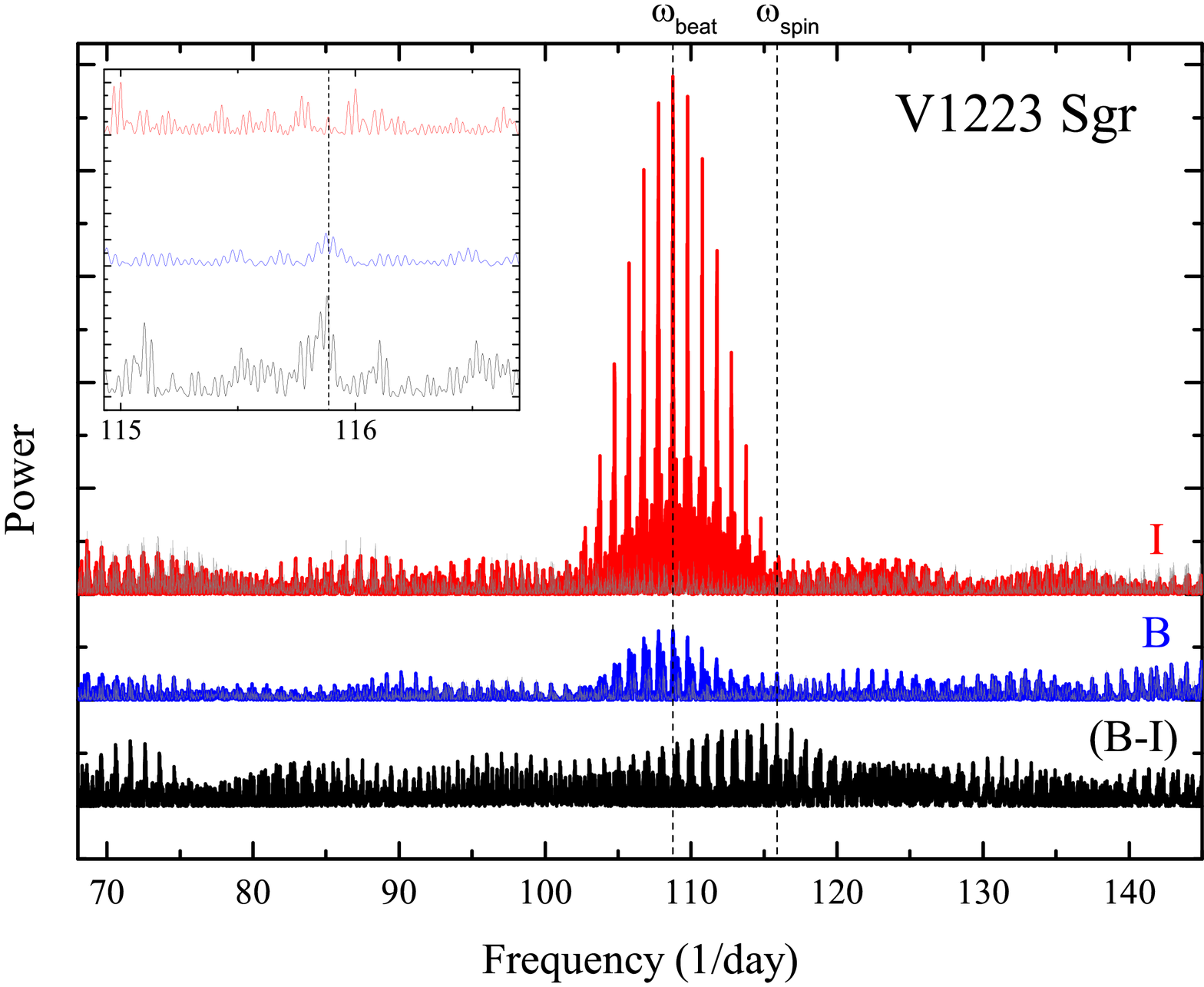}
\hspace{5 mm}
\includegraphics[width=8.5cm]{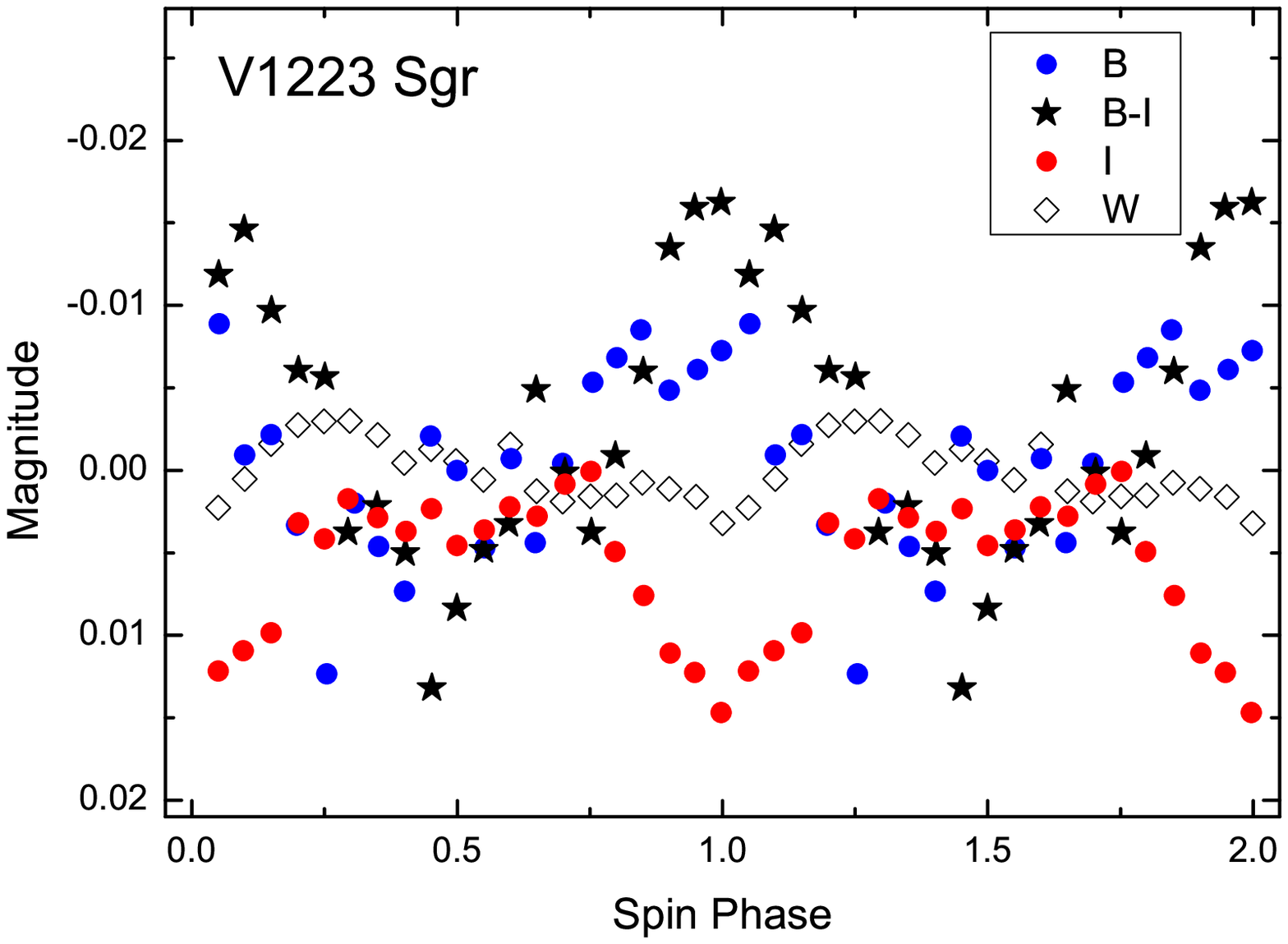}
\caption{\emph{Left panel:} The Lomb-Scargle power spectra of the $B$ (blue) and $I$
(red) light curves and the ($B-I$) colour-index curve (black) of V1223 Sgr. The gray
lines atop of the $B$ and $I$ periodograms represent corresponding pre-whitened spectra
after the removal of the orbital frequency and the pulsation at the sideband frequency
of 108.76 d$^{-1}$. The inset shows enlarged region around
the spin frequency. \emph{Right panel:} The $B$, $I$ and unfiltered (white) light curves
and the ($B-I$) colour curve of V1223~Sgr folded with the spin period. All data are
plotted twice for continuity.}
\label{fig:V1223_Sgr}
\end{figure*}

Unfortunately, such a comparison is not a simple task. Despite numerous
multicolour observations of IPs conducted and published in the past, there
are very few reports on the analysis of the colour spin variability. However,
our primary result -- the detection of the LSP variability through photometric
observations -- was obtained by means of the analysis of colour indices. Besides
this variability being most confidently seen specifically in ($B-I$), our analysis
also revealed another interesting property of the \fsaur\ colour time-series --
its power spectrum is exclusively dominated by a peak at the LSP with no power at
the LPP and the OP. We intended to test if the IPs show similar power spectra
with the strongest peak at the spin frequency.

In order to conduct such an analysis, we requested our colleagues worldwide to
provide us with their multicolour light curves. We were generously given the data
of eight ironclad and confirmed IPs which show prominent spin modulations in the
optical wavelengths alongside with different sidebands. Most of these observations
were already published: PQ~Gem \citep{Hellier1994}, V405~Aur
\citep{V405Allan}, NY~Lup and IGRJ1509--6649 \citep{Potter2012}, DQ~Her
\citep{Butters2009}, EX~Hya \citep{Ex_Hya}, FO~Aqr \citep{Chiappetti1989}.
Observations of AO~Psc were obtained by E.~L. Robinson and were extracted, alongside
with Chiappetti et al.'s data of FO~Aqr, from the collection of light curves of
Albert Bruch \citep{Bruch}.

An analysis of all these data was performed in a similar fashion. We calculated
the power spectra of the light and different colour curves, and then compared the light
and colour curves folded with the spin period. Because of space limitations, we do not
show any plots here\footnote{See http://vitaly.neustroev.net/research/intermediate-polars/ip-colours/
on each star for power spectra and pulse profiles.} but only give our conclusion on this
study. We found that in \textit{all} the IPs the power spectrum of the ($B-I$) or similar
colour curve is dominated by a peak at the spin period with little power at the beats,
in spite of the fact that in several systems the sideband variability is more prominent
in the optical light curve than the spin modulation. As for energy dependence and phasing
of the spin pulses in different wavelengths, they have not appeared to be very consistent
even for these ironclad IPs.

In conclusion, and as an example of the most extreme case, we present more detailed
analysis of V1223~Sgr. This well-known IP is known to show very strong sideband
modulations, whereas the spin pulsation is not seen in the optical. In this sense,
the behaviour of V1223~Sgr is similar to \fsaur.
Our own observations were used for the analysis.

Finally, we also performed an analysis of multicolour observations of V455~And
which shows solid evidence for being an IP \citep{BloemenV455And}.

\begin{figure*}
\begin{center}
\hbox{
\includegraphics[height=8cm]{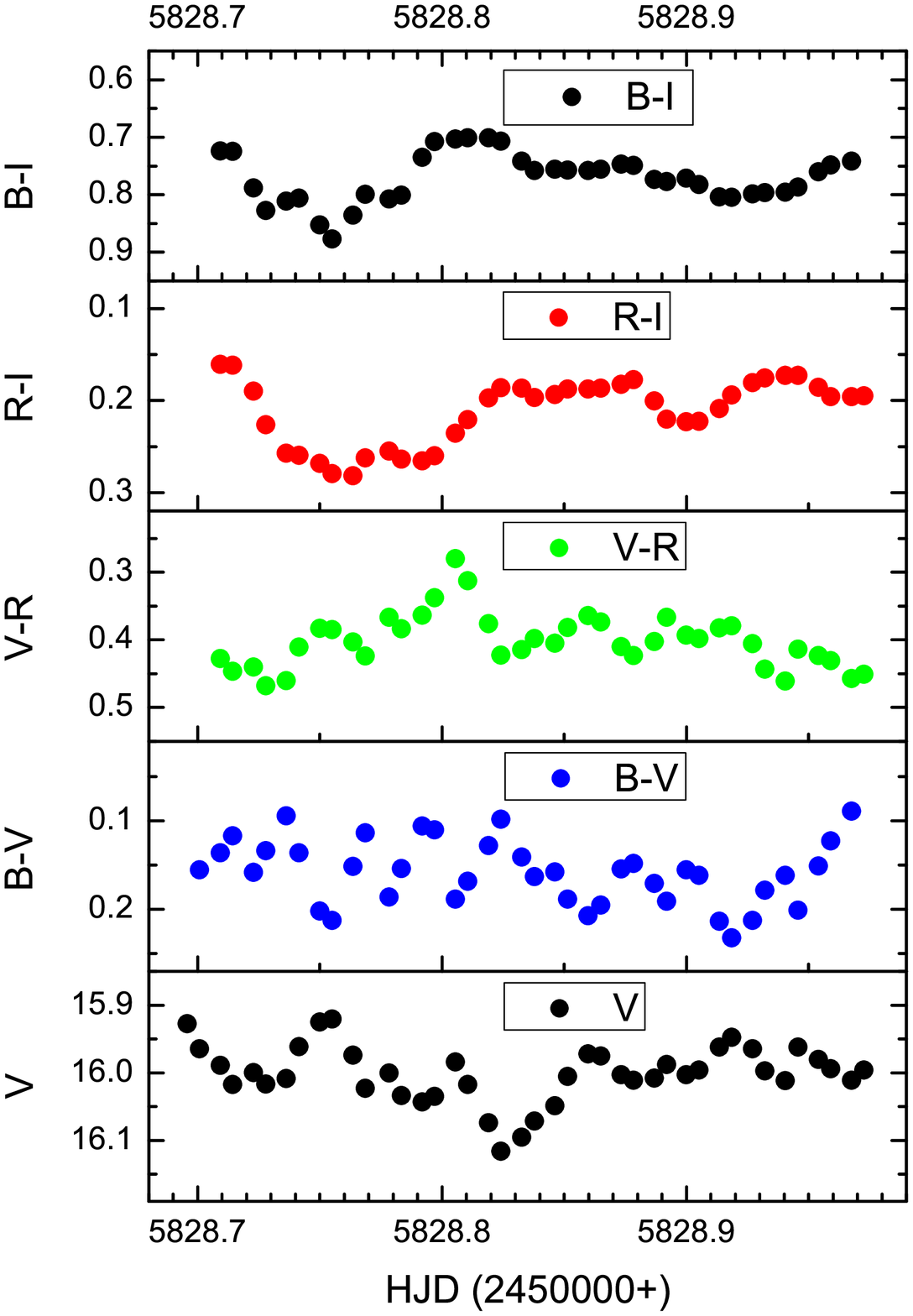}
\includegraphics[height=8cm]{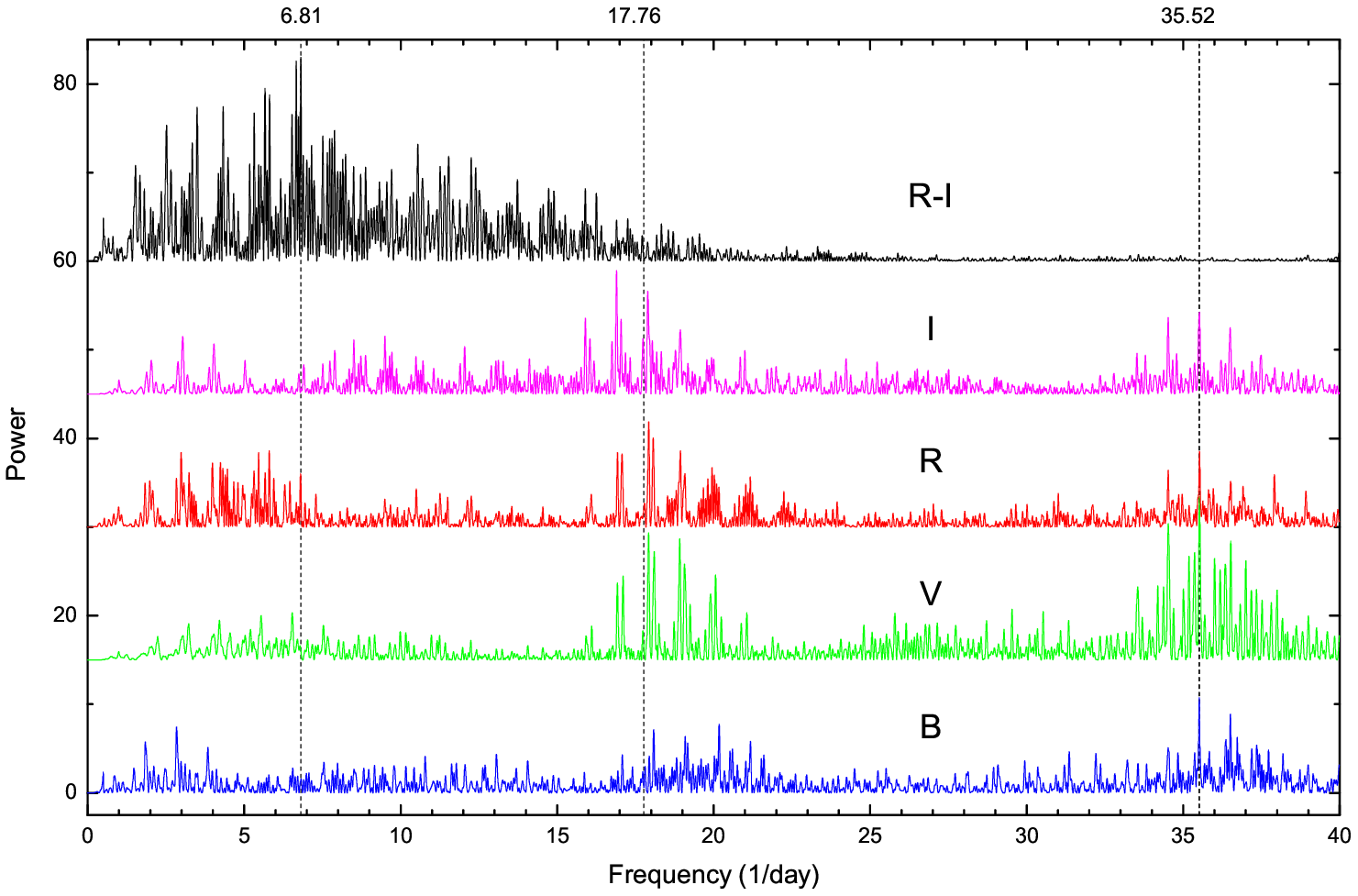}
}
\end{center}
\caption{
\emph{Left panel:} Sample $V$ light curve and ($B-V$), ($V-R$), ($R-I$)
and ($B-I$) colour curves of V455 And from JD 2\,455\,828.
\emph{Right panel:} The Lomb-Scargle power spectra of the $B\,V\,R\,I$
light curves and the ($R-I$) colour-index curve of V455 And.
No signal is detected at the orbital frequency of 17.76 d$^{-1}$ but a strong signal
is found at twice the orbital frequency of 35.52 d$^{-1}$. The ($R-I$) periodogram
shows excess of power in the low-frequency range with the strongest peaks at 6.81
d$^{-1}$ and 6.62 d$^{-1}$. These frequencies are close to the spectroscopic period
found from the radial velocity variations of the Balmer emission line wings
\citep{V455And,Tovmassian2007}.
}
\label{fig:v455}
\end{figure*}

\subsubsection{V1223 Sgr}

V1223~Sgr has an orbital period of 3.37~h and a spin period of 745.6~s. The sinusoidal
spin modulation dominates in X-rays and shows the increasing depth with decreasing
energy. In the optical, however, the object pulsates at the 794.4~s sideband and no
pulsations have been found at the spin period.

During 29 nights of September-October 2011 and 25 nights of May-June 2012 we conducted
time-resolved observations of V1223 Sgr using the same telescope and equipment as
for the observations of \fsaur. The observations of 2011 were done in unfiltered
light and with an integration time of 45 sec. Most observations of 2012 were made
with the $B$ and $I_c$ filters and with exposure times of 70 and 30 sec respectively.
A detailed analysis of these data will be discussed elsewhere, here we concentrate
on the variability with the spin period.

The 794.4~s sideband modulation was very strong in both filters. The power spectrum
of the light curve is dominated by peaks at the sideband and orbital frequencies
(108.76 and 7.13 day$^{-1}$ respectively) with no (in $I$) or negligibly
little power (in $B$) at the expected spin frequency $\Omega+\omega_{beat}=115.89$
day$^{-1}$ (Fig.~\ref{fig:V1223_Sgr}, left panel). The latter modulation has not
become apparent even in the pre-whitened light curve.
However, in the power spectrum of the ($B-I$) colour-index curve a peak
emerges at the exact spin frequency. Folding the ($B-I$) data over the spin period
results in the nearly sinusoidal modulation with the amplitude of $\sim0.03$ mag
(Fig.~\ref{fig:V1223_Sgr}, right panel). The folded $B$ and $I$ data also show
a sign of pulsations even though the latter are very weak in power spectra. It is
interesting that the character of such variability is very different from what we
detected in other IPs. During around a half of the spin period the $B$ and $I$
fluxes keep a nearly constant level, and then it increases in $B$ and decreases in
$I$. No pulsation is seen in the folded unfiltered light curve.

\subsubsection{V455 Andromedae}

V455 And (HS 2331+3905) is a high-inclination eclipsing dwarf nova.
The system shares many similar properties with FS Aur and is known to show
a variety of periodic photometric and spectroscopic variabilities.
\citet{V455And} reported that V455 And exhibits two very different spectral
periods, a short one of 81 min at the center and a longer one of 210 min in
the wings of its emission lines. Eclipses in the system's light curve allowed
an accurate and unambiguous determination of the orbital period of 81.08
minutes. However, the strongest peak in the photometric periodogram was located
at twice the orbital frequency of 35.52 d$^{-1}$ (40.54 min), while no power was
detected at the orbital frequency of 17.76 d$^{-1}$ even though a strong signal
was found at 17.27 d$^{-1}$ (83.38 min) and the one-day aliases (see Fig.~6 in
\citealt{V455And}). \citeauthor{V455And} interpreted it by a double-hump
structure of the orbital light curve and the presence of permanent superhumps.
Besides the orbital variability, V455~And exhibits a number of other periodic
photometric variations and is proved to contain a magnetic WD. The 67.6~s spin
period of the WD defines this object as an IP \citep{Gaensicke2007,BloemenV455And}.

Nevertheless, similarly to FS Aur, the LSP ($\sim$3.5~h) appearing in the far
wings of the emission lines, is not seen in the light curve of V455~And.
Also, by analogy with FS~Aur, one can expect to observe the LPP -- photometric
modulations with the beat period between the two spectral ones
($\sim$11 d$^{-1}$). \citeauthor{V455And} did not mention any variability with
such a period, even though their periodogram showed a marginal peak in this
frequency region (Fig.~6 in their paper).

For analysis of the longer-term variability,
during 11 nights of September 2011 we conducted $BVR_cI_c$ time-resolved
observations of V455~And using the same telescope and equipment as for the
observations of FS Aur. A sample $V$ light curve and ($B-V$), ($V-R$),
($R-I$) and ($B-I$) colour curves are shown in Figure~\ref{fig:v455}
(left panel) while the Lomb-Scargle power spectra of the $B, V, R, I$
light curves and the ($R-I$) colour-index curve are shown in the right panel
of Figure~\ref{fig:v455} (from bottom to top of the figure).

The light curves and the ($B-V$) and ($V-R$) colour curves display periodic
variability with a period of $\sim$80 min. When folded with the orbital period,
the double-hump structure of the light curves is clearly seen (not shown). The
individual power spectra look similar to that presented by \citeauthor{V455And}
even though they are different in some details: a) the strengths of the strongest
peaks at 35.52 d$^{-1}$ and around the orbital frequency are now compatible
(the 35.52 d$^{-1}$ is stronger in the $B$ band and weaker in $I_c$); b) the
``superhump'' peaks at 17.27 and 16.27 d$^{-1}$ disappeared but new
ones appeared at 16.93 and 17.1 d$^{-1}$ and 17.93 and 18.1 d$^{-1}$.

However, the behavior of the colour indices ($B-I$) and ($R-I$) suggests
a longer variability timescale. The corresponding periodograms show excess of
power in the low-frequency range with the strongest peaks at 6.81 d$^{-1}$ and
6.62 d$^{-1}$ (211 min and 218 min) (better seen in ($R-I$),
Fig.~\ref{fig:v455}). These frequencies are close to the spectroscopic period
found from the radial velocity variations of the Balmer emission line wings
\citep{V455And,Tovmassian2007}. We have to note that the power spectrum around
these signals is complex, suggesting a lower degree of coherence.

This is not much of a surprise as despite the many similarities between V455~And and
FS~Aur, there is a fundamental difference between them. In the case of FS~Aur,
the long periods are very strict. For V455~And where only the LSP is reported,
this period fluctuates on time scales of a few days \citep{V455And,Tovmassian2007}.

\subsection{\fsaur\ and IPs: similarities and differences}

The above analysis of multicolour optical light curves of well-known IPs has revealed
a common property of the corresponding colour power spectra -- the latter are dominated
by a peak at the spin frequency even though the sideband variability is more prominent
in the light curve than the spin modulation. The example of V1223~Sgr is the most
notable, in which the ($B-I$) power spectrum indicates the presence of the spin
pulsation which is not seen in the optical light curve at all. This closely resembles
the optical variability of \fsaur\ and V455~And with the LSP.

This effect can be naturally explained as a consequence of reprocessing of spin-modulated
X-rays by asymmetric features in the system that are locked in the rotating binary
frame, such as the secondary star or the bright spot \citep{Sideband1,Sideband2}.
The photometric modulation appears due to the variation in the visible area of the
reprocessing site which has a weak energy dependence. As a result, the sideband
modulations are not present in colour data.

Knowledge of the orbital, spin and sideband ephemerides allows the question
of the site of reprocessing in the system to be examined. Using \fsaur's ephemerides
(\ref{ephemerisOP}), (\ref{ephemerisLPP}) and (\ref{ephemerisLSP}), one can see that
the LSP and LPP modulations have coincident maxima at orbital phase $0.50\pm0.01$.
This corresponds to superior conjunction of the secondary and is likely due to heating
of the inner hemisphere of the donor star that is quite typical among the IPs.

\begin{figure}
\includegraphics[angle=270,width=8.5cm]{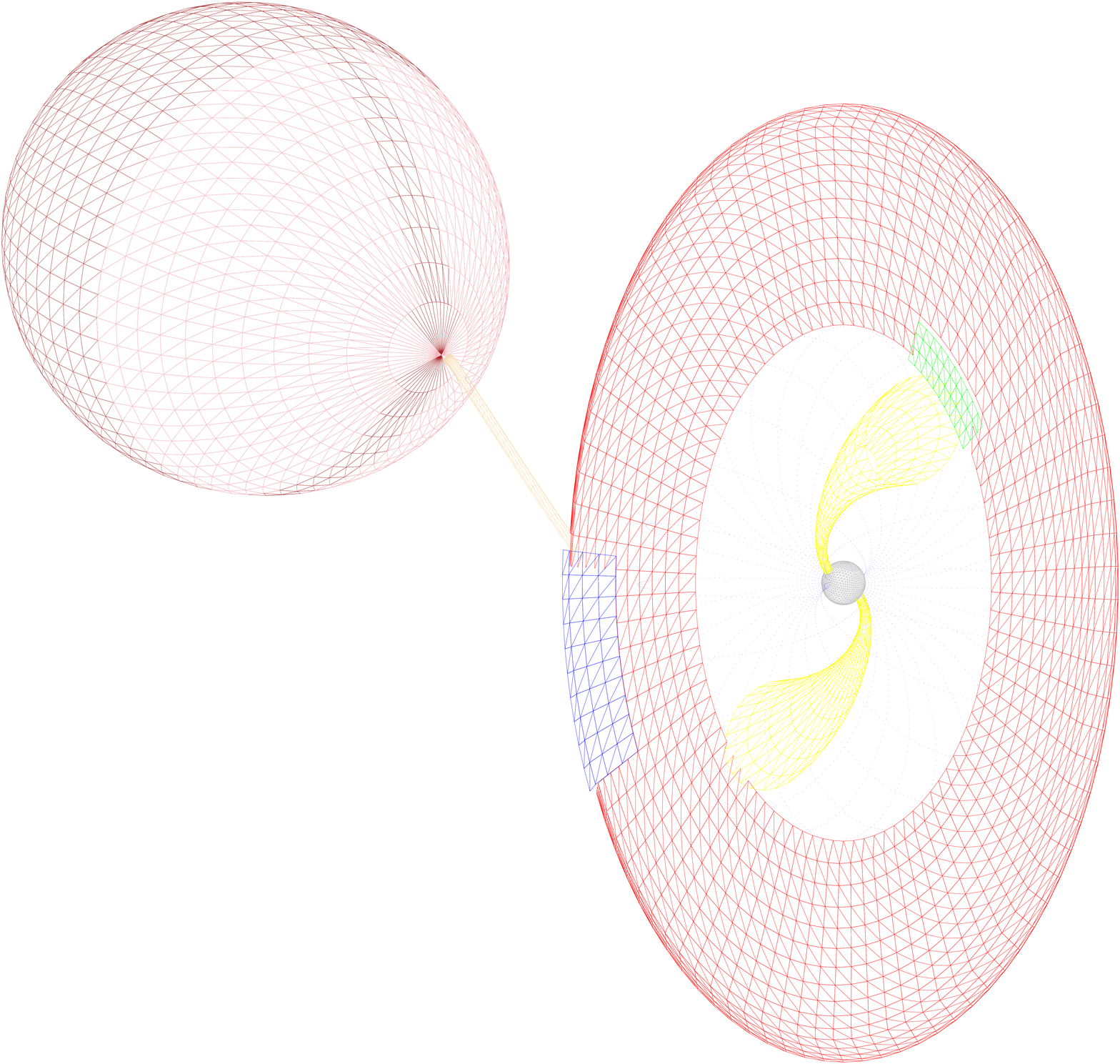}
\vspace{5 mm}
\caption{A schematic representation of the suggested geometry and components for
\fsaur. The magnetic field of the WD disrupts the inner accretion disc and forces
the accreting material to flow along field lines onto magnetic poles, forming arc-shaped,
azimuthally extended accretion curtains. The spin component of the emission lines observed
in an ordinary IP originates in a radiating gas which flows along the curtains (yellow) and
has the radial infall velocity. In \fsaur, however, the radial velocity variations
with the short spin period are smeared out during the LSP. Instead, a spin-averaged
X-ray beam from \fsaur's WD illuminates different segments of the accretion disc when
sweeping around with the precession (LSP) period. These segments  (green) reprocess
high-energy emission into the optical and have the Keplerian velocity. It crosses the line
of sight in front of the WD at the optical LSP pulse maximum when the upper precessional
pole points to the observer. The inner hemisphere
of the secondary is the most probable candidate for being both the source of the orbital
modulation and the reprocessing site of the sideband, LPP modulation.}
\label{fig:Model}
\end{figure}

Nevertheless, there is a not very common characteristic of \fsaur\ which is worth
to be mentioned: the relative phasing of the
photometric and radial velocity modulations with the LSP. In contrast to most IPs,
in which the spin component appearing in emission line wings has maximum blueshift
near optical pulse maximum -- phase 0 \citep{Hellier1999}, in \fsaur\ the LSP component
has maximum blueshift near phase 0.75. The latter results in counter-clockwise
rotation of the corresponding Doppler map (Fig.~\ref{fig:dopmap}) though otherwise
it closely resembles those of AO~Psc and FO~Aqr \citep{Hellier1999}. The location
of the emission area at ''3 o'clock'' in the IP tomograms and the occurring of maximum
blueshift at phase 0 agree well with the accretion curtain model of IPs \citep{Hellier1991}.
The different spectral phasing of the LSP modulation in \fsaur\ indicates that instead
of seeing a radiating gas flowing down the magnetic field lines on to the surface
of the WD, as follows from the accretion curtain model, in \fsaur\ we observe a
segment of the innermost region of the accretion disc moving with the local Keplerian
velocity. This disagreement should not be considered a surprise if we admit that
in fact the LSP is not the rotational period of the WD but the precession one. In
this case the radial velocity variations with the short spin period must be smeared
out during the LSP. However, a spin-averaged X-ray beam from \fsaur's WD would
illuminate different segments of the accretion disc when sweeping around with the
precession (LSP) period. These segments can reprocess
high-energy emission into the optical and should have the Keplerian velocities. Thus
the velocity modulation with the LSP would be phased with zero velocity at optical
maximum, as it is observed in \fsaur\ (Fig.~\ref{fig:Model}).

The LSP-folded trailed spectra and the corresponding Doppler map (Fig.~\ref{fig:dopmap})
allow the location of this emission source to be estimated. The brightest part of
the emission LSP-component reaches $\sim$800 \kms, with a possible extension up to
$\sim$1100 \kms (we note, however, that this higher velocity diffuse emission part
of the tomogram might be an artefact of the reconstruction). Assuming Keplerian
motion and adopting the system parameters given in \citet{Neustroev2002}, it implies
a radial distance from the WD of $8\times10^9$ cm ($\sim$8 WD radii), with a possible
extension inwards to about $4\times10^9$ cm ($\sim$4 WD radii). We suppose that this
is a characteristic distance at which the accretion disc of \fsaur\ is disrupted by
the magnetic field of the WD. It is somewhat shorter than supposed by the standard
picture of IPs ($\sim$10 WD radii). On the other hand, IPs with shorter spin periods
have smaller magnetospheres in which case the accretion discs are disrupted closer to
the WD.

\section{Summary and Conclusions}

We have presented a comprehensive study of \fsaur\ based on two extensive sets of
optical photometric observations and three X-ray data sets.

In addition to the formerly observed LPP and OP variations, these optical observations
revealed, for the first time in photometric data, the presence of the LSP modulation,
previously seen only spectroscopically. The LSP which is the presumed precession period
of the WD, is best seen in the ($B-I$) colour index derived from the 2010-2011 multicolour
observations but it was also detected in the pre-whitened light curve from the 2004-2005
observational campaign. The analysis of X-ray observations made with Chandra and Swift,
also revealed the existence of both the OP, LPP and LSP modulations.

We compared most reliable observable properties of the LSP pulsation such as energy
dependence and phasing of the LSP pulses in different wavelengths, with those of the
spin modulation of ten ironclad and confirmed IPs. We have found strong indications
that the LSP signal detected in \fsaur's X-rays and optical photometry and spectroscopy
is similar in nature to the spin modulation of the IPs. We conclude that even though
some of observational properties of \fsaur\ are not very common, they are by no means
unique to IPs.
The most serious discrepancy is the relative phasing of the photometric and radial
velocity modulations with the LSP. In \fsaur\ the LSP spectral component has zero
velocity at optical maximum, and it supports the precessional origin of the LSP
variability (Figure~\ref{fig:Model}).

In conclusion, we would like to mention a methodological aspect of this work.
There is a long lasting discussion about a necessity of multicolour observations
of CVs rather than unfiltered or single-filtered ones. Indeed, most time-resolved
observations of variable stars nowadays are taken with relatively small telescopes,
and the use of several filters is often considered as losing photons and wasting
time. However, our analysis of multicolour observations of IPs has shown that
time-series analysis of colour indices appears to be a powerful technique for revealing
hidden variabilities and shedding light on their nature. For example, the ($B-I$)
power spectrum of V1223 Sgr indicates the presence in the data of the spin pulsation
which is not seen in the optical light curve at all. Also, the analysis of the
colour indices of V455 And revealed the presence in the photometrical data of the
LSP pulsation which, similarly to \fsaur, was previously observed only spectroscopically.
Our analysis has shown that the use of colour indices of bands more distant in the
spectrum, e.g. ($B-I$), can be more efficient. We note that the ($B-I$) colour index
is a more sensitive estimator of the effective stellar temperatures with respect
to the widely used ($B-V$) \citep{B-Iindex}.

\section*{Acknowledgments}

The authors thank Seathr\'un \'O Tuairisg, Caoilfhionn Lane, Gennady Valyavin,
Byeong-Cheol Lee and the staff of the Bohyunsan Optical Astronomy Observatory (BOAO)
for assisting with the observations in 2004,
Mike Rice and the staff at New Mexico Skies for their invaluable support,
George B. and Elma Sjoberg who cared so much,
and Natalia Neustroeva for help in the preparation of this paper.
SZ and  GT acknowledge PAPIIT grants IN-109209/IN-103912  and CONACyT
grants 34521-E; 151858 for resources provided towards this research.
Our research was based on X-ray observations by NASA missions Chandra and Swift
which we acknowledge. We thank Neil Gehrels for approving the Target of Opportunity
observation with Swift and the Swift team for executing the observation. This research
has made use of data obtained through the High Energy Astrophysics Science Archive
Research Center Online Service, provided by the NASA/Goddard Space Flight Center.
We acknowledge photometric data kindly provided to us by our colleagues
Kunegunda Belle, Albert Bruch, Oliver Butters, Jerry Foote, Boris G{\"a}nsicke,
Coel Hellier, Keith Horne, Seppo Katajainen, Andrew J. Norton, Stephen Potter,
Edward L. Robinson.
We are thankful to the anonymous referee for careful reading of the manuscript.

\bsp
\label{lastpage}
\end{document}